\documentstyle[aps,psfig,bbold]{revtex}

\newcommand{\bra}[1]{\mbox{$\langle #1|$}}
\newcommand{\ket}[1]{\mbox{$|#1\rangle$}}
\newcommand{\brar}[1]{\mbox{$\langle #1\|$}}
\newcommand{\ketr}[1]{\mbox{$\|#1\rangle$}}

\newcommand{\beq}{\begin{equation}}
\newcommand{\eeq}{\end{equation}}
\newcommand{\beqa}{\begin{eqnarray}}
\newcommand{\eeqa}{\end{eqnarray}}

\begin{document}
\preprint{MKPH-T-00-24}

\title{
\hfill{\small {\bf MKPH-T-00-24}}\\
{\bf Polarization observables in elastic electron deuteron scattering 
including parity and time reversal violating contributions}\footnote[2]
{Supported by the Deutsche Forschungsgemeinschaft (SFB 443).}
}
\author
{Hartmuth Arenh\"ovel$^{(a)}$ and Shri K.\ Singh$^{(b)}$}
 \address{ $^{(a)}$Institut f\"ur Kernphysik,           
  Johannes Gutenberg-Universit\"at,  
  D-55099 Mainz, Germany }
 \address{ $^{(b)}$Department of Physics, 
 Aligarh Muslim University, 
 Aligarh, India }
\date{\today}
\maketitle
\begin{abstract}
\noindent
The general formalism for polarization 
observables in elastic electron deuteron scattering is extended to 
incorporate parity and time reversal violating contributions. Parity 
violating effects arise from the interference of $\gamma$ and $Z$ 
exchange as well as from the hadronic sector via a small 
parity violating component in the deuteron. In addition we have allowed for 
time reversal invariance violating contributions in the hadronic sector. 
Formal expressions for the additional structure functions are derived, and
their decomposition into the various multipole contributions are given 
explicitly.
\end{abstract}


\section{Introduction}\label{intro}
The study of polarization observables in electroweak (e.w.) reactions is an 
important tool in order to investigate small but interesting dynamical 
effects, which normally are buried under the dominant amplitudes in 
unpolarized total and differential cross sections, but which often may 
show up significantly in certain polarization observables. The reason 
for this feature lies in the fact that such small amplitudes or small 
contributions to large amplitudes may be amplified by interference with 
dominant amplitudes, or that dominant amplitudes interfere destructively 
leaving thus more room to the small amplitudes. For example, this fact 
has been exploited in elastic electron deuteron scattering in order to 
disentangle the charge quadrupole form factor from the monopole one 
by measuring the tensor 
asymmetry $T_{20}$ or equivalently the tensor recoil polarization $P_{20}$. 
Other prominent examples are the measurement of 
parity violation of the e.w.\ interaction, and the study of $T$-noninvariant 
form factors in the same process. 

A quite thorough discussion of polarization observables of elastic 
electron-deuteron scattering in the one-photon-approximation has been given 
by Gourdin and Piketty~\cite{GoP64} and by Schildknecht~\cite{Sch65} for 
the case of parity (P) and time reversal (T) invariant 
currents. The consequences of P violating contributions from weak
neutral currents on certain polarization observables for this process have 
been considered previously by several authors~\cite{Ram67,RaS78,MuR79,FrH91}. 
Furthermore, the influence of T-violation on the vector recoil polarization 
has been treated in~\cite{Sch66,DuC66,PrS68}. However, it seems 
that no systematic formalism for polarization observables has been 
established for electroweak scattering including weak neutral currents 
arising from $Z$ exchange. 
It is the aim of the present paper, to give a comprehensive 
and systematic derivation of all polarization observables for this reaction 
including parity and time reversal invariance violating contributions. 
To this end, we first review briefly in Sect.~\ref{formal} the basic 
ingredients for elastic electron scattering in the one-boson-exchange 
approximation. The general definition of a polarization observable is given in 
Sect.~\ref{defpolobs}, while explicit expressions in terms of structure 
functions and form factors are derived in Sect.~\ref{defstrucfun}. Also the 
corresponding beam, target and beam-target asymmetries are given there. 
Various details are presented in several appendices. 

\section{Basic Formalism}\label{formal}
In this section we briefly present the basic formalism for elastic electron 
deuteron scattering in the one-boson-exchange approximation including $Z$ 
exchange. The general expression for any observable, i.e., cross section 
and recoil polarization including the dependence on beam and target 
polarization, is given by 
\begin{eqnarray}
\label{GL1}
{\cal O}_X\,d\sigma_{fi} & = & (2\pi)^{-2} \delta^{(4)}(d'-q-d)
\,tr({\cal M}^\dagger_{fi}\,\widehat O_X\,{\cal M}_{fi}\hat \rho^e \hat 
\rho^d) \,\frac{m_e^2\,d^{3}k_2}{4k_{1,\, 0}k_{2,\,0}} 
\,\frac{d^3d'}{2M_{d}E_{d}'}\,,
\end{eqnarray}
where the observable ${\cal O}_X$ is characterized by a subscript $X$, 
which refers to the various polarizations of the final deuteron state. 
It is represented by an appropriate operator $\hat O_X$ and will 
be specified later. 
The momenta of the initial and scattered electrons (mass $m_e$)
are denoted by $k_1$ and $k_2$, respectively, and 
$q_{\mu}^2=q_0^2-\vec q^{\, 2}$ the four momentum transfer squared 
($q=k_1-k_2$). The initial and final deuteron momenta are denoted by 
$d=(E_d,\vec d)$ and $d'=(E_d',\vec d')$, respectively, and the 
deuteron mass by $M_d$. The density 
matrices $\hat \rho^e$ and $\hat \rho^d$ 
describe possible beam and target polarization. 
Covariant normalization has been assumed, i.e., $(2\pi)^{3} E/m$
for fermions and $(2\pi)^{3}2E$ for bosons. 

The amplitude ${\cal M}_{fi}$ 
contains in the lowest order, i.e., in the
one-boson-exchange approximation, contributions from both virtual $\gamma$ 
and $Z$ exchange 
with the latter naturally being 
strongly suppressed since we restrict ourselves to the low 
momentum transfer region ($-q_\mu^2 \ll M^2_Z$).  
The invariant matrix element thus contains two contributions 
\cite{MuD92}
\begin{eqnarray}
 {\cal M}_{fi}&=& \frac{e^2}{q_\mu^2}\ j^{(\gamma)\,\mu}J_{fi,\,\mu}^{(\gamma)}
 + \sqrt{2}\widetilde G_F \,j^{(Z)\,\mu}J_{fi,\,\mu}^{(Z)} \,.
\end{eqnarray}
Here and in the following, the superscripts $\gamma$ and $Z$ indicate the 
electromagnetic and weak neutral current contributions. The lepton and 
hadron currents are denoted by $j^{(\gamma/Z)}_{\mu}$ and 
$J_{fi,\,\mu}^{(\gamma/Z)}$, respectively. Furthermore, $e$ denotes the 
elementary charge with $\alpha=e^2/4\pi$ as fine structure constant, and 
$\widetilde G_F$ is related to the weak Fermi coupling constant $G_F$ by 
\begin{eqnarray} 
\widetilde G_F(q_\mu^2)& = & \frac{M_Z^2}{M_Z^2-q_\mu^2}\,G_F 
=\frac{\sqrt{2}\,g^2}{8\cos^2\theta_W \,(M_Z^2-q_\mu^2)} \,,
\end{eqnarray}
where $g$ denotes the electroweak coupling constant, $\theta_W$ the Weinberg 
angle, and $e=g\sin \theta_W$. 
 
The lepton currents are defined by 
\begin{eqnarray}
j^{(\gamma)\,\mu}&=& j^{(v)\,\mu}\,,\\
j^{(Z)\,\mu}&=&  g^e_v\,j^{(v)\,\mu}+ g^e_a\,j^{(a)\,\mu}\,,
\end{eqnarray}
where we have introduced the lepton vector and axial currents by
\begin{eqnarray}
j^{(v)\,\mu}&=& \bar u(k_2)\,\gamma^\mu\, u(k_1)\,,\\
j^{(a)\,\mu}&=& \bar u(k_2)\,\gamma^\mu \gamma_5\,u(k_1)\,.
\end{eqnarray}
Furthermore, one has 
\begin{eqnarray}
g^e_{v}&=&-\frac{1}{2} + 2\sin^2\theta_W\,,\\
g^e_{a}&=& \frac{1}{2}\,.
\end{eqnarray}
Note, that our expressions for the neutral currents contain an additional 
factor $1/2$ compared to Ref.\ \cite{MuD92}. The hadronic current $J_\mu$ is 
specified later. However, for formal reasons it is convenient to distinguish
the contributions arising from the coupling to the lepton vector and 
axial currents by introducing
\beqa
J_{fi,\,\mu}({\cal V}) &=& J_{fi,\,\mu}^{(\gamma)} 
                +J_{fi,\,\mu}^{(Z^{\cal V})}\,,\label{J_V}\\
J_{fi,\,\mu}({\cal A}) &=& J_{fi,\,\mu}^{(Z^{\cal A})}\,,
\label{J_A}
\eeqa
where
\beq
J_{fi,\,\mu}^{(Z^{{\cal V}/{\cal A}})}=\widetilde G_{v/a}\,J_{fi,\,\mu}^{(Z)}
\eeq
with
\beq\label{Gtilde}
\widetilde G_{v/a} = \sqrt{2}\,g^e_{v/a}\,\widetilde G_F \,q_\mu^2\,e^{-2}\,.
\eeq
We would like to emphasize, that the argument ${\cal V}$ and ${\cal A}$ 
merely indicates to which type of lepton current the hadronic current couples. 
Both hadronic currents, $J_{fi,\,\mu}({\cal V})$ as well as 
$J_{fi,\,\mu}({\cal A})$, contain vector and axial pieces 
(see below Eqs.~(\ref{JcalV}) and (\ref{JcalA})).
Then the invariant matrix element takes the form
\begin{eqnarray}
 {\cal M}_{fi}&=& \frac{e^2}{q_\mu^2}\Big( j^{(v)\,\mu}\,J_{fi,\,\mu}({\cal V})
 + j^{(a)\,\mu}\,J_{fi,\,\mu}({\cal A})\Big) \,.
\end{eqnarray}

Allowing for longitudinal electron polarization of degree $h$, one then finds
\begin{eqnarray}
\frac{m_e^2}{M_d^2}\,tr({\cal M}^\dagger_{fi}\,\widehat O_X\,
{\cal M}_{fi}\hat \rho^e \hat \rho^d)
&=& \Big(\frac{e^2}{q_\mu^2}\Big)^2 
\Big[\eta_{\mu\nu}^{vv}(h)\,\Big(W^{{\cal VV},\,\mu\nu}_{fi}
(\widehat O_X,\,\hat \rho^d)
+ W^{{\cal AA},\,\mu\nu}_{fi}(\widehat O_X,\,\hat \rho^d)\Big)\nonumber\\
&&+\eta_{\mu\nu}^{va}(h)\,\Big(W^{{\cal VA},\,\mu\nu}_{fi}
(\widehat O_X,\,\hat \rho^d)
+ W^{{\cal AV},\,\mu\nu}_{fi}(\widehat O_X,\,\hat \rho^d)\Big)\Big]\,,
\label{traceM}
\end{eqnarray}
where one has two types 
of lepton tensors $\eta_{\mu\nu}^{v v}$ and 
$\eta_{\mu\nu}^{v a}$, where the latter arises from the 
interference of the lepton vector with the lepton axial current, 
\begin{eqnarray}
\eta_{\mu\nu}^{v v}(h) &=& \eta_{\mu\nu}^0 + h \eta_{\mu\nu}^{\prime}\,,\\
\eta_{\mu\nu}^{v a}(h) &=& \eta_{\mu\nu}^{\prime} + h \eta_{\mu\nu}^{0}\,.
\end{eqnarray}
In the high energy limit, i.e., electron mass $m_e=0$, one has 
\begin{eqnarray}
\eta_{\mu\nu}^0 &=& (k_{1\,\mu} k_{2\,\nu} + k_{2\,\mu} k_{1\,\nu})
- g_{\mu \nu} k_1\cdot k_2\nonumber\\
&=& \frac{1}{2}(k_\mu k_\nu - q_\mu q_\nu + g_{\mu\nu} q_\rho^2)
\,,\\
\eta_{\mu\nu}^{\prime} &=& i \varepsilon_{\mu \nu \alpha \beta} 
k_2^{\alpha} k_1^{\beta}\nonumber\\
&=& \frac{i}{2} \varepsilon_{\mu \nu \alpha \beta}k^\alpha q^\beta\,,
\end{eqnarray}
where $k=k_1+k_2$. 
The hadronic tensors, appearing in (\ref{traceM}), are defined by  
\begin{eqnarray}
W^{{\cal C'C},\,\mu\nu}_{fi}(\widehat O_X,\,\hat \rho^d) 
&=& \frac{1}{M_d^2}\,tr(J_{fi}^{\mu\,\ast}({\cal C'})\,\widehat O_X\,
J_{fi}^{\nu}({\cal C})\hat \rho^d)\,,
\end{eqnarray}
where ${\cal C'},\,{\cal C} \in \{{\cal V},{\cal A}\}$, and the trace refers 
to the deuteron spin quantum numbers. 

\section{Definition of a general polarization observable}\label{defpolobs}

Proceeding as in the electromagnetic case by switching to the usual 
three-dimensional representation of the lepton tensors in terms of virtual 
boson density matrices, one obtains in analogy to the pure electromagnetic 
process the following expression for an observable 
\begin{eqnarray}
{\cal O}_X\,\frac{d\sigma^{\gamma +Z}}{d \Omega_{k_2}^{lab} 
} =  \frac{2\alpha^2}{q_\mu^4} \,
\Big(\frac{ k_2^{lab}}{k_1^{lab}}\Big)^2\, 
\sum_{\lambda, \lambda^{\prime}} \sum_{ m',m, n',n} 
\rho_{m n}^d  
&&\Big[ (\rho_{\lambda \lambda^{\prime}}^{0}+
h\rho_{\lambda \lambda^{\prime}}^{\prime})  
\sum_{{\cal C} \in \{{\cal V},{\cal A}\}}
t_{n'\lambda' n}^{\ast}({\cal C})(\widehat O_X)_{n'm'} 
t_{m'\lambda m}({\cal C})
\nonumber \\
 & & +(h\rho_{\lambda \lambda^{\prime}}^{0}+
\rho_{\lambda \lambda^{\prime}}^{\prime})\,  
\sum_{{\cal C'}\neq {\cal C}\in \{{\cal V},{\cal A}\}}
t_{n'\lambda' n}^{\ast}({\cal C'}) (\widehat O_X)_{n'm'}
t_{m'\lambda m}({\cal C})\Big]\,.
\label{obs_general}
\end{eqnarray}
Here, we have introduced the $t$-matrices, which are 
related to the various current matrix elements between 
the intrinsic deuteron states by
\begin{eqnarray}
 t_{m'\lambda m}({\cal C}) & = &  \frac{\sqrt{E_d' E_d}}{M_d}\,
 \bra{m'} J_{\lambda}({\cal C}) \ket{m} \,.\label{tmatrix}
\end{eqnarray}
The current components refer to a coordinate system with $z$-axis along 
$\vec q$, $y$-axis along $\vec k_1\times \vec k_2$, i.e., perpendicular 
to the scattering plane, and $x$-axis chosen as to form a right-handed 
system, i.e., $\hat x=\hat y\times\hat z$. Also the deuteron spin states
refer to this system with $\vec q$ as quantization axis. 
Thus $\lambda = \pm 1$ refers to the transverse current components (with 
respect to $\vec q\,$), while the $\lambda = 0$ component is given by a 
combination of charge and longitudinal current component
\begin{eqnarray}
J_0 &=& -\frac{|\vec q\,|^2}{q_\mu^2}(\rho - \frac{\omega}{|\vec q\,|^2} 
\vec q \cdot \vec J)\nonumber\\
&=& \rho - \frac{\omega }{q_\mu^2}(\omega\rho - \vec q \cdot \vec J)\,,
\end{eqnarray}
which reduces to the charge density $\rho$ for a conserved current. 
Furthermore, $E_d$ and $E_d'$ denote the initial and final deuteron energies, 
respectively. The c.m.\ motion of the 
initial and final deuteron states with c.m.\ momenta $\vec d$ and 
$\vec d'$, respectively, has been eliminated and we have switched to 
noncovariant normalization. 

The spherical components of the two types of virtual boson density matrices 
obey the symmetry relations
\begin{eqnarray}
 \rho_{\lambda \lambda^{\prime}}^{0/\prime} & = & 
\rho_{\lambda^{\prime} \lambda}^{0/\prime} \ ,\\
 \rho_{-\lambda -\lambda^{\prime}}^0 & = & (-)^{\lambda +\lambda^{\prime}}
    \rho_{\lambda \lambda^{\prime}}^0 \ , \\
 \rho_{-\lambda -\lambda^{\prime}}^{\prime} & = & 
 (-)^{\lambda +\lambda^{\prime}+1}\rho_{\lambda \lambda^{\prime}}^{\prime} \ .
\end{eqnarray}
Here, $\rho^{0/\prime}$ can be expanded into independent components with 
respect to diagonal longitudinal ($L$) and transverse ($T$) 
contributions, and interference terms ($LT$ and $TT$)
\beqa
\rho_{\lambda \lambda'}^{0/\prime}&=&\sum_{\alpha \in\{ L,\, T,\, LT,\, TT\}}
\delta^{(\prime)\,\alpha}_{\lambda \lambda'}\rho^{(\prime)}_\alpha\,,
\eeqa
with
\begin{equation}
\begin{array}{ll}
\delta^{L}_{\lambda \lambda'}=\delta_{\lambda \lambda'}\delta_{\lambda 0}\,,
\quad &\delta^{LT}_{\lambda \lambda'}=\lambda'\delta_{\lambda 0}
+\lambda\delta_{\lambda' 0}\,,\cr
 & \cr
\delta^{T}_{\lambda \lambda'}=\delta_{\lambda \lambda'}|\lambda|\,, &
\delta^{TT}_{\lambda \lambda'}=\delta_{\lambda,\, -\lambda'}|\lambda|\,,\cr
 & \cr
\delta^{\prime\,L}_{\lambda \lambda'}=0\,, &
\delta^{\prime\,LT}_{\lambda \lambda'}=|\lambda'|\delta_{\lambda 0}
+|\lambda|\delta_{\lambda' 0}\,,\cr
 & \cr
\delta^{\prime\,T}_{\lambda \lambda'}=\delta_{\lambda \lambda'}\lambda\,, &
\delta^{\prime\,TT}_{\lambda \lambda'}=0\,.\cr
\end{array}
\end{equation}
The nonvanishing components are 
\begin{eqnarray}
\begin{array}{ll}
 \rho_L=\rho_{00}^0=-\beta^2 q_{\nu}^2\frac{\xi^2}{2\zeta} 
\,,\quad& \rho_T=\rho_{11}^0
  =-\frac{1}{2}q_{\nu}^2\,\Big(1+\frac{\xi}{2 \zeta} \Big) \,,\cr
 \rho_{LT}=\rho_{01}^0=-\beta q_{\nu}^2 \frac{\xi}{\zeta}\,
 \sqrt{\frac{\zeta+ \xi}{8}}
\, ,& \rho_{TT}=\rho_{-11}^0=q_{\nu}^2\frac{\xi}{4 \zeta} \,,\cr
 \rho_{LT}^{\prime}=\rho_{01}^{\prime}=
 -\frac{1}{2}\,\beta\frac{q_{\nu}^2}{\sqrt{2\zeta}}\,\xi \,,\quad&
 \rho_T^{\prime}=\rho_{11}^{\prime}=
  -\frac{1}{2}q_{\nu}^2\, \sqrt{\frac{\zeta+\xi}{\zeta}} \, ,\cr
\end{array}
\end{eqnarray}
with 
\begin{eqnarray}
\beta = {|{\vec q}^{\,lab}| \over |{\vec q}^{\,c}|}\,,\quad
\xi = -\frac{q_{\nu}^2}{|{\vec q}^{\,lab}|^2} \,, \quad 
\zeta = \tan^2\frac{\theta_e}{2}\ ,
\end{eqnarray}
where $\beta$ expresses the boost from the lab system to the frame in which 
the hadronic tensor is evaluated and ${\vec q}^{\,c}$ denotes the 
momentum transfer in this frame. In order to make contact to 
the kinematic functions $v_{\alpha^{(\prime)}}$ in the review of Musolf et 
al.\ \cite{MuD94}, we note the simple relation (for $\beta=1$)
\begin{eqnarray}
\rho_\alpha^{(\prime)} &=& -\frac{q_\mu^2}{2\zeta}\, v_{\alpha^{(\prime)}}\,,
\label{valpha}
\end{eqnarray}
where $\alpha \in\{ L,\, T,\, LT,\, TT\}$.

Now we will discuss the various hadronic tensors of (\ref{obs_general}) 
in detail. The hadronic currents can be classified according to 
their vector and axial current contributions. The e.m.\ current contains 
only a vector piece $J_{fi,\,\mu}^{\gamma}$ while the neutral current 
consists of both, vector and axial parts, $J_{fi,\,\mu}^{Z_v}$ and 
$J_{fi,\,\mu}^{Z_a}$, respectively. Thus for the hadron current interacting 
with the lepton vector current $J_{fi,\,\mu}({\cal V})$ one has 
$J_{fi,\,\mu}^{\gamma/Z^{\cal V}_v}$ as vector part and 
$J_{fi,\,\mu}^{Z^{\cal V}_a}$ as axial part, i.e., 
\beqa
J_{fi,\,\mu}({\cal V})
&=&J_{fi,\,\mu}^{\gamma}+\widetilde G_v\,(J_{fi,\,\mu}^{Z_v}
                       +J_{fi,\,\mu}^{Z_a})\nonumber\\
&=&J_{fi,\,\mu}^{\gamma}+J_{fi,\,\mu}^{Z^{\cal V}_v}
                       +J_{fi,\,\mu}^{Z^{\cal V}_a}\,.\label{JcalV}
\eeqa
The corresponding contributions to the hadron current 
$J_{fi,\,\mu}({\cal A})$ interacting with the lepton axial current are 
$J_{fi,\,\mu}^{Z^{\cal A}_v}$ and $J_{fi,\,\mu}^{Z^{\cal A}_a}$, respectively, 
\beqa
J_{fi,\,\mu}({\cal A})
&=&\widetilde G_a\,(J_{fi,\,\mu}^{Z_v} +J_{fi,\,\mu}^{Z_a})\nonumber\\
&=&J_{fi,\,\mu}^{Z^{\cal A}_v}+J_{fi,\,\mu}^{Z^{\cal A}_a}\,.\label{JcalA}
\eeqa
Note, that $J_{fi,\,\mu}^{Z^{\cal V}_{v/a}}$ and 
$J_{fi,\,\mu}^{Z^{\cal A}_{v/a}}$ are related by the ratio of $g^e_v/g^e_a$, i.e.,
\beq
J_{fi,\,\mu}^{Z^{\cal V}_v}= g^e_v/g^e_a\,J_{fi,\,\mu}^{Z^{\cal A}_v}
\quad\mbox{ and }\quad
J_{fi,\,\mu}^{Z^{\cal V}_a}= g^e_v/g^e_a\,J_{fi,\,\mu}^{Z^{\cal A}_a}\,.
\eeq
Thus $J_{fi,\,\mu}^{Z^{\cal V}_{v/a}}$ will be suppressed compared to 
$J_{fi,\,\mu}^{Z^{\cal A}_{v/a}}$.
Since we allow also for parity violation in 
the hadronic states, any current matrix element can be split into two 
contributions with opposite parity transformation properties, i.e.,
\beq
J_{fi}^c= J_{fi}^{c_{pc}}+J_{fi}^{c_{pnc}}\,,
\eeq
where, denoting the dominant component by an upper index ``$pc$'' and the 
small, parity violating component of opposite parity by ``$pnc$''
\beqa
J_{fi}^{c_{pc}}&=& _{pc}\bra{f} J \ket{i}_{pc}
+ {_{pnc}\bra{f}} J \ket{i}_{pnc}\,,\\
J_{fi}^{c_{pnc}}&=& _{pnc}\bra{f} J \ket{i}_{pc}
+ {_{pc}\bra{f}} J \ket{i}_{pnc}\,,
\eeqa
where $|\rangle_{pc}$ denotes the dominant parity conserving wave function 
component and $|\rangle_{pnc}$ the small parity violating component.
Thus, in order to classify the various contributions, we will define two 
symbolic index sets ${\cal C}_{\cal V}$ and ${\cal C}_{\cal A}$ according to 
the interaction with the lepton vector and axial currents, respectively, by
\beqa
{\cal C}_{\cal V}&=&\{\gamma_{pc},\,\gamma_{pnc},\,Z^{\cal V}_{v,\,pc},\,
Z^{\cal V}_{v,\,pnc},\,Z^{\cal V}_{a,\,pc},\,Z^{\cal V}_{a,\,pnc}\}\,,\\
{\cal C}_{\cal A}&=&\{Z^{\cal A}_{v,\,pc},\,Z^{\cal A}_{v,\,pnc},\,
Z^{\cal A}_{a,\,pc},\,Z^{\cal A}_{a,\,pnc}\}\,.
\eeqa
It is also convenient to introduce two other sets of current contributions 
according to their behaviour under parity transformations, whether they 
transform like a vector or like an axial current. They are defined by
\beqa
{\cal C}_{pc}&=&\{\gamma_{pc},\,Z^{\cal V}_{v,\,pc},\,Z^{\cal V}_{a,\,pnc},\,
             Z^{\cal A}_{v,\,pc},\,Z^{\cal A}_{a,\,pnc}\}\,,\\
{\cal C}_{pnc}&=&\{\gamma_{pnc},\,Z^{\cal V}_{v,\,pnc},\,Z^{\cal V}_{a,\,pc},\,
             Z^{\cal A}_{v,\,pnc},\,Z^{\cal A}_{a,\,pc}\}\,.
\eeqa
In order to characterize the opposite behaviour with respect to parity, 
we will introduce a symbolic $\delta$-function by
\beq
\delta^P_{c}:=\left\{\matrix{0 & \mbox{for}\;
c\in {\cal C}_{pc} \cr 1 &
\mbox{for}\; c\in {\cal C}_{pnc} \cr} \right\}\,.
\eeq
Furthermore, in order to be more general we will also allow for violation 
of time reversal invariance. Consequently, we will split each of the two 
sets ${\cal C}_{pc}$ and ${\cal C}_{pnc}$ into two subsets, one containing the 
contributions which respect time reversal invariance and the other 
those violating it, labeled in addition by ``$tc$'' and ``$tnc$'', 
respectively,
\beqa
{\cal C}_{pc}&=& {\cal C}_{pc,\, tc}\cup{\cal C}_{pc,\,tnc}\,,\\
{\cal C}_{pnc}&=& {\cal C}_{pnc,\,tc}\cup{\cal C}_{pnc,\,tnc}\,,
\eeqa
where the four different sets are given by 
\beqa
{\cal C}_{pc,\, tc}&=&\{\gamma_{pc,\, tc},\,Z^{\cal V}_{v,\,pc,\, tc},\,
Z^{\cal V}_{a,\,pnc,\, tc,}\,Z^{\cal A}_{v,\,pc,\, tc},\,
Z^{\cal A}_{a,\,pnc,\, tc}\}\,,\label{setpctc}\\
{\cal C}_{pnc,\, tc}&=&\{\gamma_{pnc,\, tc},\,Z^{\cal V}_{v,\,pnc,\, tc},\,
Z^{\cal V}_{a,\,pc,\, tc},\,Z^{\cal A}_{v,\,pnc,\, tc},\,
Z^{\cal A}_{a,\,pc,\, tc}\}\,,\label{setpnctc}\\
{\cal C}_{pc,\, tnc}&=&\{\gamma_{pc,\, tnc},\,Z^{\cal V}_{v,\,pc,\, tnc},\,
Z^{\cal V}_{a,\,pnc,\, tnc,}\,Z^{\cal A}_{v,\,pc,\, tnc},\,
Z^{\cal A}_{a,\,pnc,\, tc}\}\,,\label{setpctnc}\\
{\cal C}_{pnc,\, tnc}&=&\{\gamma_{pnc,\, tnc},\,Z^{\cal V}_{v,\,pnc,\, tnc},\,
Z^{\cal V}_{a,\,pc,\, tnc},\,Z^{\cal A}_{v,\,pnc,\, tnc},\,
Z^{\cal A}_{a,\,pc,\, tnc}\}\,.\label{setpnctnc}
\eeqa
Correspondingly, in order to characterize the opposite transformation 
behaviour under time reversal we introduce
\beq
\delta^T_{c}:=\left\{\matrix{0 & \mbox{for}\;
c\in {\cal C}_{pc,\, tc}\cup{\cal C}_{pnc,\, tc} \cr 1 &
\mbox{for}\; c\in {\cal C}_{pc,\,tnc}\cup{\cal C}_{pnc,\, tnc} \cr} \right\}\,.
\eeq
As a shorthand, we will use
\beq
\delta_c^{PT} = \delta_c^P+\delta_c^T\,.
\eeq
Now we write the $t$-matrix element of (\ref{tmatrix}) as a sum of the 
various current contributions labeled by a superscript ``$c$''
\beqa
t_{m'\lambda m}({\cal V}/{\cal A}) & = & 
\sum_{c\in {\cal C}_{{\cal V}/{\cal A}}} 
t^{c}_{m'\lambda m}\,,
\eeqa
and obtain for the hadronic current tensors in (\ref{obs_general})
\beqa
\sum_{{\cal C} \in \{{\cal V},{\cal A}\}}
t_{n'\lambda n}^{\ast}({\cal C})\,t_{m'\lambda ^{\prime} m}({\cal C})
&=& \sum_{{\cal C} = ({\cal C}_{\cal V},\,{\cal C}_{\cal A})}
\sum_{c',\,c\in {\cal C}}t^{c'\,\ast}_{n'\lambda' n}\,t^{c}_{m'\lambda m}\,
\nonumber \\
\sum_{{\cal C'}\neq {\cal C}\in \{{\cal V},{\cal A}\}}
t_{n'\lambda n}^{\ast}({\cal C'})\,t_{m'\lambda ^{\prime} m}({\cal C})
&=& \sum_{c'\in {\cal C}_{\cal V}}
\sum_{c\in {\cal C}_{\cal A}}\Big(t^{c'\,\ast}_{n'\lambda' n}\,
t^{c}_{m'\lambda m}+(c\leftrightarrow c')\Big)\,
\,.\label{t*t}
\eeqa
Any of these current matrix elements $t^{c}_{m'\lambda m}$ can be expanded 
into multipoles
\beqa
t^{c}_{m'\lambda m}&=& (-)^\lambda\,a_\lambda\,\frac{\sqrt{E_d' E_d}}{M_d}
 \sum_L i^L\hat L \bra{1m'}{\cal O}^\lambda_{L\lambda}(c)\ket{1m}\nonumber\\
&=&(-)^{1-m'+\lambda}\,a_\lambda\, \sum_L i^L\hat L 
\left(\matrix{1 & L & 1 \cr -m' & \lambda & m \cr}\right)
{\cal O}^\lambda_{L}(c)\,,
\label{multipole}
\eeqa
where $a_\lambda=\sqrt{2\pi(1+\delta_{\lambda 0})}$, and 
\beq
{\cal O}^\lambda_{LM} = \delta_{\lambda 0}\,{\cal C}_{LM} + 
 \delta_{|\lambda| 1}\,({\cal E}_{LM} + \lambda\,{\cal M}_{LM})\,
\eeq
denotes a general multipole. 
The argument ``$c$'' of the multipole ${\cal O}^\lambda_{L}(c)$ in 
(\ref{multipole}) 
indicates the current contribution. In (\ref{multipole}) we have chosen the 
direction of the momentum transfer $\vec q$ as quantization axis for the 
deuteron spin states and have introduced for the reduced matrix elements of 
the multipole operators betwee the deuteron states the notation 
\beqa
{\cal O}^\lambda_{L}(c) &=&\frac{\sqrt{E_d' E_d}}{M_d}\,
\brar{1}{\cal O}^\lambda_{L}(c)\ketr{1}
\nonumber\\
&=& \delta_{\lambda 0}\,C_{L}(c) + 
 i\,\delta_{|\lambda| 1}\,(E_{L}(c) + \lambda\,M_{L}(c))\,.
\eeqa
Here the factor $\sqrt{E_d' E_d}/M_d$ has been included for convenience in the 
definition of the reduced charge $(C_{L}(c))$ and transverse 
$(E_{L}(c),\,M_{L}(c))$ matrix elements. Furthermore, 
a factor ``$i$'' has been separated from the transverse multipoles in 
order to have $E_L$ and $M_L$ as real quantities, because one has 
$({\cal O}^\lambda_{L}(c))^*=(-)^\lambda{\cal O}^\lambda_{L}(c)$ 
(see Appendix A).
From time reversal one has the following selection rules for the multipoles 
\beq
C_{L}(c): (-)^{L+\delta^T_{c}}=1\,,\quad E/M_{L}(c): (-)^{L+\delta^T_{c}}=-1\,.
\eeq
On the other hand, the parity transformation yields as selection rules
\beqa
(C/E)_{L}(c): (-)^{L+\delta^P_{c}}=1,\,\quad 
M_{L}(c): (-)^{L+\delta^P_{c}}=-1 \,.
\eeqa
Combining these selection rules, one finds as nonvanishing multipole 
contributions
\beq
\begin{array}{ll}
C_0(c),\,C_2(c),\,M_1(c) & \quad \mbox{for}\quad  c\in {\cal C}_{pc,\, tc}
\,, \cr 
E_2(c) & \quad \mbox{for}\quad c\in {\cal C}_{pc,\,tnc}\,, \cr
E_1(c) & \quad \mbox{for}\quad c\in {\cal C}_{pnc,\,tc}\,, \cr
C_1(c),\,M_2(c) & \quad \mbox{for}\quad  c\in {\cal C}_{pnc,\,tnc}
\,. \cr 
\end{array}
\eeq

Before proceeding further, we have to specify the observable $X$ in 
(\ref{GL1}) describing any observable for the analysis 
of the final target spin state. We choose the representation 
$X= (IM\pm)$ $(I=0,1,2,\,M\ge 0)$ with a corresponding hermitean 
operator in deuteron spin space
\beq
\widehat O_{IM{\rm sig}_M}=c_{M{\rm sig}_M}(\tau^{[I]}_M 
+{\rm sig}_M (-)^M \tau^{[I]}_{-M})\,,
\eeq
with 
\beq
c_{M{\rm sig}_M}=\left\{\matrix{
\frac{1}{1+\delta_{M0}} & \mbox{ for }{\rm sig}_M=+\,,\cr
i & \mbox{ for }{\rm sig}_M= -\,.\cr}\right.\label{cMsigM}
\eeq
Here we have introduced a sign function by ${\rm sig}_M:=\pm$, where 
the subscript $M$ merely indicates to which variable it refers. One 
should note, that obviously for $(IM{\rm sig}_M)=(I0-)$ the operator 
vanishes, i.e., $\widehat O_{I0-}=0$. 

The irreducible tensors $\tau^{[I]}$ are the usual statistical tensors
for the parametrization of the density matrix of a spin-one particle
\beq
\rho^d = \frac{1}{3}\sum_{I=0}^2\sum_{M=-I}^I 
\tau^{[I]}_{M}\,P^{d\,\ast}_{IM}\,,
\eeq
where $P^{d}_{IM}$ characterizes the initial state polarization with 
$P^d_{00}=1$. 

The tensors $\tau^{[I]}$ are normalized as 
$\langle 1 ||\tau^{[I]}||1\rangle = \sqrt{3}\,\widehat I$, where 
$\widehat I=\sqrt{2I+1}$, i.e., in detail
\beq
\tau^{[I]} = \left\{
\matrix{ {\mathbb 1}_3 & \mbox{ no polarization,}\cr
     \sqrt{\frac{3}{2}}\,S^{[1]} & \mbox{ vector polarization,}\cr
     \sqrt{3}\,S^{[2]} & \mbox{ tensor polarization,}\cr}\right.
\label{deuttensor}
\eeq
where ${\mathbb 1}_3$ is the unit matrix, $S^{[1]}$ the spin-one operator, 
and $S^{[2]}=[S^{[1]}\times S^{[1]}]^{[2]}$ the tensor operator whose 
cartesian components are 
\beq
S^{[2]}_{kl}=\frac{1}{2}(S_kS_l+S_lS_k)-\frac{2}{3}\,\delta_{kl}\,.
\eeq
Using the relation of the $\widehat O_{IM{\rm sig}_M}$ to the cartesian 
spin operators
\beq
\begin{array}{ll}
S_{x/y}=\mp \frac{1}{\sqrt{3}}\widehat O_{11\pm}\,,&
S_{z}= \sqrt{\frac{2}{3}}\widehat O_{10+}\,,\cr
S^{[2]}_{xx/yy}=\pm\frac{1}{2\sqrt{3}}\widehat O_{22+}
- \frac{1}{3\,\sqrt{2}}\widehat O_{20+}\,,\qquad&
S^{[2]}_{zz}=\frac{\sqrt{2}}{3}\,\widehat O_{20+}\,,\cr
S^{[2]}_{xy}=-\frac{1}{2\sqrt{3}}\widehat O_{22-}\,,&
S^{[2]}_{zx/zy}=\mp\frac{1}{2\sqrt{3}}\widehat O_{21\pm}\,,
\end{array}
\eeq
one finds for the relation of the above defined observables 
${\cal O}_{IM{\rm sig}_M}$ to the cartesian spin observables $P_{k}$ and 
$P_{kl}$
\beq
\begin{array}{ll}
P_{x/y}=\mp \frac{1}{\sqrt{3}}{\cal O}_{11\pm}\,,&
P_{z}= \sqrt{\frac{2}{3}}{\cal O}_{10+}\,,\cr
P_{xx/yy}=\pm\frac{1}{2\sqrt{3}}{\cal O}_{22+}
- \frac{1}{3\,\sqrt{2}}{\cal O}_{20+}\,,\qquad&
P_{zz}=\frac{\sqrt{2}}{3}\,{\cal O}_{20+}\,,\cr
P_{xy}=-\frac{1}{2\sqrt{3}}{\cal O}_{22-}\,,&
P_{zx/zy}=\mp\frac{1}{2\sqrt{3}}{\cal O}_{21\pm}\,,
\end{array}
\eeq
where the cartesian observables are defined by the deuteron density matrix 
in the form
\beq
\rho^d=\frac{1}{3}\,({\mathbb 1}_3+\vec P\cdot\vec S 
+\sum_{kl}P_{kl}S^{[2]}_{kl})\,.
\eeq

From now on we will 
assume that the density matrix is diagonal with respect to a certain 
orientation axis, characterized by spherical angles $\theta_d$ and $\phi_d$. 
Then one can write
\beq
P^{d}_{IM}=P_I^d e^{iM\phi_d}d^I_{M0}(\theta_d)\,,
\eeq
with the deuteron vector ($P_1^d$) and tensor ($P_2^d$) polarization 
parameters which are related to the occupation probabilities $p_m$ of the 
different spin projection states of the deuteron with respect to the 
orientation axis as quantization axis by
\beqa
 P^d_1 &=& P^d_{10} = \sqrt{\frac{3}{2}}(p_1-p_{-1})\,, \\
 P^d_2 &=& P^d_{20} = \frac{1}{\sqrt{2}}\left(3(p_1+p_{-1})-2\right)\,. 
\eeqa

\section{Structure Functions and Asymmetries}\label{defstrucfun}

Inserting these various expressions into (\ref{obs_general}), one obtains 
finally for an observable $X=(I'M'{\rm sig}_{M'})$ in terms of four
types of structure functions $F_\alpha^{(\prime)\,IM{\rm sig}_M}(X)$ 
and $\widetilde F_\alpha^{(\prime)\,IM{\rm sig}_M}(X)$
\beqa
{\cal O}_{X}\,\frac{d\sigma^{\gamma +Z}}{d \Omega_{k_2}^{lab} 
} &=&  \sigma_{\rm{Mott}}\,\sum _{I=0}^2 P_I^d \sum _{M=0}^I\,
\sum_{{\rm sig}_{M}=\pm}\,
\cos\Big(M\phi_d+\frac{\pi}{4}(1-{\rm sig}_M 1)\Big)\,
d^I_{M0}(\theta_d)\nonumber\\
&&\times \sum_{\alpha\in\{L,\,T,\,LT,\,TT\}}\,
\Big[v_\alpha\Big( F_\alpha^{IM{\rm sig}_M}(X)
+h\widetilde F_\alpha^{IM{\rm sig}_M}(X)\Big) 
+v_\alpha'\Big(h F_\alpha^{\prime\,IM{\rm sig}_M}(X)+
\widetilde F_\alpha^{\prime\,IM{\rm sig}_M}(X)\Big)
\Big]\,,\label{genobs} 
\eeqa
where we have introduced the Mott cross section as given by
\beq
\sigma_{\rm{Mott}}=\frac{\alpha^2\,\cos^2\frac{\theta_e^{lab}}{2}}{4\,
\sin^4\frac{\theta_e^{lab}}{2}}\,\frac{k_2^{lab}}{(k_1^{lab})^3}\,,
\eeq
and switched to the $v_\alpha^{(\prime)}$'s instead of the 
$\rho_\alpha^{(\prime)}$'s according to (\ref{valpha}). Their explicit form is
\begin{eqnarray}
\begin{array}{ll}
 v_L=\frac{\beta^2}{(1+\eta)^2} 
\,,\quad& v_T=\frac{1}{2\,(1+\eta)}+\tan^2\frac{\theta_e^{lab}}{2} \,,\cr
 v_{LT}=\frac{1}{\sqrt{2}}\,\frac{ \beta}{ 1+\eta}\,\sec\frac{\theta_e^{lab}}{2}\,\sqrt{\frac{1+\eta\,\sin^2\frac{\theta_e^{lab}}{2}}{1+\eta}}\,,
& v_{TT}=- \frac{1}{2\,(1+\eta)}\,,\cr
 v_{LT}^{\prime}=\frac{1}{\sqrt{2}}\,
\frac{ \beta}{ 1+\eta}\,\tan\frac{\theta_e^{lab}}{2}\,
,\quad&
 v_T^{\prime}=\sec\frac{\theta_e^{lab}}{2}\,\tan\frac{\theta_e^{lab}}{2}\,
\sqrt{\frac{1+\eta\,\sin^2\frac{\theta_e^{lab}}{2}}{1+\eta}} \, ,\cr
\end{array}
\end{eqnarray}
with 
\beq
\eta=\frac{Q^2}{4\,M_d^2}\,\qquad q^2=-Q^2\,.
\eeq
The structure functions are defined by
\beqa
F_\alpha^{(\prime)\,IM{\rm sig}_M}(X)&=&
\sum_{{\cal C} \in \{{\cal C}_{\cal V},\,{\cal C}_{\cal A}\}}
\sum_{c',\,c\in {\cal C}}f_\alpha^{(\prime)\,IM{\rm sig}_M}(X;\,c',c)\,,
\label{defF}\\
\widetilde F_\alpha^{(\prime)\,IM{\rm sig}_M}(X)&=&
2\sum_{c'\in {\cal C_V},\,c\in {\cal C_A}}
f_\alpha^{(\prime)\,IM{\rm sig}_M}(X;\,c',c)\,.\label{defFbar}
\eeqa
Here the various current contributions 
$f_\alpha^{(\prime)\,IM{\rm sig}_M}(X;\,c',c)$ are given in terms of 
the quantities 
${\cal O}_{\alpha,\,I'M'}^{(\prime)\,IM}(c',c)$ defined below by 
\beq
f_\alpha^{(\prime)\,IM{\rm sig}_M}(X;\,c',c)=c_{M'{\rm sig}_{M'}}
\Big({\cal O}_{\alpha,\,I'M'}^{(\prime)\,IM{\rm sig}_M}(c',c)
+{\rm sig}_{M'}\,(-)^{M'}
{\cal O}_{\alpha,\,I'-M'}^{(\prime)\,IM{\rm sig}_M}(c',c)
\Big)\,,
\eeq 
with
\beq
{\cal O}_{\alpha,\,I'M'}^{(\prime)\,IM{\rm sig}_M}(c',c)=c_{M{\rm sig}_M}
\Big({\cal O}_{\alpha,\,I'M'}^{(\prime)\,IM}(c',c)
+{\rm sig}_M\,(-)^M{\cal O}_{\alpha,\,I'M'}^{(\prime)\,I-M}(c',c)
\Big)\,.
\eeq
The basic quantities are related to the $t$-matrix elements according to
\beq
{\cal O}_{\alpha,\,I'M'}^{(\prime)\,IM}(c',c)= 
\sum_{\lambda',\lambda}\delta^{(\prime)\,\alpha}_{\lambda\lambda'}
{\cal U}^{\lambda'\lambda\,IM}_{I'M'}(c',c)\,,
\eeq
where the ${\cal U}$'s are quadratic hermitean forms in the $t$-matrix elements
\beq
{\cal U}^{\lambda'\lambda\,IM}_{I'M'}(c',c)=\frac{1}{6}\,
\sum_{n',n,m',m} \Big(
t^{c'\,\ast}_{n'\lambda' n}\,(\tau^{[I']}_{M'})_{n'm'}\,t^{c}_{m'\lambda m}\,
(\tau^{[I]}_{M})_{m n} + (c'\leftrightarrow c)\Big)\,.
\eeq
Angular momentum conservation leads to the selection rule 
\beq
\lambda'-\lambda=M'+M\,. \label{srprojections}
\eeq
Note, that by definition ${\cal U}$ and thus the structure functions are 
symmetric under the interchange $(c\leftrightarrow c')$. Furthermore, one 
has the following symmetry properties 
\beqa
\Big({\cal U}^{\lambda'\lambda\,IM}_{I'M'}(c',c)\Big)^\ast&=&
(-)^{\delta^T_{c}+\delta^T_{c'}}\,{\cal U}^{\lambda'\lambda\,IM}_{I'M'}(c',c)
\,,\label{uast}\\
{\cal U}^{\lambda\lambda'\,IM}_{I'M'}(c',c)&=&(-)^{M+M'}\,
\Big({\cal U}^{\lambda'\lambda\,I-M}_{I'-M'}(c',c)\Big)^\ast\,,
\label{symA}\\
{\cal U}^{-\lambda'-\lambda\,I-M}_{I'-M'}(c',c)&=&
(-)^{\delta^{PT}(c',c)+I+I'}\,
\Big({\cal U}^{\lambda'\lambda\,IM}_{I'M'}(c',c)\Big)^\ast\,,\label{symB}\\
{\cal U}^{\lambda'\lambda\,IM}_{I'M'}(c',c)&=&
(-)^{I+M+I'+M'+\delta^T_{c}+\delta^T_{c'}}\,
{\cal U}^{\lambda'\lambda\,I'M'}_{IM}(c',c)\,,\label{symD}
\eeqa
where we have introduced
\beq
\delta^{PT}(c',c)=\delta^{PT}_{c'}+\delta^{PT}_{c}\,.
\eeq
The first two can be combined to yield the symmetry
\beqa
{\cal U}^{-\lambda'-\lambda\,IM}_{I'M'}(c',c)&=&
(-)^{\delta^{PT}(c',c)+I+M+I'+M'}\,
{\cal U}^{\lambda\lambda'\,IM}_{I'M'}(c',c)\,.\label{symC}
\eeqa
These symmetries are derived in the Appendix A, where we also give a closed
expression for ${\cal U}^{\lambda\lambda'\,IM}_{I'M'}(c',c)$ in terms of the 
reduced multipole matrix elements. Furthermore, by a proper choice of the 
phases for the state vectors in order to have simple time reversal properties 
one can make all ${\cal U}^{\lambda\lambda'\,IM}_{I'M'}(c',c)$'s real or 
imaginary depending on whether $(-)^{\delta^T_{c}+\delta^T_{c'}}=\pm 1$, 
respectively, as also shown in the Appendix A. Then one finds corresponding 
simple symmetries for the ${\cal O}_{\alpha,\,I'M'}^{(\prime)\,IM}(c',c)$ 
\beqa
\Big({\cal O}^{(\prime)\,IM}_{\alpha,\,I'M'}(c',c)\Big)^\ast&=&
(-)^{\delta^T_{c}+\delta^T_{c'}}\,{\cal O}^{(\prime)\,IM}_{\alpha,\,I'M'}(c',c)
\,,\label{oast}\\
{\cal O}^{(\prime)\,IM}_{\alpha,\,I'M'}(c',c)&=&(-)^{M+M'}\,
\Big({\cal O}^{\,I-M}_{\alpha,\,I'-M'}(c',c)\Big)^\ast\,,
\label{symAa}\\
{\cal O}^{(\prime)\,IM}_{\alpha,\,I'M'}(c',c)&=&\pm
(-)^{\delta^{PT}(c',c)+I+I'}\,
{\cal O}^{(\prime)\,IM}_{\alpha,\,I'M'}(c',c)\,,\label{symCc}
\eeqa
where the minus sign in (\ref{symCc}) refers to the primed quantity 
${\cal O}^{\prime\,IM}_{\alpha,\,I'M'}(c',c)$.
For the interchange $(IM)\leftrightarrow (I'M')$ one has 
\beqa
{\cal O}^{(\prime)\,IM}_{\alpha,\,I'M'}(c',c)&=&
(-)^{I+M+I'+M'+\delta^T_{c}+\delta^T_{c'}}\,
{\cal O}^{(\prime)\,I'M'}_{\alpha,\,IM}(c',c)\,.\label{symDd}
\eeqa
The symmetry property of (\ref{symCc}) leads to the interesting selection rule
\beq
{\cal O}^{(\prime)\,IM}_{\alpha,\,I'M'}(c',c)=\frac{1}{2}\,
\Big(1\pm(-)^{\delta^{PT}(c',c)+I+I'}\Big)\,
{\cal O}^{(\prime)\,IM}_{\alpha,\,I'M'}(c',c)\,,
\eeq
which means that 
\beq
\begin{array}{ll}
f_\alpha^{IM{\rm sig}_M}(I'M'{\rm sig}_{M'};\,c',c) =0\quad & \mbox{ for }  
(-)^{\delta^{PT}(c',c)+I+I'}= -1\,,\cr
f_\alpha^{\prime\,IM{\rm sig}_M}(I'M'{\rm sig}_{M'};\,c',c) =0 \quad & 
\mbox{ for } (-)^{\delta^{PT}(c',c)+I+I'}= 1\,.\cr
\end{array}
\eeq 
Another selection rule follows from (\ref{oast}) and (\ref{symAa}) yielding
\beq
f_\alpha^{(\prime)\,IM{\rm sig}_M}(I'M'{\rm sig}_{M'};\,c',c)=
c_{M'{\rm sig}_{M'}}
\Big(1+{\rm sig}_{M}{\rm sig}_{M'} (-)^{\delta^T_{c}+\delta^T_{c'}}\Big)\,
{\cal O}_{\alpha,\,I'M'}^{(\prime)\,IM{\rm sig}_M}(c',c)\,.
\eeq
Therefore, only for 
${\rm sig}_M={\rm sig}_{M'}(-)^{\delta^T_{c}+\delta^T_{c'}}$ one has a 
nonvanishing contribution. Combining these two selection rules and 
introducing as a shorthand
\beq
\delta^{(\prime)\,I,\,I'}_{{\rm sig}_{M}{\rm sig}_{M'}}(c',c)=\frac{1}{4}\,
\Big(1+{\rm sig}_{M}{\rm sig}_{M'} (-)^{\delta^T_{c}+\delta^T_{c'}}\Big)
\Big(1\pm(-)^{\delta^{PT}(c',c)+I+I'}\Big)\,,\label{deltaIc}
\eeq 
one obtains
\beq
f_\alpha^{(\prime)\,IM{\rm sig}_M}(I'M'{\rm sig}_{M'};\,c',c)=
2\,\delta^{(\prime)\,I,\,I'}_{{\rm sig}_{M}{\rm sig}_{M'}}(c',c)\,
c_{M'{\rm sig}_{M'}}
{\cal O}_{\alpha,\,I'M'}^{(\prime)\,IM{\rm sig}_M}(c',c)\,.
\eeq
The remaining nonvanishing functions are listed in Table~\ref{nonvanishing_f}.
In detail, one finds for them (note that by definition 
$M,\,M'\ge 0$)
\beqa
f_L^{IM{\rm sig}_M}(I'M'{\rm sig}_{M'};\,c',c)&=&
2\,\delta^{I,\,I'}_{{\rm sig}_{M}{\rm sig}_{M'}}(c',c)\,
c_{M{\rm sig}_M}\,c_{M'{\rm sig}_{M'}}\, 
\delta_{M'M}\Big(\delta_{M0}+{\rm sig}_M(-)^M\Big)\,\,
{\cal U}^{00\,I-M}_{I'M}(c',c)\,,\label{f_L}\\
f_T^{(\prime)\,IM{\rm sig}_M}(I'M'{\rm sig}_{M'};\,c',c)&=&
4\,\delta^{(\prime)\,I,\,I'}_{{\rm sig}_{M}{\rm sig}_{M'}}(c',c)\,
c_{M{\rm sig}_M}\,c_{M'{\rm sig}_{M'}}\, 
\delta_{M'M}\Big(\delta_{M0}+{\rm sig}_M(-)^M\Big)\,\,
{\cal U}^{11\,I-M}_{I'M}(c',c)\,,\label{f_T}\\
f_{LT}^{(\prime)\,IM{\rm sig}_M}(I'M'{\rm sig}_{M'};\,c',c)&=&
-4\,\delta^{(\prime)\,I,\,I'}_{{\rm sig}_{M}{\rm sig}_{M'}}(c',c)\,
c_{M{\rm sig}_M}\,c_{M'{\rm sig}_{M'}}\, 
\Big({\rm sig}_M(-)^{M+\delta^T_{c}+\delta^T_{c'}}\delta_{M',\,M+1}\,\,
{\cal U}^{01\,IM}_{I'-M-1}(c',c)\nonumber\\
&&
+((-)^{\delta^T_{c}+\delta^T_{c'}}\delta_{M',\,1-M}
-{\rm sig}_M(-)^M\delta_{M',\,M-1})\,\,
{\cal U}^{01\,I-M}_{I'M-1}(c',c)\Big)\,,\label{f_LT}\\
f_{TT}^{IM{\rm sig}_M}(I'M'{\rm sig}_{M'};\,c',c)&=&
2\,\delta^{I,\,I'}_{{\rm sig}_{M}{\rm sig}_{M'}}(c',c)\,
c_{M{\rm sig}_M}\,c_{M'{\rm sig}_{M'}}\, 
\Big({\rm sig}_M(-)^{M+\delta^T_{c}+\delta^T_{c'}}\delta_{M',\,M+2}\,\,
{\cal U}^{-11\,IM}_{I'-M-2}(c',c)\nonumber\\
&&
+((-)^{\delta^T_{c}+\delta^T_{c'}}\delta_{M',\,2-M}
+{\rm sig}_M(-)^M\delta_{M',\,M-2})\,\,
{\cal U}^{-11\,I-M}_{I'M-2}(c',c)\Big)\,.\label{f_TT}
\eeqa

The symmetry property in (\ref{symDd}) leads to a simple relation 
for the interchange $(IM{\rm sig}_M\leftrightarrow I'M'{\rm sig}_{M'})$ 
\beq
f_\alpha^{(\prime)\,IM{\rm sig}_M}(I'M'{\rm sig}_{M'};\,c',c)=
(-)^{I+M+I'+M'+\delta^T_{c}+\delta^T_{c'}}
f_\alpha^{(\prime)\,I'M'{\rm sig}_{M'}}(IM{\rm sig}_M;\,c',c)\,,
\label{intchangef}
\eeq
which relates the $f$-functions for a given target and recoil polarization 
to the corresponding ones, where the target and recoil polarizations have 
been interchanged. Thus this symmetry reduces the number of independent 
structure functions considerably and gives an additional selection rule 
for $(IM{\rm sig}_M)=(I'M'{\rm sig}_{M'})$
\beq
f_\alpha^{(\prime)\,IM{\rm sig}_M}(IM{\rm sig}_{M};\,c',c)=0
\eeq
for $\delta^T_{c}+\delta^T_{c'}=1$. 
Another symmetry exists for the structure functions with $M> 0$ and 
$M'> 0$ for the simultaneous sign change 
${\rm sig}_M \rightarrow -{\rm sig}_M$ and 
${\rm sig}_{M'}\rightarrow -{\rm sig}_{M'}$. First we note from (\ref{cMsigM}) 
the property
\beq
c_{M\,-{\rm sig}_M}= i\,{\rm sig}_M\,c_{M{\rm sig}_M}\,,
\eeq
from which follows for the above transformation
\beq
c_{M{\rm sig}_M}\,c_{M'{\rm sig}_{M'}}\rightarrow -{\rm sig}_M\,{\rm sig}_{M'}
\,c_{M{\rm sig}_M}\,c_{M'{\rm sig}_{M'}}\,.\label{transcMsigM}
\eeq
Secondly, the invariance of 
$\delta^{I,\,I'}_{{\rm sig}_{M}{\rm sig}_{M'}}(c',c)$ is evident. Finally, from
the formal expressions in (\ref{f_L}) through (\ref{f_TT}) one notes that 
for $M> 0$ and $M'> 0$ 
$f_\alpha^{(\prime)\,IM{\rm sig}_M}(I'M'{\rm sig}_{M'};\,c',c)$ is proportional
to ${\rm sig}_M$ for $\alpha\in\{L,T,LT\}$, whereas for $\alpha=TT$ it is 
independent because in this case only $M=M'=1$ gives a nonvanishing 
contribution to (\ref{f_TT}) according to Table~\ref{M-values}. Thus with 
(\ref{transcMsigM}) and the equivalence ${\rm sig}_M\,{\rm sig}_{M'}\equiv 
(-)^{\delta_{c'}^T+\delta_{c}^T}$ implied by 
$\delta^{I,\,I'}_{{\rm sig}_{M}{\rm sig}_{M'}}(c',c)$, one finds
\beqa
f_\alpha^{(\prime)\,IM\,-{\rm sig}_M}(I'M'\,-{\rm sig}_{M'};\,c',c)&=&
(-)^{\delta_{c'}^T+\delta_{c}^T}\,
f_\alpha^{(\prime)\,IM{\rm sig}_M}(I'M'{\rm sig}_{M'};\,c',c)\,,\label{symsiga}
\eeqa
for $\alpha\in\{L,T,LT\}$ and 
\beqa
f_{TT}^{(\prime)\,IM\,-{\rm sig}_M}(I'M'\,-{\rm sig}_{M'};\,c',c)&=&
-(-)^{\delta_{c'}^T+\delta_{c}^T}\,
f_{TT}^{(\prime)\,IM{\rm sig}_M}(I'M'{\rm sig}_{M'};\,c',c)\,.\label{symsigb}
\eeqa
Explicit expressions for the nonvanishing $f$-functions for the various 
current contributions for the case of recoil polarization without target 
polarization are listed in the Appendix B. The ones for target polarization 
without analysis of the recoil polarization can be obtained from these by 
using the symmetry in (\ref{intchangef}).

Besides the dominant $P$- and $T$-invariance obeying contribution 
$c=\gamma_{pc,\,tc}$, we now will restrict ourselves to the first order 
contributions with respect to the weak coupling constant and to leading 
order $T$-violation. In other words, 
of all the contributions of ${\cal C}_{\cal V}$ and 
${\cal C}_{\cal A}$ the following zero and first order contributions remain
\beq
\begin{array}{ll}
{\cal C}_{\cal V}^{(0)}=\Big\{\gamma_{pc,\,tc}\Big\}\,,&
{\cal C}_{\cal A}^{(0)}=\emptyset\,,\cr
{\cal C}_{\cal V}^{(1)}=\Big\{\gamma_{pnc,\,tc},\,\gamma_{pc,\,tnc},\,
Z^{\cal V}_{v,\,pc,\,tc},\,Z^{\cal V}_{a,\,pc,\,tc}\Big\}\,,&
{\cal C}_{\cal A}^{(1)}=\Big\{Z^{\cal A}_{v,\,pc,\,tc},\,
Z^{\cal A}_{a,\,pc,\,tc}\Big\}\,,\cr
\end{array}
\eeq
or with respect to the other classification of (\ref{setpctc}) through 
(\ref{setpnctnc})
\beq
\begin{array}{ll}
{\cal C}_{pc,\,tc}^{(0)}=\Big\{\gamma_{pc,\,tc}\Big\}\,,&
{\cal C}_{pc,\,tc}^{(1)}=\Big\{Z^{\cal V}_{v,\,pc,\,tc},\,
Z^{\cal A}_{v,\,pc,\,tc}\Big\}\,,\cr
{\cal C}_{pnc,\,tc}^{(1)}=\Big\{\gamma_{pnc,\,tc},\,Z^{\cal V}_{a,\,pc,\,tc},\,
Z^{\cal A}_{a,\,pc,\,tc}\Big\}\,,&
{\cal C}_{pc,\,tnc}^{(1)}=\Big\{\gamma_{pc,\,tnc}\Big\}\,,\cr 
{\cal C}_{pnc,\,tnc}^{(1)}=\emptyset\,. & \cr
\end{array}
\eeq
This means, one has to make the following substitutions in the general 
expression for the structure functions in (\ref{defF}) and (\ref{defFbar}) 
\beqa
\sum_{{\cal C} = ({\cal C}_{\cal V},\,{\cal C}_{\cal A})}
\sum_{c',\,c\in {\cal C}}&\rightarrow&\,
\sum_{c'=c=\gamma_{pc,\,tc}}+
\,2\sum_{c'=\gamma_{pc,\,tc},\,c\in {\cal C}_{\cal V}^{(1)}}\,,\\
\sum_{c'\in {\cal C_V},\,c\in {\cal C_A}}&\rightarrow&
\sum_{c'=\gamma_{pc,\,tc},\,c\in \{Z^{\cal A}_{v,\,pc,\,tc},
Z^{\cal A}_{a,\,pc,\,tc}\}}\,,
\eeqa
and obtains for the structure functions one diagonal $P$- and $T$-conserved 
contribution ($c=\gamma_{pc,\,tc}$) to 
$F_\alpha^{(\prime)\,IM{\rm sig}_M}(X)$ and three nondiagonal $P$- or 
$T$-violating ones, namely two $P$-violating contributions from the 
hadronic $P$-violation ($c=\gamma_{pnc,\,tc}$) and from the hadronic axial 
current coupled to the lepton vector current ($c=Z^{\cal V}_{a,\,pc,\,tc}$), 
and one hadronic $T$-violating contribution ($\gamma_{pc,\,tnc}$). Here 
and in the following, ``$X$'' stands always for an observable 
``$I'M'{\rm sig}_{M'}$''. The 
corresponding structure functions are determined either by the $P$- and 
$T$-conserved $f$-functions or by the $P$- or $T$-violating ones.  
In view of the selection rules for 
$\delta^{I,\,I'}_{{\rm sig}_{M}{\rm sig}_{M'}}(c',c)$ 
contained in (\ref{deltaIc}), one finds in detail 
for the $P$- and $T$-conserving structure functions for which 
${\rm sig}_M={\rm sig}_{M'}$ applies,
\beqa
F_\alpha^{\,IM{\rm sig}_M}(X)&=& 
f_\alpha^{IM{\rm sig}_M}(X;\,\gamma_{pc,\,tc},\gamma_{pc,\,tc})\nonumber\\&&
+2\,f_\alpha^{\,IM{\rm sig}_M}(X;\,\gamma_{pc,\,tc},
Z^{\cal V}_{v,\,pc,\,tc})
\,,
\quad \mbox{ for } I+I' \mbox{ even},\\
F_\alpha^{\,\prime\,IM{\rm sig}_M}(X)&=& 
f_\alpha^{\prime\,IM{\rm sig}_M}(X;\,\gamma_{pc,\,tc},\gamma_{pc,\,tc})
\nonumber\\&&
+2\,f_\alpha^{\prime\,IM{\rm sig}_M}(X;\,\gamma_{pc,\,tc},
Z^{\cal V}_{v,\,pc,\,tc})
\,,
\quad \mbox{ for } I+I' \mbox{ odd}\,.
\eeqa
Taking into account the proportionality of the neutral vector current to 
the e.m. current
\beq
J^{Z_v}_\mu =g_v^d\,J^{\gamma }_\mu\,,\label{neutral-em}
\eeq
where $g_v^d=-2\sin^2\theta_W$,
then one obtains
\beqa
F_\alpha^{\,IM{\rm sig}_M}(X)&=& (1+2\,g_v^d\,\widetilde G_v)\,
f_\alpha^{IM{\rm sig}_M}(X;\,\gamma_{pc,\,tc},\gamma_{pc,\,tc})
\,,
\quad \mbox{ for } I+I' \mbox{ even},\label{renormffa}\\
F_\alpha^{\,\prime\,IM{\rm sig}_M}(X)&=& (1+2\,g_v^d\,\widetilde G_v)\,
f_\alpha^{\prime\,IM{\rm sig}_M}(X;\,\gamma_{pc,\,tc},\gamma_{pc,\,tc})
\,,
\quad \mbox{ for } I+I' \mbox{ odd}\,,\label{renormffb}
\eeqa
which means a simple renormalization by a factor almost unity.
Furthermore, for the $P$-violating structure functions, for which also 
${\rm sig}_M={\rm sig}_{M'}$ applies, one has
\beqa
F_\alpha^{\,IM{\rm sig}_M}(X)&=& 
2\,f_\alpha^{\,IM{\rm sig}_M}(X;\,\gamma_{pc,\,tc},\gamma_{pnc,\,tc})
\nonumber\\&&
+2\,f_\alpha^{\,IM{\rm sig}_M}(X;\,\gamma_{pc,\,tc},
Z^{\cal V}_{a,\,pc,\,tc})
\quad \mbox{ for } I+I'  \mbox{ odd},\\
F_\alpha^{\,\prime\,IM{\rm sig}_M}(X)&=& 
2\,f_\alpha^{\prime\,IM{\rm sig}_M}(X;\,\gamma_{pc,\,tc},\gamma_{pnc,\,tc})
\nonumber\\&&
+2\,f_\alpha^{\prime\,IM{\rm sig}_M}(X;\,\gamma_{pc,\,tc},
Z^{\cal V}_{a,\,pc,\,tc})
\quad \mbox{ for } I+I' \mbox{ even}.
\eeqa
Finally, for the $T$-violating structure functions, for which 
${\rm sig}_M=-{\rm sig}_{M'}$ applies, one finds
\beqa
F_\alpha^{\,IM{\rm sig}_M}(X)&=& 
2\,f_\alpha^{\,IM{\rm sig}_M}(X;\,\gamma_{pc,\,tc},\gamma_{pc,\,tnc})
\quad \mbox{ for } I+I'  \mbox{ odd},\\
F_\alpha^{\,\prime\,IM{\rm sig}_M}(X)&=& 
2\,f_\alpha^{\prime\,IM{\rm sig}_M}(X;\,\gamma_{pc,\,tc},\gamma_{pc,\,tnc})
\quad \mbox{ for } I+I' \mbox{ even}.
\eeqa
To $\widetilde F_\alpha^{(\prime)\,IM{\rm sig}_M}(X)$ one has two 
nondiagonal $P$-violating contributions from the neutral hadron 
current, containing vector and axial pieces, coupled to the axial lepton 
current, i.e.,
\beqa
\widetilde F_\alpha^{\,IM{\rm sig}_M}(X)&=&
2\,f_\alpha^{IM{\rm sig}_M}(X;\,\gamma_{pc,\,tc},Z^{\cal A}_{v,\,pc,\,tc})
\quad \mbox{ for } I+I'  \mbox{ even},\\
\widetilde F_\alpha^{\,IM{\rm sig}_M}(X)&=&
2\,f_\alpha^{IM{\rm sig}_M}(X;\,\gamma_{pc,\,tc},Z^{\cal A}_{a,\,pc,\,tc})
\quad \mbox{ for } I+I'  \mbox{ odd},\\
\widetilde F_\alpha^{\,\prime\,IM{\rm sig}_M}(X)&=&
2\,f_\alpha^{\prime\,IM{\rm sig}_M}(X;\,\gamma_{pc,\,tc},
Z^{\cal A}_{v,\,pc,\,tc})
\quad \mbox{ for } I+I' \mbox{ odd},\\
\widetilde F_\alpha^{\,\prime\,IM{\rm sig}_M}(X)&=&
2\,f_\alpha^{\prime\,IM{\rm sig}_M}(X;\,\gamma_{pc,\,tc},
Z^{\cal A}_{a,\,pc,\,tc})
\quad \mbox{ for } I+I' \mbox{ even},
\eeqa
where again ${\rm sig}_M={\rm sig}_{M'}$ applies. Explicit expressions for 
the nonvanishing structure functions are listed in Appendix C.

It is useful to introduce scalar, vector and tensor target 
asymmetries $A_d^I(X)$ ($I=0,1,2$) and corresponding beam-target asymmetries 
$A_{ed}^I(X)$ which can be separated by a proper variation of the electron
polarization parameter $h$ and the target polarization parameters $P^d_I$. 
They are defined by 
\beqa
{\cal O}_{X}\,\frac{d\sigma^{\gamma +Z}}{d \Omega_{k_2}^{lab}} &=& 
\sigma_{\rm{Mott}}\,S_0\,
\Big[ A_d^0(X) + P_1^d\,A_d^1(X)+ P_2^d\,A_d^2(X) 
+h\,\Big(A_{ed}^0(X) + P_1^d\,A_{ed}^1(X)+ P_2^d\,A_{ed}^2(X)\Big)\Big]\,,
\eeqa
where 
\beq
S_0=v_L\,F_L^{00+}(1) + v_T\,F_T^{00+}(1)\,.
\eeq
Note that $A_d^0(X=1)=A_d^0(00+)=1$. 
Comparison with (\ref{genobs}) yields the following expressions
\beqa
A_{d/ed}^I(X)&=& \sum _{M=0}^I\,d^I_{M0}(\theta_d)\,\sum_{{\rm sig}_{M}}\,
\cos\Big(M\phi_d+\frac{\pi}{4}(1-{\rm sig}_M 1)\Big)\,A_d^{IM{\rm sig}_M}(X)\,,
\eeqa
or in detail
\beqa
A_{d/ed}^0(X)&=& A_{d/ed}^{00+}(X)\,,\\
A_{d/ed}^1(X)&=&\cos \theta_d\,A_{d/ed}^{10+}(X)
   -\frac{\sin \theta_d}{\sqrt{2}}\,
  \Big( \cos \phi_d\,A_{d/ed}^{11+}(X) - \sin \phi_d\,A_{d/ed}^{11-}(X)\Big)\,,
\\
A_{d/ed}^2(X)&=&\frac{1}{2}\,(3\,\cos \theta_d -1)\,A_{d/ed}^{20+}(X)
   -\sqrt{\frac{3}{2}}\,\sin \theta_d\,\cos \theta_d\,
   \Big( \cos \phi_d\,A_{d/ed}^{21+}(X) - \sin \phi_d\,A_{d/ed}^{21-}(X)\Big)
\nonumber\\
&&   +\frac{1}{2}\sqrt{\frac{3}{2}}\,\sin^2 \theta_d\,
   \Big( \cos 2\phi_d\,A_{d/ed}^{22+}(X) - \sin 2\phi_d\,A_{d/ed}^{22-}(X)\Big)
\,,
\eeqa
where we have separated explicitly the dependence on the angles of 
the deuteron orientation axis by introducing  
\beqa
A_d^{IM{\rm sig}_M}(X)&=&\frac{1}{S_0}\,
\sum_{\alpha\in\{L,\,T,\,LT,\,TT\}}\,\Big[v_\alpha
F_\alpha^{IM{\rm sig}_M}(X)+v_\alpha^{\,\prime}
\widetilde F_\alpha^{\prime\,IM{\rm sig}_M}(X)\Big]\,,\label{Asym-d}\\
A_{ed}^{IM{\rm sig}_M}(X)&=&\frac{1}{S_0}\,
\sum_{\alpha\in\{L,\,T,\,LT,\,TT\}}\,\Big[v_\alpha
\widetilde F_\alpha^{IM{\rm sig}_M}(X)+v_\alpha^{\,\prime}
F_\alpha^{\prime\,IM{\rm sig}_M}(X)\Big]\,.\label{Asym-ed}
\eeqa

The latter asymmetries $A_{e/ed}^{IM{\rm sig}_M}(X)$ can be separated by a 
proper choice of the orientation angles $\theta_d$ and $\phi_d$. 
We list explicit expressions of all asymmetries for ${\rm sig}_M=+$ 
in the Appendix D except for those which can be obtained from the symmetry in
(\ref{intchangef}), i.e, from
\beq
A_{d/ed}^{(\prime)\,IM{\rm sig}_M}(I'M'{\rm sig}_{M'})=
(-)^{I+M+I'+M'+\delta_t}
A_{d/ed}^{(\prime)\,I'M'{\rm sig}_{M'}}(IM{\rm sig}_M)\,,
\eeq
where $\delta_t=1$ for the $T$-violating contributions, and $\delta_t=0$ else.
The ones for ${\rm sig}_M=-$ can be obtained with the help 
of the relations in (\ref{symsiga}) and (\ref{symsigb}) yielding
\beqa
A_d^{IM-{\rm sig}_M}(I'M'{\rm sig}_{M'})&=&\frac{1}{S_0}\,
\sum_{\alpha\in\{L,\,T,\,LT,\,TT\}}\,(-)^{\delta_{\alpha,TT}+\delta_t}\,
\Big[v_\alpha
F_\alpha^{IM{\rm sig}_M}(I'M'-{\rm sig}_{M'})
+v_\alpha^{\,\prime}
\widetilde F_\alpha^{\prime\,IM{\rm sig}_M}(I'M'-{\rm sig}_{M'})\Big]
\nonumber\\
&=&(-)^{\delta_t}\,\Big(A_d^{IM{\rm sig}_M}(I'M'-{\rm sig}_{M'}) 
-\frac{2}{S_0}\,v_{TT}\,F_{TT}^{IM{\rm sig}_M}(I'M'-{\rm sig}_{M'})\Big)
\,,\label{asymsiga}\\
A_{ed}^{IM-{\rm sig}_M}(I'M'{\rm sig}_{M'})
&=&(-)^{\delta_t}\,\Big(A_{ed}^{IM{\rm sig}_M}(I'M'-{\rm sig}_{M'}) 
-\frac{2}{S_0}\,v_{TT}\,
\widetilde F_{TT}^{IM{\rm sig}_M}(I'M'-{\rm sig}_{M'})\Big)
\,.\label{asymsigb}
\eeqa

With respect to the explicit expressions of Appendix~D one should keep in mind 
the relation (\ref{neutral-em}) of the neutral hadronic vector current of the 
deuteron $J^{Z_v}_\mu$, which 
means that the $P$- and $T$-conserving form factors of $J^{Z_v}_\mu$ are 
proportional to the corresponding e.m.\ form factors with $g_v^d$ as 
proportionality constant. In particular, this means for the neutral current 
form factors appearing in the $P$-violating asymmetries of Appendix~D 
according to (\ref{JcalV}) and (\ref{JcalA}) 
\beq
C_L^{Z_v^{\cal V/A}} = g_v^d\,\widetilde G_{v/a}\,
C_L^{\gamma}\quad\mbox{and}\quad
M_1^{Z_v^{\cal V/A}} = g_v^d\,\widetilde G_{v/a}\,
M_1^{\gamma}\,,\label{ptcweak}
\eeq
where 
$\widetilde G_{v/a} = \sqrt{2}\,g^e_{v/a}\,\widetilde G_F \,q_\mu^2\,e^{-2}$
(see (\ref{Gtilde})).
Furthermore, in the $P$- and $T$-conserving asymmetries the e.m.\ form factors 
become renormalized by a factor very close to unity according to 
(\ref{renormffa}) and (\ref{renormffb}). Finally, the $P$-violating 
$E1$-multipole contributions 
$E_1^{Z^{\cal V/A}_{a}}$ of the axial part of the hadronic neutral current 
$J_\mu^{Z_a}$ are related to the deuteron axial form factor 
\beq
F_{E1}^A = \frac{\sqrt{E_d' E_d}}{M_d}\,
\brar{1}{\cal E}_{1}(J_\mu^{Z_a})\ketr{1}
\eeq
by
\beq
E_1^{Z^{\cal V/A}_{a}} = \widetilde G_{v/a}\,F_{E1}^A\,.\label{ptncweak}
\eeq

At the end of this section, we will furthermore introduce the usual 
invariant multipole form factors and structure functions depending 
on $Q^2$ alone by 
\beqa
G_{C}&=&\sqrt{\frac{4\,\pi}{3}}\,\frac{\beta}{1+\eta}\,C_0\,,\\
G_{Q}&=&\sqrt{\frac{3\,\pi}{2}}\,\frac{\beta}{\eta(1+\eta)}\,C_2\,,\\
G_{(E/M)L}&=&\sqrt{\frac{\pi}{\eta\,(1+\eta)}}\,(E/M)_L\,.
\eeqa
and
\beqa
G_{L}^{\,IM{\rm sig}_M}(X)&=& \frac{\beta^2}{(1+\eta)^2}\,
                              F_{L}^{\,IM{\rm sig}_M}(X)\,,\\
G_{T}^{(\prime)\,IM{\rm sig}_M}(X)&=& \frac{1}{2\,\eta\,(1+\eta)}\,
                              F_{T}^{(\prime)\,IM{\rm sig}_M}(X)\,,\\
G_{LT}^{(\prime)\,IM{\rm sig}_M}(X)&=& 
                  \frac{\beta}{(1+\eta)\sqrt{2\,\eta\,(1+\eta)}}\,
                   F_{LT}^{(\prime)\,IM{\rm sig}_M}(X)\,,\\
G_{TT}^{\,IM{\rm sig}_M}(X)&=& \frac{1}{2\,\eta\,(1+\eta)}\,
                              F_{TT}^{\,IM{\rm sig}_M}(X)\,,
\eeqa
and corresponding relations for the 
$\widetilde G_{\alpha}^{(\prime)\,IM{\rm sig}_M}(X)$ structure functions.
In terms of these invariant structure functions the asymmetries in 
(\ref{Asym-d}) and (\ref{Asym-ed}) become 
\beqa
A_d^{IM{\rm sig}_M}(X)&=&\frac{1}{S_0}\,
\sum_{\alpha\in\{L,\,T,\,LT,\,TT\}}\,\Big[\widetilde v_\alpha
G_\alpha^{IM{\rm sig}_M}(X)+\widetilde v_\alpha^{\,\prime}
\widetilde G_\alpha^{\prime\,IM{\rm sig}_M}(X)\Big]\,,\label{GAsym-d}\\
A_{ed}^{IM{\rm sig}_M}(X)&=&\frac{1}{S_0}\,
\sum_{\alpha\in\{L,\,T,\,LT,\,TT\}}\,\Big[\widetilde v_\alpha
\widetilde G_\alpha^{IM{\rm sig}_M}(X)+\widetilde v_\alpha^{\,\prime}
G_\alpha^{\prime\,IM{\rm sig}_M}(X)\Big]\,,\label{GAsym-ed}
\eeqa
with
\beq
 S_0 = {G_{C}}^2 + \frac{8}{9}\,{\eta }^2\,{G_{Q}}^2 + 
   \frac{2}{3} \,\eta \, \Big( 1 + 2\,( 1 + \eta  ) \,
         {\tan^2 \frac{\theta }{2}} \Big)\,{G_{M}}^2\,, 
\eeq
and
\begin{eqnarray}
\begin{array}{ll}
 \widetilde v_L=\frac{(1+\eta)^2}{\beta^2}\,v_L\,,\quad& 
 \widetilde v_T^{\,(\prime)}={2\,\eta\,(1+\eta)}\,v_T^{\,(\prime)}\,,\cr
 \widetilde v_{LT}^{\,(\prime)}=\frac{1}{\beta}\,(1+\eta)\sqrt{2\,\eta\,(1+\eta)}\,v_{LT}^{\,(\prime)}\,,\quad
&  \widetilde v_{TT}={2\,\eta\,(1+\eta)}\,v_{TT}\,,\cr
\end{array}
\end{eqnarray}
or in explicit form
\begin{eqnarray}
\begin{array}{ll}
 \widetilde v_L=1
\,,\quad& 
 \widetilde v_T=\eta\,\Big(1+2\,(1+\eta)\,\tan^2\frac{\theta_e^{lab}}{2}\Big)
\,,\cr
 \widetilde v_{LT}=\sec\frac{\theta_e^{lab}}{2}\,
\sqrt{\eta\,(1+\eta\,\sin^2\frac{\theta_e^{lab}}{2})}\,,\quad
&  \widetilde v_{TT}=-\eta\,,\cr
 \widetilde v_{LT}^{\,\prime}=
\tan\frac{\theta_e^{lab}}{2}\,\sqrt{\eta\,(1+\eta)}\,,&
 \widetilde v_T^{\,\prime}=2\,\sec\frac{\theta_e^{lab}}{2}\,
\tan\frac{\theta_e^{lab}}{2}\,\eta\,
\sqrt{(1+\eta)(1+\eta\,\sin^2\frac{\theta_e^{lab}}{2})}\, .\cr
\end{array}
\end{eqnarray}
Detailed expressions of the resulting asymmetries are listed in Appendix E. 
Similarly to what has been said with respect to Appendix D above, we would 
like to remind the reader that relations analogous to (\ref{ptcweak}) exist
also for the $P$- and $T$-conserving neutral invariant form factors, 
namely 
\beq
G_{C/Q}^{Z_v^{\cal V/A}} = g_v^d\,\widetilde G_{v/a}\,
G_{C/Q}\quad\mbox{and}\quad
G_M^{Z_v^{\cal V/A}} = g_v^d\,\widetilde G_{v/a}\,
G_M\,,\label{Gptcweak}
\eeq
and that
one has the relation of the $P$-violating invariant form 
factors $G_{E1}^{Z^{\cal V/A}_{a}}$ to the deuteron invariant axial form 
factor 
\beq
G_{E1}^A = \sqrt{\frac{\pi}{\eta\,(1+\eta)}}\,F_{E1}^A\,,
\eeq
which reads
\beq
G_{E1}^{Z^{\cal V/A}_{a}} = \widetilde G_{v/a}\,G_{E1}^A\,.
\eeq

\section{Discussion and Summary}\label{summary}
A schematic survey of the nonvanishing asymmetries is given in Tables
\ref{scalarasym} through \ref{tensorasym} where we have not listed those 
which are related to the listed ones by the above mentioned symmetries. 
The simplest asymmetries to measure are the scalar asymmetries in 
Table~\ref{scalarasym} involving the determination of the 
deuteron recoil polarization for an unpolarized deuteron target without 
or with longitudinal electron polarization, or for the equivalent situation 
using an oriented deuteron target but not measuring the recoil polarization. 
We will discuss these scalar asymmetries in some detail. The vector and tensor 
asymmetries do not provide additional information but they may be used for 
independent checks.

\subsection{$P$- and $T$-conserving contributions} 
For the
$P$- and $T$-conserving currents one finds as scalar asymmetries 
only tensor recoil polarization 
components, if the electrons are unpolarized as is well known, and our 
results for them agree with the ones given in the literature, 
\beqa
S_0\,A_{d}^{ 0 0 \,+ }( 2 0 \,+ ) &=& S_0\,T_{20}=
  -\frac{\eta}{3\,{\sqrt{2}}}\,
      \Big( 8(G_{C} + \frac{\eta}{3} \,{G_{Q}})\,\,G_{Q}  + 
   ( 1 + 2\,( 1 + \eta ) \,{\tan^2 \frac{\theta }{2}})
   \,{G_{M}}^2 \Big)\,
,\\ S_0\,A_{d}^{ 0 0 \,+ }( 2 1 \,+ ) &=& S_0\,T_{21}=
  \frac{4}{{\sqrt{3}}} \,\sec \frac{\theta }{2}\,{\eta }\,
     {\sqrt{\eta \,\Big( 1 + \eta\,{\sin^2 \frac{\theta }{2}} 
      \Big) }}\,G_{M}\,G_{Q}
,\\ S_0\,A_{d}^{ 0 0 \,+ }( 2 2 \,+ ) &=& S_0\,T_{22}=
  -\frac{\eta }{{\sqrt{3}}}\,{G_{M}}^2 \,.
\eeqa 
In particular, with respect to the expressions given in Eq.~(5.11) 
of~\cite{Sch65}, using Schildknecht's notation, one finds
\beqa
s^{\prime \,11}= P_{zz}&=&\frac{\sqrt{2}}{3}\,{\cal O}_{20+}=
\frac{\sqrt{2}}{3}\,A_{d}^{00+}(20+)\,,\\
s^{\prime\, 22}= P_{xx}&=&\frac{1}{2\,\sqrt{3}}\,{\cal O}_{22+}
-\frac{1}{3\,\sqrt{2}}\,{\cal O}_{20+}=
\frac{1}{2\,\sqrt{3}}\,A_{d}^{00+}(22+)
-\frac{1}{3\,\sqrt{2}}\,A_{d}^{00+}(20+)\,,\\
s^{\prime\, 12}= P_{zx}&=&-\frac{1}{2\,\sqrt{3}}\,{\cal O}_{21+}=
-\frac{1}{2\,\sqrt{3}}\,A_{d}^{00+}(21+)\,.
\eeqa
The tensor component $T_{20}$ is used to separate the charge from the 
quadrupole form factor, while $T_{21}$ allows to determine the relative 
phase between the magnetic and quadrupole form factor. The component $T_{22}$
does not provide new information, it could only be taken as an independent 
check of the structure function $B(Q^2)$ because one would not need to 
perform a Rosenbluth separation.

With additional longitudinal electron polarization one finds as scalar 
asymmetries for the 
leading order $P$- and $T$-conserving currents two vector recoil polarization 
components, again in agreement with the ones given in Eq.~(5.16) 
of~\cite{Sch65}, taking into account the relations
\beqa
\frac{^2c_{11}}{^1a}= P_z&=&\sqrt{\frac{2}{3}}\,{\cal O}_{10+}=
\sqrt{\frac{2}{3\,S_0}}\,A_{ed}^{00+}(10+)\nonumber\\
&=& 
  \frac{2}{3}\,\sec \frac{\theta }{2}\,
   \tan \frac{\theta }{2} \,{\eta }\,
   {\sqrt{( 1 + \eta ) \,
       \Big( 1 + \eta\,{\sin^2 \frac{\theta }{2}}  \Big) }}\,G_{M}^2
\,,\label{asymPz}\\
\frac{^2c_{21}}{^1a}= P_x&=&-\frac{1}{\sqrt{3}}\,{\cal O}_{11+}=
-\frac{1}{\sqrt{3}}\,A_{ed}^{00+}(11+)\nonumber\\
&=& - \frac{4}{3\,S_0} \,
     \tan \frac{\theta }{2}\,{\sqrt{\eta \,( 1 + \eta ) }}\,
     \Big( G_{C} + \frac{\eta}{3} \,G_{Q} \Big)\,G_{M}\,.
\eeqa
The first one, $P_z$, is proportional to $G_M^2$, whereas the component 
perpendicular to the momentum transfer but in the scattering plane, $P_x$, 
contains interference of $G_M$ with $G_C$ and $G_Q$. The vector 
and tensor asymmetries listed in the Appendix~\ref{appE} do not contain 
additional information but they could be used for consistency checks.

\subsection{Parity violating contributions}
Parity violation gives a small contribution to the unpolarized cross 
section from the $E1$ contribution 
$G_{E1}^{Z^{\cal A}_a}=\widetilde G_{a}\,G_{E1}^{A}$ to the hadronic 
neutral axial current 
\beq
 S_0\,A_{d}^{ 0 0 \,+ }( 0 0 \,+ ) = \frac{8}{3}\,\sec \frac{\theta }{2}\,
     \tan \frac{\theta }{2} \,{\eta }\,{\sqrt{( 1 + \eta ) \,
         \Big( 1 + \eta\,{\sin^2 \frac{\theta }{2}}  \Big) }}\,
     \widetilde G_{a}\,G_{E1}^{A}\,G_{M}\,,
\eeq
and also to some recoil tensor polarization components 
(see Appendix~\ref{appE}) which, however, will be very difficult to 
disentangle from the leading order contribution. One has to look for 
observables for which the leading order contribution vanishes.
According to Table~\ref{scalarasym}, the vector polarization components 
provide such observables. 
The axial form factor $G_{E1}^{Z^{\cal V}_a}=\widetilde G_{v}\,G_{E1}^{A}$ 
of the hadronic neutral axial 
current as well as parity violation in the hadronic structure, manifest in 
a nonvanishing axial form factor $G_{E1}^{\gamma}$, induce vector polarization 
components in the scattering plane, $P_x$ and $P_z$~\cite{FrH91}. They are 
given by
\beqa
S_0\,P_z&=& \sqrt{\frac{2}{3}}\,S_0\,A_{d}^{ 0 0 \,+ }( 1 0 \,+ )\nonumber\\ 
  &=& \frac{2}{3} \,\eta \,
   \Big( 1 + 2\,( 1 + \eta ) \,{\tan^2 \frac{\theta }{2}} \Big) 
   \,\Big(G_{E1}^\gamma+\widetilde G_{v}\,G_{E1}^{A}\Big)\,G_{M}\nonumber\\&&+ 
   \frac{4}{3}\,\sec\frac{\theta }{2}\,
   \tan\frac{\theta }{2}\,\eta\,
   \sqrt{(1+\eta)(1+\eta\,\sin^2\frac{\theta }{2})}\,g_v^d\,\widetilde G_{a}\,
   G_{M}^2 \,,\label{pz}\\ 
S_0\,P_x&=&-\frac{1}{\sqrt{3}}\,S_0\,A_{d}^{ 0 0 \,+ }( 1 1 \,+ )\nonumber\\ 
&=& -\frac{4}{3}\,\sec \frac{\theta }{2}\,
     {\sqrt{\eta \,\Big( 1 + \eta\,{\sin^2 \frac{\theta }{2}}  \Big) }}\,
     {\Big(G_{E1}^\gamma+\widetilde G_{v}\,G_{E1}^{A}\Big)}\,
     \Big( G_{C} + \frac{\eta}{3} \,G_{Q} \Big)\nonumber\\&&- 
   \frac{8}{3}\, \tan\frac{\theta }{2}\,\sqrt{\eta\,(1+\eta)}\, 
   \,g_v^d\,\widetilde G_{a}\Big(G_{C} + \frac{\eta}{3}\,G_{Q}\Big)\,G_{M}\,.  
\label{px}
\eeqa
Obviously, these observables allow one to determine only the combination
of the axial form factors $G_{E1}^\gamma+\widetilde G_{v}\,G_{E1}^{A}$. 
However, one has to keep in mind that contributions proportional to 
$\widetilde G_{v}$ are suppressed by $(4\,\sin^2 \theta_W-1)$ compared to 
those proportional to $\widetilde G_{a}$. 

The same combination of the axial form factors $G_{E1}^\gamma$ and 
$G_{E1}^{A}$ leads also to a 
nonvanishing asymmetry of the differential cross section with respect to 
longitudinally polarized electrons without deuteron 
polarization~\cite{MuR79,FrH91} according to 
\beqa
 S_0\,A_{ed}^{ 0 0 \,+ }( 0 0 \,+ ) &=& 
  2\,g_v^d\,\widetilde G_{a}\,S_0+
  \frac{8}{3}\,\sec \frac{\theta }{2}\, \tan \frac{\theta }{2} \,
  \eta \,{\sqrt{( 1 + \eta ) \,
  \Big( 1 + \eta\,{\sin^2 \frac{\theta }{2}}  \Big) }}\,
  {\Big(G_{E1}^\gamma+\widetilde G_{v}\,G_{E1}^{A}\Big)}\,G_{M}\,.\label{p0e}
\eeqa
With respect to the neutral hadron current contributions to the asymmetries 
in (\ref{pz}), (\ref{px}), and (\ref{p0e}), these expressions agree with those 
of~\cite{FrH91} if one makes the following identifications
\beq
G_0\equiv G_C,\quad G_2\equiv \frac{2\,\sqrt{2}}{3}\,\eta\,G_Q,\quad 
G_1\equiv G_M,\quad F_A\equiv \sqrt{\frac{\eta}{1+\eta}}\,G_{E1}^{A},\quad 
\mbox{and}\quad g_V^n \equiv 2\,g_v^d\,.
\eeq 

Another contribution from $P$-violation via the larger form factor 
$G_{E1}^{Z^{\cal A}_a}=\widetilde G_{a}\,G_{E1}^{A}$ to observables, 
depending on the electron 
polarization, appears for the recoil vector polarization $P_z$
\beqa
S_0\,A_{ed}^{ 0 0 \,+ }( 1 0 \,+ ) &=& 
  {\sqrt{\frac{2}{3}}}\,\eta \,\Big( 1 + 2\,( 1 + \eta ) \,
      {\tan^2 \frac{\theta }{2}} \Big)\,\widetilde G_{a}\,G_{E1}^{A}\,G_{M}\,,
\eeqa
which, in principle, would allow one to determine separately the neutral 
current axial form factor $G_{E1}^{A}$. However, like 
$A_{d}^{ 0 0 \,+ }( 0 0 \,+ )$ this observable 
will be buried by the leading order of (\ref{asymPz}). This 
is a general feature as a closer inspection of Appendix~\ref{appE} shows, 
whenever $G_{E1}^{Z^{\cal A}_a}=\widetilde G_{a}\,G_{E1}^{A}$ 
contributes to a polarization observable there is also a leading order 
contribution. The reason for this feature is that these terms arise from 
the interaction of the axial lepton current with the 
axial hadron current which is equivalent to the interaction of the lepton 
and hadron vector currents. 

Finally, the tensor recoil polarizations offer another possibility of 
obtaining a clean access to $P$-violation via the axial form factors, i.e.
\beqa
S_0\,A_{ed}^{ 0 0 \,+ }( 2 0 \,+ ) &=& 
   - \frac{8\,{\sqrt{2}}}{3}\,g_v^d\,\widetilde G_{a}\, 
      \Big( G_{C} + \frac{\eta}{3} \,G_{Q}\Big)\,G_{Q} - 
   \frac{{\sqrt{2}}}{3}\,
      \eta\,\Big(1+2\,(1+\eta)\,\tan^2\frac{\theta }{2}\Big)
  \,g_v^d\,\widetilde G_{a}\,G_{M}^2
  \nonumber\\&&-\frac{2\,{\sqrt{2}}}{3}\,\eta \,\sec \frac{\theta }{2}\,
     \tan \frac{\theta }{2} \,{\sqrt{( 1 + \eta ) \,
         \Big( 1 + \eta\,{\sin^2 \frac{\theta }{2}}  \Big) }}\,
     {\Big(G_{E1}^\gamma+\widetilde G_{v}\,G_{E1}^{A}\Big)}\,G_{M}
,\\ S_0\,A_{ed}^{ 0 0 \,+ }( 2 1 \,+ ) &=& 
  \frac{8}{{\sqrt{3}}}\,\eta \,
\sec\frac{\theta}{2}\,\sqrt{\eta\,(1+\eta\,\sin^2\frac{\theta}{2})}
\,g_v^d\,\widetilde G_{a}\,G_{M}\,G_{Q}\,
 \nonumber\\&& 
  +\frac{4}{{\sqrt{3}}}\,\eta \,
     \tan \frac{\theta }{2} \,{\sqrt{\eta \,( 1 + \eta ) }}
  \,\Big(G_{E1}^\gamma+\widetilde G_{v}\,G_{E1}^{A}\Big)\,G_{Q}\,
,\\ S_0\,A_{ed}^{ 0 0 \,+ }( 2 2 \,+ ) &=& 
  -\frac{2}{{\sqrt{3}}}\,\eta\,g_v^d\,\widetilde G_{a}\,G_{M}^2\,.
\eeqa

\subsection{$T$-violating contributions}
Looking at the Tables~\ref{scalarasym} through \ref{tensorasym}, one notes 
that $T$-violation induces very few nonvanishing observables. However, these 
appear always isolated, that means, they do not have to compete with leading 
order contributions or those from $P$-violation. The simplest candidate is the 
recoil vector polarization component $P_y$, perpendicular to the scattering 
plane~\cite{Sch66,DuC66,PrS68}, which is given by 
\beqa
S_0\,P_y&=& \frac{1}{\sqrt{3}}\,S_0\,A_{d}^{ 0 0 \,+ }( 1 1 \,- ) \nonumber\\
&=&  \frac{4}{3}\,\sec \frac{\theta }{2}\,{\eta }\,
     {\sqrt{\eta \,\Big( 1 + \eta\,{\sin^2 \frac{\theta }{2}} \Big) }}
    \,G_{E2}^\gamma\,G_{Q}\,.
\eeqa
The latter result corresponds to the one given in~\cite{Sch66,PrS68} if one 
identifies the additional form factor $G$ of \cite{Sch66,PrS68} with 
$\frac{1}{2\,\eta}\,G_{E2}^\gamma$.
With electron polarization one finds only one contribution from 
$T$-violation to the scalar asymmetries, namely to the tensor recoil 
polarization
\beqa
S_0\,A_{ed}^{ 0 0 \,+ }( 2 1 \,- ) &=& 
  \frac{4}{{\sqrt{3}}}\,
     \tan \frac{\theta }{2} \,{\sqrt{\eta \,( 1 + \eta ) }}\,
     G_{E2}^\gamma\,\Big( G_{C} + \frac{1}{3}\,\eta \,G_{Q} \Big)\,. 
\eeqa
With this we will conclude the formal study of polarization observables in 
elastic electron deuteron scattering.

\begin{table}
\caption{Listing of the ${\rm sig}_{M'}$- and $(I+I')$-values for the 
nonvanishing 
$f_\alpha^{(\prime)\,IM{\rm sig}_M}(I'M'{\rm sig}_{M'};\,c',c)$ 
for various current contributions.}
\begin{center}
\begin{tabular}{crrrcc}
&&&&\multicolumn{2}{c}{$I'+I$}\\
\mbox{$c'$-$c$} & $(-)^{\delta_{c'}^T+\delta_{c}^T}$ & $(-)^{\delta(c',c)}$ &
${\rm sig}_{M'}$ & unprimed & primed \\
\hline\rule{0pt}{10pt}
$pctc$-$pctc$     & 1 & 1 & ${\rm sig}_{M}$ & even & odd \\
$pnctc$-$pctc$    & 1 & $-1$ & ${\rm sig}_{M}$ & odd & even \\
$pctnc$-$pctc$    &$-1$ & $-1$ & $-{\rm sig}_{M}$ & odd & even \\
$pnctnc$-$pctc$   &$-1$ & 1 & $-{\rm sig}_{M}$ & even & odd \\
$pctnc$-$pnctc$   &$-1$ & 1 & $-{\rm sig}_{M}$ & even & odd \\
$pnctnc$-$pnctc$  &$-1$ & $-1$ & $-{\rm sig}_{M}$ & odd & even \\
$pnctnc$-$pctnc$  & 1 & $-1$ & ${\rm sig}_{M}$ & odd & even \\
\end{tabular}
\end{center}
 \label{nonvanishing_f}
\end{table}

\begin{table}
\caption{Listing of the $(M,M')$-values for the 
$f_\alpha^{(\prime)\,IM{\rm sig}_M}(I'M'{\rm sig}_{M'};\,c',c)$.}
\begin{center}
\begin{tabular}{cc}
$\alpha$ & $(M,M')$ \\
\hline\rule{0pt}{10pt}
$L,\,T$ & (0,0),\, (1,1),\,(2,2)\\\rule{0pt}{10pt}
$LT$    & (0,1),\,(1,0),\,(1,2),\,(2,1)\\\rule{0pt}{10pt}
$TT$    & (0,2),\, (1,1),\,(2,0)
\end{tabular}
\end{center}
 \label{M-values}
\end{table}

\begin{table}
\caption{Schematic survey of nonvanishing scalar asymmetries 
$A_{d/ed}^{00+}(I'M'{\rm sig}_{M'})$.}
\begin{center}
\begin{tabular}{ccccccccccc}
type & current & 00+ & 10+ & 11+ & 11-- & 20+ & 21+ & 21-- & 22+ & 22-- \\
\hline\rule{0pt}{10pt}
 & $PT$-conserving & $\surd$ & & & &$\surd$ &$\surd$ & &$\surd$ & \\
 $A^{00+}_d(I'M'{\rm sig}_{M'})$ 
& $P$-violating &$\surd$ &$\surd$ &$\surd$ & &$\surd$ &$\surd$ & & & \\
\rule[-2mm]{0mm}{3mm}
 & $T$-violating & & & &$\surd$ & & & & & \\ 
\hline\rule{0pt}{10pt}
 & $PT$-conserving & &$\surd$ &$\surd$ & & & & & & \\
 $A^{00+}_{ed}(I'M'{\rm sig}_{M'})$ 
& $P$-violating &$\surd$ &$\surd$ &$\surd$ & &$\surd$ &$\surd$ & &$\surd$ & \\
 & $T$-violating & & & & & & &$\surd$ & & \\
\end{tabular}
\end{center}
 \label{scalarasym}
\end{table}

\begin{table}
\caption{Schematic survey of nonvanishing vector asymmetries
$A_{d/ed}^{1M+}(I'M'{\rm sig}_{M'})$ for $I'\ge 1$.}
\begin{center}
\begin{tabular}{cccccccccc}
type & current & 10+ & 11+ & 11-- & 20+ & 21+ & 21-- & 22+ & 22-- \\
\hline\rule{0pt}{10pt}
     & $PT$-conserving  &$\surd$ &$\surd$ & & & & & & \\
 $A^{10+}_{d}(I'M'{\rm sig}_{M'})$ 
& $P$-violating    & &$\surd$ & &$\surd$ &$\surd$ & & & \\
\rule[-2mm]{0mm}{3mm}
     & $T$-violating    & & & & & &$\surd$ & & \\
\hline\rule{0pt}{10pt}
     & $PT$-conserving  & &$\surd$ & & & & & & \\
 $A^{11+}_{d}(I'M'{\rm sig}_{M'})$ & 
$P$-violating    & &$\surd$ & & &$\surd$ & &$\surd$ & \\
\rule[-2mm]{0mm}{3mm}
     & $T$-violating    & & & & & &$\surd$ & &$\surd$ \\
\hline
\hline\rule{0pt}{10pt}
      & $PT$-conserving & & & &$\surd$ &$\surd$ & & & \\
 $A^{10+}_{ed}(I'M'{\rm sig}_{M'})$ 
& $P$-violating &$\surd$ &$\surd$ & &$\surd$ &$\surd$ & & & \\
\rule[-2mm]{0mm}{3mm}
      & $T$-violating & & &$\surd$ & & & & & \\
\hline\rule{0pt}{10pt}
      & $PT$-conserving & & & & &$\surd$ & &$\surd$ & \\
 $A^{11+}_{ed}(I'M'{\rm sig}_{M'})$ 
& $P$-violating & &$\surd$ & & &$\surd$ & &$\surd$ & \\
      & $T$-violating & & &$\surd$ & & & & & \\
\end{tabular}
\end{center}
 \label{vectorasym}
\end{table}

\begin{table}
\caption{Schematic survey of nonvanishing tensor asymmetries
$A_{d/ed}^{2M+}(I'M'{\rm sig}_{M'})$ for $I'\ge 2$.}
\begin{center}
\begin{tabular}{cccccccc}
type & current & 20+ & 21+ & 21-- & 22+ & 22-- \\
\hline\rule[-2mm]{0mm}{6mm}
 \raisebox{-1.5ex}[1.5ex]{$A^{20+}_{d}(I'M'{\rm sig}_{M'})$} 
     & $PT$-conserving  &$\surd$ &$\surd$ & &$\surd$ & \\
\rule[-2mm]{0mm}{3mm}
& $P$-violating  &$\surd$ &$\surd$ & & & \\
\hline\rule[-2mm]{0mm}{6mm}
 \raisebox{-1.5ex}[1.5ex]{$A^{21+}_{d}(I'M'{\rm sig}_{M'})$} 
     & $PT$-conserving     & &$\surd$ & &$\surd$ & \\
\rule[-2mm]{0mm}{3mm}
& $P$-violating & &$\surd$ & &$\surd$ & \\
\hline\rule[-2mm]{0mm}{6mm}
 $A^{22+}_{d}(I'M'{\rm sig}_{M'})$ 
& $PT$-conserving  & & & &$\surd$ & \\
\hline
\hline\rule[-2mm]{0mm}{6mm}
 \raisebox{-1.5ex}[1.5ex]{$A^{20+}_{ed}(I'M'{\rm sig}_{M'})$} 
& $P$-violating &$\surd$ &$\surd$ & &$\surd$ & \\
\rule[-2mm]{0mm}{3mm}
& $T$-violating    & & &$\surd$ & & \\
\hline\rule[-2mm]{0mm}{6mm}
 \raisebox{-1.5ex}[1.5ex]{$A^{21+}_{ed}(I'M'{\rm sig}_{M'})$} 
& $P$-violating & &$\surd$ & &$\surd$ & \\
\rule[-2mm]{0mm}{3mm}
& $T$-violating    & & &$\surd$ & &$\surd$ \\
\hline\rule[-2mm]{0mm}{6mm}
 $A^{22+}_{ed}(I'M'{\rm sig}_{M'})$ 
& $P$-violating  & & & &$\surd$ & \\
\end{tabular}
\end{center}
 \label{tensorasym}
\end{table}

\newpage

\begin{appendix}
\renewcommand{\theequation}{A\arabic{equation}}
\setcounter{equation}{0}

\section{Symmetries and closed form of 
${\cal U}^{\lambda'\lambda\,IM}_{I'M'}(\lowercase{c',c})$}\label{app1}

Here we will derive the various symmetries listed in (\ref{symA}) through
(\ref{symD}). We will start by considering first the symmetries of the 
$t$-matrix elements given in (\ref{multipole}). 
For the reduced multipole matrix elements one finds as symmetry properties 
\beqa
({\cal O}^{\lambda}_{L}(c))^\ast&=& (-)^{\lambda}\,{\cal O}^{\lambda}_{L}(c)
\,,\label{symtb}\\
{\cal O}^\lambda_{L}(c)&=& (-)^{L+\delta^P_{c}}\,{\cal O}^{-\lambda}_{L}(c)\,,
\label{symta}\\
{\cal O}^\lambda_{L}(c)&=& (-)^{L+\lambda+\delta^T_{c}}\,
{\cal O}^{\lambda}_{L}(c)\,,
\label{symtc}
\eeqa
which follow from hermiticity, and from parity and time reversal 
transformations, respectively. First we note
\beq
(t^{c}_{m'\lambda m})^\ast= (-)^{1-m'+\lambda}\,a_\lambda\,
 \sum_L i^L(-)^L\hat L 
\left(\matrix{1 & L & 1 \cr -m' & \lambda & m \cr}\right)
({\cal O}^\lambda_{L}(c))^\ast\,. \label{tstar0}
\eeq
Using hermiticity and time reversal properties from (\ref{symtb}) and 
(\ref{symtc}), yielding
$({\cal O}^{\lambda}_{L}(c))^\ast=(-)^{L+\delta^T_{c}}\,
{\cal O}^{\lambda}_{L}(c)$, one finds
\beq
(t^{c}_{m'\lambda m})^\ast= (-)^{\delta^T_{c}}\,
t^{c}_{m'\lambda m}\,,\label{tstar}
\eeq
which means that all $t$-matrix elements are real or imaginary quantities 
depending on whether $(-)^{\delta^T_{c}}=\pm 1$, respectively. 
From this relation and the fact that the matrix elements of the statistical 
tensors are real follows directly (\ref{uast}), which means that the 
${\cal U}$'s are real or imaginary depending on whether 
$(-)^{\delta^T_{c}+\delta^T_{c'}}=\pm 1$, respectively.
Second we consider
\beqa
t^{c}_{-m'-\lambda -m}&=& (-)^{1+m'-\lambda}\,a_\lambda\,
 \sum_L i^L\hat L 
\left(\matrix{1 & L & 1 \cr m' & -\lambda & -m \cr}\right)
{\cal O}^{-\lambda}_{L}(c)\nonumber\\
&=& \Big((-)^{1-m'+\lambda}\,a_\lambda\,
 \sum_L i^L\hat L 
\left(\matrix{1 & L & 1 \cr -m' & \lambda & m \cr}\right)
({\cal O}^{-\lambda}_{L}(c))^\ast\Big)^\ast\,,
\eeqa
where in the second expression we have made use of the symmetry of the 
$3j$-symbol. This then gives the relation 
\beq
t^{c}_{-m'-\lambda -m}=(-)^{\delta_{c}}\,(t^{c}_{m'\lambda m})^\ast= 
(-)^{\delta^P_{c}}\,t^{c}_{m'\lambda m}\,,\label{tminus}
\eeq
using $({\cal O}^{-\lambda}_{L}(c))^\ast=
 (-)^{\delta_{c}}\,{\cal O}^{\lambda}_{L}(c)$ 
from (\ref{symta}) and (\ref{symtc}). The same relation can be applied 
to (\ref{tstar0}) together with the symmetry of the $3j$-symbol with respect 
to a sign change of all projections, resulting in
\beq
(t^{c}_{m'\lambda m})^\ast= 
(-)^{\delta^P_{c}}\,t^{c}_{m -\lambda m'}\,.\label{tminusl}
\eeq
Now we are ready to prove the symmetries of the 
${\cal U}^{\lambda'\lambda\,IM}_{I'M'}(c',c)$. First we consider the 
interchange $\lambda\leftrightarrow\lambda'$ which gives
\beq
{\cal U}^{\lambda\lambda'\,IM}_{I'M'}(c',c)=\frac{1}{6}\,\sum_{n',n,m',m} 
\Big(t^{c'\,\ast}_{n'\lambda n}\,(\tau^{[I']}_{M'})_{n'm'}\,
t^{c}_{m'\lambda' m}\,
(\tau^{[I]}_{M})_{m n} + (c'\leftrightarrow c)\Big)\,.
\eeq
Using 
\beq
(\tau^{[I]}_{M})_{m n}=(-)^M(\tau^{[I]}_{-M})_{n m}\label{symtau1}
\eeq
and renaming the indices, one obtains the symmetry of (\ref{symA}) 
\beqa
{\cal U}^{\lambda\lambda'\,IM}_{I'M'}(c',c)&=&(-)^{M'+M}
\Big({\cal U}^{\lambda'\lambda\,I-M}_{I'-M'}(c',c)\Big)^\ast\\
&=&(-)^{\lambda'+\lambda}{\cal U}^{\lambda'\lambda\,I-M}_{I'-M'}(c',c)\,,
\eeqa
where the latter follows from (\ref{srprojections}) and (\ref{uast}).
The second symmetry refers to the sign change of the various projections
\beqa
{\cal U}^{-\lambda'-\lambda\,I-M}_{I'-M'}(c',c)&=&\frac{1}{6}\,\sum_{n',n,m',m} 
\Big(t^{c'\,\ast}_{n'-\lambda' n}\,(\tau^{[I']}_{-M'})_{n'm'}\,
t^{c}_{m'-\lambda m}\,(\tau^{[I]}_{-M})_{m n} + (c'\leftrightarrow c)\Big)\,.
\eeqa
Changing the signs of all summation indices, using (\ref{tminus}) and the 
property 
\beq
(\tau^{[I]}_{-M})_{-m -n}=(-)^I(\tau^{[I]}_{M})_{m n}\label{symtau2}
\eeq
results in (\ref{symB}). Finally, considering
\beqa
{\cal U}^{\lambda'\lambda\,I'M'}_{IM}(c',c)&=&\frac{1}{6}\,\sum_{n',n,m',m} 
\Big(t^{c'\,\ast}_{n'\lambda' n}\,(\tau^{[I]}_{M})_{n'm'}\,
t^{c}_{m'\lambda m}\,(\tau^{[I']}_{M'})_{m n} + (c'\leftrightarrow c)\Big)\,,
\eeqa
making for the summation indices the interchanges $m'\leftrightarrow m$ and 
$n'\leftrightarrow n$, using (\ref{symtau1}) and (\ref{tminus}), one first 
finds
\beqa
{\cal U}^{\lambda'\lambda\,I'M'}_{IM}(c',c)&=&(-)^{\delta^{PT}(c',c)+I+I'}
{\cal U}^{-\lambda'-\lambda\,I-M}_{I'-M'}(c',c)\,,
\eeqa
which gives combined with (\ref{symB}) the symmetry of (\ref{symD}).

At the end of this appendix, we will derive a closed expression for 
${\cal U}^{\lambda'\lambda\,I'M'}_{IM}(c',c)$ in terms of reduced multipole 
matrix elements. To this end we use the 
multipole expansion of the $t$-matrix and the Wigner-Eckart theorem for 
the occurring matrix elements of the multipole operators and statistical 
tensors
\beqa
\bra{1m'}{\cal O}^\lambda_{LM}\ket{1m}&=& (-)^{1-m'}\,
\left(\matrix{1 & L & 1 \cr -m' & M & m \cr}\right)\,{\cal O}^\lambda_L\,,\\
\bra{1m'}\tau^{[I]}_{M}\ket{1m}&=& (-)^{1-m'}\,
\left(\matrix{1 & I & 1 \cr -m' & M & m \cr}\right)\,\sqrt{3}\,\hat I\,.
\eeqa
With the help of a sum rule for a sum over four $3j$-symbols~\cite{RoB59}
\beqa
{\cal S}\left[\small{\matrix{L & L' & I & I' \cr \lambda & 
\lambda' & M & M' \cr}}
\right]
&=&\sum_{n',n,m',m}(-)^{\lambda+L'+I'+m'+m}
\left(\matrix{1 & L & 1 \cr -m' & \lambda & m \cr}\right)
\left(\matrix{1 & L' & 1 \cr -n' & \lambda' & n \cr}\right) 
\nonumber\\&&\hspace*{2cm}
\times\left(\matrix{1 & I & 1 \cr -m & M & n \cr}\right)
\left(\matrix{1 & I' & 1 \cr -n' & M' & m' \cr}\right)
\nonumber\\
&=&\sum_{J, m} {\hat J}^2 
\left(\matrix{L & L' & J \cr \lambda & -\lambda' & m \cr}\right)
\left(\matrix{I & I' & J \cr -M & -M' & m \cr}\right)
\left\{\matrix{L & L' & J \cr 1 & 1 & I \cr 1 & 1 & I' \cr}\right\}\,,
\eeqa
one obtains in closed form 
\beqa
{\cal U}^{\lambda'\lambda\,IM}_{I'M'}(c',c)&=&
(-)^{\lambda'+I'}\frac{1}{2}\,a_{\lambda'}\,a_{\lambda}\,\hat I
\,\hat I'\,
\sum_{L',L}i^{L'+L}\,\hat L\,\hat L'\,
{\cal S}\left[\small{\matrix{L & L' & I & I' \cr \lambda & 
\lambda' & M & M' \cr}}
\right]\,
\Big({\cal O}^{\lambda'\,\ast}_{L'}(c')\,{\cal O}^{\lambda}_{L}(c)
+ (c'\leftrightarrow c)\Big)\,.
\eeqa

\renewcommand{\theequation}{B\arabic{equation}}
\setcounter{equation}{0}

\section{General expressions for the $\lowercase{f}$-functions}

Here we list all nonvanishing $f$-functions for the case of recoil 
polarization without target polarization, i.e., 
$f_\alpha^{00+}(I'M'{\rm sig}_{M'};c',c)$ for the various 
diagonal and interference contributions.

(A) Diagonal contributions:

(i) $c',\,c \in {\cal C}_{pc,\,tc}$:

\beqa
f_{L}^{ 0 0 \,+ }( 0 0 \,+ ;\,c', c) &=& 
  \frac{4\,\pi}{3} \,\Big( C_0(c)\,C_0(c') + C_2(c)\,C_2(c') \Big)  
,\\f_{L}^{ 0 0 \,+ }( 2 0 \,+ ;\,c', c) &=& 
  -\frac{2\,\pi }{3} \,\Big( 2\,C_0(c')\,C_2(c) + 2\,C_0(c)\,C_2(c') + 
       {\sqrt{2}}\,C_2(c)\,C_2(c') \Big) 
,\\f_{T}^{ 0 0 \,+ }( 0 0 \,+ ;\,c', c) &=& 
  \frac{4\,\pi}{3} \,M_1(c)\,M_1(c') 
,\\f_{T}^{ 0 0 \,+ }( 2 0 \,+ ;\,c', c) &=& 
  -\frac{{\sqrt{2}}\,\pi}{3} \,M_1(c)\,M_1(c')  
,\\f_{T}^{\prime\, 0 0 \,+ }( 1 0 \,+ ;\,c', c) &=& 
  {\sqrt{\frac{2}{3}}}\,\pi \,M_1(c)\,M_1(c') 
,\\f_{LT}^{ 0 0 \,+ }( 2 1 \,+ ;\,c', c) &=& 
  2\,\pi \,\Big( C_2(c')\,M_1(c) + C_2(c)\,M_1(c') \Big)  
,\\f_{LT}^{\prime\, 0 0 \,+ }( 1 1 \,+ ;\,c', c) &=& 
  \frac{2\,\pi}{3} \,\Big( (2\,{\sqrt{2}}\,C_0(c') + C_2(c'))\,M_1(c) + 
       (2\,{\sqrt{2}}\,C_0(c) + C_2(c))\,M_1(c') \Big)
,\\f_{TT}^{ 0 0 \,+ }( 2 2 \,+ ;\,c', c) &=& 
  \frac{2\,\pi }{{\sqrt{3}}}\,M_1(c)\,M_1(c'). 
\eeqa

(ii) $c',\,c \in {\cal C}_{pc,\,tnc}$:

\beqa
f_{T}^{ 0 0 \,+ }( 0 0 \,+ ;\,c', c) &=& 
  \frac{4\,\pi}{3} \,E_2(c)\,E_2(c') 
,\\f_{T}^{ 0 0 \,+ }( 2 0 \,+ ;\,c', c) &=& 
  -\frac{{\sqrt{2}}\,\pi }{3} \,E_2(c)\,E_2(c')  
,\\f_{T}^{\prime\, 0 0 \,+ }( 1 0 \,+ ;\,c', c) &=& 
  {\sqrt{\frac{2}{3}}}\,\pi \,E_2(c)\,E_2(c') 
,\\f_{TT}^{ 0 0 \,+ }( 2 2 \,+ ;\,c', c) &=& 
  \frac{2\,\pi }{{\sqrt{3}}}\,E_2(c)\,E_2(c'). 
\eeqa

(iii) $c',\,c \in {\cal C}_{pnc,\,tc}$:

\beqa
f_{T}^{ 0 0 \,+ }( 0 0 \,+ ;\,c', c) &=& 
  \frac{4\,\pi }{3}\,E_1(c)\,E_1(c') 
,\\f_{T}^{ 0 0 \,+ }( 2 0 \,+ ;\,c', c) &=& 
  -\frac{{\sqrt{2}}\,\pi}{3} \,E_1(c)\,E_1(c')  
,\\f_{T}^{\prime\, 0 0 \,+ }( 1 0 \,+ ;\,c', c) &=& 
  {\sqrt{\frac{2}{3}}}\,\pi \,E_1(c)\,E_1(c') 
,\\f_{TT}^{ 0 0 \,+ }( 2 2 \,+ ;\,c', c) &=& 
  -\frac{2\,\pi }{{\sqrt{3}}}\,E_1(c)\,E_1(c'). 
\eeqa

(iv) $c',\,c \in {\cal C}_{pnc,\,tnc}$:

\beqa
f_{L}^{ 0 0 \,+ }( 0 0 \,+ ;\,c', c) &=& 
  \frac{4\,\pi}{3} \,C_1(c)\,C_1(c') 
,\\f_{L}^{ 0 0 \,+ }( 2 0 \,+ ;\,c', c) &=& 
  \frac{2\,{\sqrt{2}}\,\pi }{3}\,C_1(c)\,C_1(c') 
,\\f_{T}^{ 0 0 \,+ }( 0 0 \,+ ;\,c', c) &=& 
  \frac{4\,\pi }{3} \,M_2(c)\,M_2(c')
,\\f_{T}^{ 0 0 \,+ }( 2 0 \,+ ;\,c', c) &=& 
  -\frac{{\sqrt{2}}\,\pi}{3}  \,M_2(c)\,M_2(c') 
,\\f_{T}^{\prime\, 0 0 \,+ }( 1 0 \,+ ;\,c', c) &=& 
  {\sqrt{\frac{2}{3}}}\,\pi \,M_2(c)\,M_2(c') 
,\\f_{LT}^{ 0 0 \,+ }( 2 1 \,+ ;\,c', c) &=& 
 - \frac{2\,\pi }{{\sqrt{3}}}\,\Big( C_1(c')\,M_2(c) + C_1(c)\,M_2(c') \Big)   
,\\f_{LT}^{\prime\, 0 0 \,+ }( 1 1 \,+ ;\,c', c) &=& 
  \frac{2\,\pi }
   {{\sqrt{3}}} \,\Big( C_1(c')\,M_2(c) + C_1(c)\,M_2(c') \Big) 
,\\f_{TT}^{ 0 0 \,+ }( 2 2 \,+ ;\,c', c) &=& 
  -\frac{2\,\pi}{{\sqrt{3}}} \,M_2(c)\,M_2(c'). 
\eeqa

(B) Interference contributions: 

(i) $c' \in {\cal C}_{pc,\,tc}$ and $c \in {\cal C}_{pc,\,tnc}$:
\beqa
f_{LT}^{ 0 0 \,+ }( 1 1 \,- ;\,c', c) &=& 
  2\,\pi \,C_2(c')\,E_2(c) 
,\\f_{LT}^{\prime\, 0 0 \,+ }( 2 1 \,- ;\,c', c) &=& 
  \frac{2\,\pi}{3}  \,\Big( 2\,{\sqrt{2}}\,C_0(c') + C_2(c') \Big) \,E_2(c).
\eeqa

(ii) $c' \in {\cal C}_{pc,\,tc}$ and $c \in {\cal C}_{pnc,\,tc}$:
\beqa
f_{T}^{ 0 0 \,+ }( 1 0 \,+ ;\,c', c) &=& 
  {\sqrt{\frac{2}{3}}}\,\pi \,E_1(c)\,M_1(c') 
,\\f_{T}^{\prime\, 0 0 \,+ }( 0 0 \,+ ;\,c', c) &=& 
  \frac{4\,\pi}{3} \,E_1(c)\,M_1(c') 
,\\f_{T}^{\prime\, 0 0 \,+ }( 2 0 \,+ ;\,c', c) &=& 
  -\frac{{\sqrt{2}}\,\pi}{3} \,E_1(c)\,M_1(c')  
,\\f_{LT}^{ 0 0 \,+ }( 1 1 \,+ ;\,c', c) &=& 
  \frac{2\,\pi }{3}\,\Big( 2\,{\sqrt{2}}\,C_0(c') + C_2(c') \Big) \,E_1(c) 
,\\f_{LT}^{\prime\, 0 0 \,+ }( 2 1 \,+ ;\,c', c) &=& 
  2\,\pi \,C_2(c')\,E_1(c). 
\eeqa

(iii) $c' \in {\cal C}_{pc,\,tc}$ and $c \in {\cal C}_{pnc,\,tnc}$:

\beqa
f_{LT}^{ 0 0 \,+ }( 2 1 \,- ;\,c', c) &=& 
  -\frac{2\,\pi}{3}  \,\Big( {\sqrt{3}}\,C_1(c)\,M_1(c') - 
         \Big( 2\,{\sqrt{2}}\,C_0(c') + C_2(c') \Big) \,M_2(c) \Big) 
,\\f_{LT}^{\prime\, 0 0 \,+ }( 1 1 \,- ;\,c', c) &=& 
  \frac{2\,\pi}{3} \,\Big( {\sqrt{3}}\,C_1(c)\,M_1(c') + 
       3\,C_2(c')\,M_2(c) \Big)  
,\\f_{TT}^{ 0 0 \,+ }( 2 2 \,- ;\,c', c) &=& 
  \frac{2\,\pi}{{\sqrt{3}}} \,M_1(c')\,M_2(c). 
\eeqa

(iv) $c' \in {\cal C}_{pc,\,tnc}$ and $c \in {\cal C}_{pnc,\,tc}$:

\beqa
f_{TT}^{ 0 0 \,+ }( 2 2 \,- ;\,c', c) &=& 
  -\frac{2\,\pi}{{\sqrt{3}}} \,E_1(c)\,E_2(c'). 
\eeqa

(v) $c' \in {\cal C}_{pc,\,tnc}$ and $c \in {\cal C}_{pnc,\,tnc}$:

\beqa
f_{T}^{ 0 0 \,+ }( 1 0 \,+ ;\,c', c) &=& 
  {\sqrt{\frac{2}{3}}}\,\pi \,E_2(c')\,M_2(c) 
,\\f_{T}^{\prime\, 0 0 \,+ }( 0 0 \,+ ;\,c', c) &=& 
  \frac{4\,\pi}{3} \,E_2(c')\,M_2(c) 
,\\f_{T}^{\prime\, 0 0 \,+ }( 2 0 \,+ ;\,c', c) &=& 
  -\frac{{\sqrt{2}}\,\pi}{3} \,E_2(c')\,M_2(c) 
,\\f_{LT}^{ 0 0 \,+ }( 1 1 \,+ ;\,c', c) &=& 
  \frac{2\,\pi}{{\sqrt{3}}} \,C_1(c)\,E_2(c') 
,\\f_{LT}^{\prime\, 0 0 \,+ }( 2 1 \,+ ;\,c', c) &=& 
  -\frac{2\,\pi}{{\sqrt{3}}} \,C_1(c)\,E_2(c').  
\eeqa

(vi) $c' \in {\cal C}_{pnc,\,tc}$ and $c \in {\cal C}_{pnc,\,tnc}$:

\beqa
f_{LT}^{ 0 0 \,+ }( 1 1 \,- ;\,c', c) &=& 
  \frac{2\,\pi}{{\sqrt{3}}} \,C_1(c)\,E_1(c') 
,\\f_{LT}^{\prime\, 0 0 \,+ }( 2 1 \,- ;\,c', c) &=& 
  -\frac{2\,\pi}{{\sqrt{3}}} \,C_1(c)\,E_1(c'). 
\eeqa

\renewcommand{\theequation}{C\arabic{equation}}
\setcounter{equation}{0}

\section{Listing of structure functions including $P$- and 
$T$-violation}

Here we list all nonvanishing structure functions 
$F_\alpha^{IM{\rm sig}_M}(I'M'{\rm sig}_{M'})$ and 
$\widetilde F_\alpha^{IM{\rm sig}_M}(I'M'{\rm sig}_{M'})$ for ${\rm sig}_M=+$, 
$I'\ge I$, and $M'\ge M$. Those for $I'< I$, and $M'< M$ as well as the ones 
for ${\rm sig}_M=-$ can be obtained from the listed ones using the symmetry 
relations in (\ref{intchangef}), (\ref{symsiga}) and (\ref{symsigb}). Note 
that ${\rm sig}_{M'}$ is fixed uniquely with the choice of ${\rm sig}_{M}$.

(i) $P$- and $T$-conserved structure functions:

\beqa 
F_{L}^{ 0 0 \,+ }( 0 0 \,+ ) &=& 
  \frac{4\,\pi }{3} \,\Big( {(C_0^\gamma})^2 + 
       {(C_2^\gamma})^2 \Big) 
,\\F_{L}^{ 0 0 \,+ }( 2 0 \,+ ) &=& 
  -\frac{2\,\pi  }{3}\,C_2^\gamma\,
     \Big( 4\,C_0^\gamma + {\sqrt{2}}\,C_2^\gamma \
\Big) 
,\\F_{L}^{ 1 0 \,+ }( 1 0 \,+ ) &=& 
  \frac{2\,\pi }{3} \,\Big( {\sqrt{2}}\,C_0^\gamma-
  C_2^\gamma \Big)^2 
,\\F_{L}^{ 1 1 \,+ }( 1 1 \,+ ) &=& 
  \frac{4\,\pi  }{3}\,\Big( 2\,{(C_0^\gamma})^2 + 
       {\sqrt{2}}\,C_0^\gamma\,C_2^\gamma - 
       2\,{(C_2^\gamma})^2 \Big) 
,\\F_{L}^{ 2 0 \,+ }( 2 0 \,+ ) &=& 
  \frac{2\,\pi}{3} \,\Big( 2\,{(C_0^\gamma})^2 + 
       2\,{\sqrt{2}}\,C_0^\gamma\,C_2^\gamma + 
       3\,{(C_2^\gamma})^2 \Big)  
,\\F_{L}^{ 2 1 \,+ }( 2 1 \,+ ) &=& 
  \frac{4\,\pi }{3}\,\Big( 2\,{(C_0^\gamma})^2 + 
       {\sqrt{2}}\,C_0^\gamma\,C_2^\gamma - 
       2\,{(C_2^\gamma})^2 \Big) 
,\\F_{L}^{ 2 2 \,+ }( 2 2 \,+ ) &=& 
  \frac{4\,\pi }{3} \,\Big( {\sqrt{2}}\,C_0^\gamma-
  C_2^\gamma \Big)^2 , \\
\rule{0pt}{30pt}
F_{T}^{ 0 0 \,+ }( 0 0 \,+ ) &=& 
  \frac{4\,\pi}{3} \,{(M_1^\gamma})^2 
,\\F_{T}^{ 0 0 \,+ }( 2 0 \,+ ) &=& 
  -\frac{{\sqrt{2}}\,\pi}{3} \,{(M_1^\gamma})^2   
,\\F_{T}^{ 1 1 \,+ }( 1 1 \,+ ) &=& 
  2\,\pi \,{(M_1^\gamma})^2 
,\\F_{T}^{ 2 0 \,+ }( 2 0 \,+ ) &=& 
  -\frac{4\,\pi}{3} \,{(M_1^\gamma})^2 
,\\F_{T}^{ 2 1 \,+ }( 2 1 \,+ ) &=& 
  -2\,\pi \,{(M_1^\gamma})^2, \\
\rule{0pt}{30pt}
F_{T}^{\,\prime\, 0 0 \,+ }( 1 0 \,+ ) &=& 
  {\sqrt{\frac{2}{3}}}\,\pi \,{(M_1^\gamma})^2 
,\\F_{T}^{\,\prime\, 1 0 \,+ }( 2 0 \,+ ) &=& 
  \frac{2\,\pi}{{\sqrt{3}}} \,{(M_1^\gamma})^2 
,\\F_{T}^{\,\prime\, 1 1 \,+ }( 2 1 \,+ ) &=& 
  2\,\pi \,{(M_1^\gamma})^2, \\
\rule{0pt}{30pt}
F_{LT}^{ 0 0 \,+ }( 2 1 \,+ ) &=& 
  4\,\pi \,C_2^\gamma\,M_1^\gamma 
,\\F_{LT}^{ 1 0 \,+ }( 1 1 \,+ ) &=& 
  \frac{2\,\pi}{{\sqrt{3}}}  \,\Big( -2\,C_0^\gamma + 
       {\sqrt{2}}\,C_2^\gamma \Big) \,M_1^\gamma
,\\F_{LT}^{ 2 0 \,+ }( 2 1 \,+ ) &=& 
  - 2\,\pi \,\Big( 2\,C_0^\gamma + 
       {\sqrt{2}}\,C_2^\gamma \Big) \,M_1^\gamma \
,\\F_{LT}^{ 2 1 \,+ }( 2 2 \,+ ) &=& 2\,
  {\sqrt{\frac{2}{3}}}\,\pi \,\Big( -2\,C_0^\gamma + 
     {\sqrt{2}}\,C_2^\gamma \Big) \,M_1^\gamma, \\
\rule{0pt}{30pt}
F_{LT}^{\,\prime\, 0 0 \,+ }( 1 1 \,+ ) &=& 
  \frac{4\,\pi}{3} \,\Big( 2\,{\sqrt{2}}\,C_0^\gamma + 
       C_2^\gamma \Big) \,M_1^\gamma 
,\\F_{LT}^{\,\prime\, 1 0 \,+ }( 2 1 \,+ ) &=& 
  2\,{\sqrt{\frac{2}{3}}}\,\pi \,\Big( {\sqrt{2}}\,C_0^\gamma - 
     C_2^\gamma \Big) \,M_1^\gamma 
,\\F_{LT}^{\,\prime\, 1 1 \,+ }( 2 0 \,+ ) &=& 
  -\frac{ 2\,\pi }{3} \,\Big( 2\,C_0^\gamma + 
         5\,{\sqrt{2}}\,C_2^\gamma \Big) \,M_1^\gamma
,\\F_{LT}^{\,\prime\, 1 1 \,+ }( 2 2 \,+ ) &=& 
  - 2\,{\sqrt{\frac{2}{3}}}\,\pi \,
     \Big( -2\,C_0^\gamma + 
       {\sqrt{2}}\,C_2^\gamma \Big) \,M_1^\gamma,\\
\rule{0pt}{30pt}
F_{TT}^{ 0 0 \,+ }( 2 2 \,+ ) &=& 
  \frac{2\,\pi}{{\sqrt{3}}} \,{(M_1^\gamma})^2 
,\\F_{TT}^{ 1 1 \,+ }( 1 1 \,+ ) &=& 
  2\,\pi \,{(M_1^\gamma})^2 
,\\F_{TT}^{ 2 0 \,+ }( 2 2 \,+ ) &=& 
  -2\,{\sqrt{\frac{2}{3}}}\,\pi \,{(M_1^\gamma})^2 
,\\F_{TT}^{ 2 1 \,+ }( 2 1 \,+ ) &=& 
  2\,\pi \,{(M_1^\gamma})^2. 
\eeqa

(ii) $P$-violating structure functions:

\beqa 
F_{T}^{ 0 0 \,+ }( 1 0 \,+ ) &=& 
  2\,{\sqrt{\frac{2}{3}}}\,\pi \,\Big(E_1^\gamma+
E_1^{Z^{\cal A}_a}\Big)\,M_1^\gamma 
,\\F_{T}^{ 1 0 \,+ }( 2 0 \,+ ) &=& 
  \frac{4\,\pi}{{\sqrt{3}}} \,\Big(E_1^\gamma+
E_1^{Z^{\cal A}_a}\Big)\,M_1^\gamma 
,\\F_{T}^{ 1 1 \,+ }( 2 1 \,+ ) &=& 
  4\,\pi \,\Big(E_1^\gamma+E_1^{Z^{\cal A}_a}\Big)\,M_1^\gamma, \\
\rule{0pt}{30pt}
F_{T}^{\,\prime\, 0 0 \,+ }( 0 0 \,+ ) &=& 
  \frac{8\,\pi}{3} \,\Big(E_1^\gamma+E_1^{Z^{\cal A}_a}\Big)\,M_1^\gamma 
,\\F_{T}^{\,\prime\, 0 0 \,+ }( 2 0 \,+ ) &=& 
  -2\,\frac{{\sqrt{2}}\,\pi}{3} \,
       \Big(E_1^\gamma+E_1^{Z^{\cal A}_a}\Big)\,M_1^\gamma  
,\\F_{T}^{\,\prime\, 1 1 \,+ }( 1 1 \,+ ) &=& 
  4\,\pi \,\Big(E_1^\gamma+E_1^{Z^{\cal A}_a}\Big)\,M_1^\gamma 
,\\F_{T}^{\,\prime\, 2 0 \,+ }( 2 0 \,+ ) &=& 
  -\frac{8\,\pi}{3} \,\Big(E_1^\gamma+E_1^{Z^{\cal A}_a}\Big)\,M_1^\gamma 
,\\F_{T}^{\,\prime\, 2 1 \,+ }( 2 1 \,+ ) &=& 
  -4\,\pi \,\Big(E_1^\gamma+E_1^{Z^{\cal A}_a}\Big)\,M_1^\gamma, \\
\rule{0pt}{30pt}
F_{LT}^{ 0 0 \,+ }( 1 1 \,+ ) &=& 
  \frac{4\,\pi}{3}  \,\Big(E_1^\gamma+E_1^{Z^{\cal A}_a}\Big)\,\Big( 2\,{\sqrt{2}}\,C_0^\gamma + 
       C_2^\gamma \Big) 
,\\F_{LT}^{ 1 0 \,+ }( 2 1 \,+ ) &=& 
  \frac{4\,\pi}{{\sqrt{6}}} \,\Big(E_1^\gamma+E_1^{Z^{\cal A}_a}\Big)\,\Big( {\sqrt{2}}\,C_0^\gamma - 
       C_2^\gamma \Big)  
,\\F_{LT}^{ 1 1 \,+ }( 2 0 \,+ ) &=& 
  -\frac{2\,\pi}{3} \,\Big(E_1^\gamma+E_1^{Z^{\cal A}_a}\Big)\,\Big( 2\,C_0^\gamma + 
         5\,{\sqrt{2}}\,C_2^\gamma \Big)   
,\\F_{LT}^{ 1 1 \,+ }( 2 2 \,+ ) &=& 
  \frac{4\,\pi}{{\sqrt{3}}} \,\Big(E_1^\gamma+E_1^{Z^{\cal A}_a}\Big)\,\Big( 
  {\sqrt{2}}\,C_0^\gamma - C_2^\gamma \Big), \\
\rule{0pt}{30pt}
F_{LT}^{\,\prime\, 0 0 \,+ }( 2 1 \,+ ) &=& 
  4\,\pi \,\Big(E_1^\gamma+E_1^{Z^{\cal A}_a}\Big)\,C_2^\gamma 
,\\F_{LT}^{\,\prime\, 1 0 \,+ }( 1 1 \,+ ) &=& 
  -\frac{4\,\pi}{{\sqrt{6}}} \,\Big(E_1^\gamma+E_1^{Z^{\cal A}_a}\Big)\,\Big( {\sqrt{2}}\,C_0^\gamma - 
         C_2^\gamma \Big)  
,\\F_{LT}^{\,\prime\, 2 0 \,+ }( 2 1 \,+ ) &=& 
  -2\,\sqrt{2}\,\pi \,\Big(E_1^\gamma+E_1^{Z^{\cal A}_a}\Big)\,\Big( {\sqrt{2}}\,C_0^\gamma + 
         C_2^\gamma \Big)  
,\\F_{LT}^{\,\prime\, 2 1 \,+ }( 2 2 \,+ ) &=& 
  -\frac{4\,\pi}{{\sqrt{3}}} \,\Big(E_1^\gamma+E_1^{Z^{\cal A}_a}\Big)\,
\Big( {\sqrt{2}}\,C_0^\gamma - C_2^\gamma \Big).  
\eeqa

(iii) $T$-violating structure functions:

\beqa 
F_{T}^{ 1 1 \,+ }( 2 1 \,- ) &=& 
  4\,\pi \,E_2^\gamma\,M_1^\gamma, \\
\rule{0pt}{30pt}
F_{T}^{\,\prime\, 1 1 \,+ }( 1 1 \,- ) &=& 
  4\,\pi \,E_2^\gamma\,M_1^\gamma 
,\\F_{T}^{\,\prime\, 2 1 \,+ }( 2 1 \,- ) &=& 
  -4\,\pi \,E_2^\gamma\,M_1^\gamma, \\
\rule{0pt}{30pt}
F_{LT}^{ 0 0 \,+ }( 1 1 \,- ) &=& 
  4\,\pi \,C_2^\gamma\,E_2^\gamma 
,\\F_{LT}^{ 1 0 \,+ }( 2 1 \,- ) &=& 
  -\frac{4\,\pi}{{\sqrt{6}}} \,\Big( {\sqrt{2}}\,C_0^\gamma - 
         C_2^\gamma \Big) \,E_2^\gamma 
,\\F_{LT}^{ 1 1 \,+ }( 2 2 \,- ) &=& 
  \frac{4\,\pi}{{\sqrt{3}}}  \,\Big( {\sqrt{2}}\,C_0^\gamma - 
       C_2^\gamma \Big) \,E_2^\gamma,\\
\rule{0pt}{30pt}
F_{LT}^{\,\prime\, 0 0 \,+ }( 2 1 \,- ) &=& 
  \frac{4\,\pi}{3} \,\Big( 2\,{\sqrt{2}}\,C_0^\gamma + 
       C_2^\gamma \Big) \,E_2^\gamma 
,\\F_{LT}^{\,\prime\, 1 0 \,+ }( 1 1 \,- ) &=& 
  \frac{4\,\pi}{{\sqrt{6}}} \,\Big( {\sqrt{2}}\,C_0^\gamma - 
       C_2^\gamma \Big) \,E_2^\gamma 
,\\F_{LT}^{\,\prime\, 2 0 \,+ }( 2 1 \,- ) &=& 
  -\frac{2\,{\sqrt{2}}}{3}\,\pi  \,\Big( {\sqrt{2}}\,C_0^\gamma + 
         5\,C_2^\gamma \Big) \,E_2^\gamma  
,\\F_{LT}^{\,\prime\, 2 1 \,+ }( 2 2 \,- ) &=& 
  - \frac{4\,\pi}{{\sqrt{3}}} \,\Big( {\sqrt{2}}\,C_0^\gamma - 
         C_2^\gamma \Big) \,E_2^\gamma, \\
\rule{0pt}{30pt}
F_{TT}^{ 1 1 \,+ }( 2 1 \,- ) &=& 
  4\,\pi \,E_2^\gamma\,M_1^\gamma. 
\eeqa

(iv) $P$-violating structure functions $\widetilde F$:

\beqa  
\widetilde F_{L}^{ 0 0 \,+ }( 0 0 \,+ ) &=& 
  \frac{8\,\pi }{3} \,\Big( C_0^{\gamma}\,
        C_0^{Z^{\cal A}_{v}} + 
       C_2^{\gamma}\,C_2^{Z^{\cal A}_{v}} \Big) 
,\\\widetilde F_{L}^{ 0 0 \,+ }( 2 0 \,+ ) &=& 
  -\frac{4\,\pi }{3} \,\Big( 2\,C_0^{Z^{\cal A}_{v}}\,
        C_2^{\gamma} + 
       \Big( 2\,C_0^{\gamma} + 
          {\sqrt{2}}\,C_2^{\gamma} \Big) \,
        C_2^{Z^{\cal A}_{v}} \Big) 
,\\\widetilde F_{L}^{ 1 0 \,+ }( 1 0 \,+ ) &=& 
  \frac{4\,\pi }{3} \,\Big( {\sqrt{2}}\,C_0^{\gamma}- C_2^{\gamma}\Big)\,
        \Big( {\sqrt{2}}\,C_0^{Z^{\cal A}_{v}} - 
          C_2^{Z^{\cal A}_{v}} \Big)  
,\\\widetilde F_{L}^{ 1 1 \,+ }( 1 1 \,+ ) &=& 
  \frac{4\,\pi }{3}\,\Big( C_2^{\gamma}\,
        \Big( {\sqrt{2}}\,C_0^{Z^{\cal A}_{v}} - 
          4\,C_2^{Z^{\cal A}_{v}} \Big)  + 
       C_0^{\gamma}\,
        \Big( 4\,C_0^{Z^{\cal A}_{v}} + 
          {\sqrt{2}}\,C_2^{Z^{\cal A}_{v}} \Big) \Big)  
,\\\widetilde F_{L}^{ 2 0 \,+ }( 2 0 \,+ ) &=& 
  \frac{4\,\pi }{3} \,\Big( C_2^{\gamma}\,
        \Big( {\sqrt{2}}\,C_0^{Z^{\cal A}_{v}} + 
          3\,C_2^{Z^{\cal A}_{v}} \Big)  + 
       C_0^{\gamma}\,
        \Big( 2\,C_0^{Z^{\cal A}_{v}} + 
          {\sqrt{2}}\,C_2^{Z^{\cal A}_{v}} \Big) \Big) 
,\\\widetilde F_{L}^{ 2 1 \,+ }( 2 1 \,+ ) &=& 
  \frac{4\,\pi}{3} \,\Big( C_2^{\gamma}\,
        \Big( {\sqrt{2}}\,C_0^{Z^{\cal A}_{v}} - 
          4\,C_2^{Z^{\cal A}_{v}} \Big)  + 
       C_0^{\gamma}\,
        \Big( 4\,C_0^{Z^{\cal A}_{v}} + 
          {\sqrt{2}}\,C_2^{Z^{\cal A}_{v}} \Big) \Big)  
,\\\widetilde F_{L}^{ 2 2 \,+ }( 2 2 \,+ ) &=& 
  \frac{8\,\pi}{3} \,\Big( {\sqrt{2}}\,C_0^{\gamma}- C_2^{\gamma}\Big)\,
        \Big( {\sqrt{2}}\,C_0^{Z^{\cal A}_{v}} - 
          C_2^{Z^{\cal A}_{v}} \Big), \\
\rule{0pt}{30pt} 
\widetilde F_{T}^{ 0 0 \,+ }( 0 0 \,+ ) &=& 
  \frac{8\,\pi}{3} \,M_1^{\gamma}\,
     M_1^{Z^{\cal A}_{v}} 
,\\\widetilde F_{T}^{ 0 0 \,+ }( 1 0 \,+ ) &=& 
  2\,{\sqrt{\frac{2}{3}}}\,\pi \,E_1^{Z^{\cal A}_a}\,
   M_1^\gamma
,\\\widetilde F_{T}^{ 0 0 \,+ }( 2 0 \,+ ) &=& 
  -\frac{2\,{\sqrt{2}}}{3}\,\pi \,M_1^{\gamma}\,
     M_1^{Z^{\cal A}_{v}} 
,\\\widetilde F_{T}^{ 1 0 \,+ }( 2 0 \,+ ) &=& 
  \frac{4\,\pi}{{\sqrt{3}}}  \,E_1^{Z^{\cal A}_a}\,M_1^\gamma
,\\\widetilde F_{T}^{ 1 1 \,+ }( 1 1 \,+ ) &=& 
  4\,\pi \,M_1^{\gamma}\,M_1^{Z^{\cal A}_{v}} 
,\\\widetilde F_{T}^{ 1 1 \,+ }( 2 1 \,+ ) &=& 
  4\,\pi \,E_1^{Z^{\cal A}_a}\,M_1^\gamma 
,\\\widetilde F_{T}^{ 2 0 \,+ }( 2 0 \,+ ) &=& 
  -\frac{8\,\pi}{3} \,M_1^{\gamma}\,
     M_1^{Z^{\cal A}_{v}} 
,\\\widetilde F_{T}^{ 2 1 \,+ }( 2 1 \,+ ) &=& 
  -4\,\pi \,M_1^{\gamma}\,M_1^{Z^{\cal A}_{v}} 
, \\\rule{0pt}{30pt}
\widetilde F_{T}^{\,\prime\, 0 0 \,+ }( 0 0 \,+ ) &=& 
  \frac{8\,\pi}{3}  \,E_1^{Z^{\cal A}_a}\,M_1^\gamma
,\\\widetilde F_{T}^{\,\prime\, 0 0 \,+ }( 1 0 \,+ ) &=& 
  2\,{\sqrt{\frac{2}{3}}}\,\pi \,M_1^{\gamma}\,
   M_1^{Z^{\cal A}_{v}} 
,\\\widetilde F_{T}^{\,\prime\, 0 0 \,+ }( 2 0 \,+ ) &=& 
  -\frac{2\,{\sqrt{2}}\,\pi }{3} \,E_1^{Z^{\cal A}_a}\,
       M_1^\gamma 
,\\\widetilde F_{T}^{\,\prime\, 1 0 \,+ }( 2 0 \,+ ) &=& 
  \frac{4\,\pi}{\sqrt{3}} \,M_1^{\gamma}\,M_1^{Z^{\cal A}_{v}} 
,\\\widetilde F_{T}^{\,\prime\, 1 1 \,+ }( 1 1 \,+ ) &=& 
  4\,\pi \,E_1^{Z^{\cal A}_a}\,M_1^\gamma 
,\\\widetilde F_{T}^{\,\prime\, 1 1 \,+ }( 2 1 \,+ ) &=& 
  4\,\pi \,M_1^{\gamma}\,M_1^{Z^{\cal A}_{v}} 
,\\\widetilde F_{T}^{\,\prime\, 2 0 \,+ }( 2 0 \,+ ) &=& 
  -\frac{8\,\pi}{3}  \,E_1^{Z^{\cal A}_a}\,M_1^\gamma
,\\\widetilde F_{T}^{\,\prime\, 2 1 \,+ }( 2 1 \,+ ) &=& 
  -4\,\pi \,E_1^{Z^{\cal A}_a}\,M_1^\gamma, \\
\rule{0pt}{30pt}
\widetilde F_{LT}^{ 0 0 \,+ }( 1 1 \,+ ) &=& 
  \frac{4\,\pi}{3}  \,\Big( 2\,{\sqrt{2}}\,C_0^\gamma + 
       C_2^\gamma \Big) \,E_1^{Z^{\cal A}_a}
,\\\widetilde F_{LT}^{ 0 0 \,+ }( 2 1 \,+ ) &=& 
  4\,\pi \,\Big( C_2^{Z^{\cal A}_{v}}\,
      M_1^{\gamma} + C_2^{\gamma}\,
      M_1^{Z^{\cal A}_{v}} \Big)  
,\\\widetilde F_{LT}^{ 1 0 \,+ }( 1 1 \,+ ) &=& 
  -2\,{\sqrt{\frac{2}{3}}}\,\pi \,
   \Big( \Big({\sqrt{2}}\,C_0^{Z^{\cal A}_{v}} - C_2^{Z^{\cal A}_{v}}\Big)\,
      M_1^{\gamma} + \Big( {\sqrt{2}}\,C_0^{\gamma} - 
        C_2^{\gamma} \Big) \,
      M_1^{Z^{\cal A}_{v}} \Big)  
,\\\widetilde F_{LT}^{ 1 0 \,+ }( 2 1 \,+ ) &=& 
  \frac{4\,\pi}{{\sqrt{6}}}  \,\Big( {\sqrt{2}}\,C_0^\gamma - 
       C_2^\gamma \Big) \,E_1^{Z^{\cal A}_a}
,\\\widetilde F_{LT}^{ 1 1 \,+ }( 2 0 \,+ ) &=& 
  -\frac{2\,\pi}{3} \,\Big( 2\,C_0^\gamma + 
         5\,{\sqrt{2}}\,C_2^\gamma \Big) \,
       E_1^{Z^{\cal A}_a}
,\\\widetilde F_{LT}^{ 1 1 \,+ }( 2 2 \,+ ) &=& 
  \frac{4\,\pi}{{\sqrt{6}}} \,\Big( 2\,C_0^\gamma - 
       {\sqrt{2}}\,C_2^\gamma \Big) \,
     E_1^{Z^{\cal A}_a}
,\\\widetilde F_{LT}^{ 2 0 \,+ }( 2 1 \,+ ) &=& 
  -2\,{\sqrt{2}}\,\pi \,\Big(\Big( {\sqrt{2}}\,
      C_0^{Z^{\cal A}_{v}} + 
     C_2^{Z^{\cal A}_{v}}\Big)\,M_1^{\gamma} + 
     \Big( {\sqrt{2}}\,C_0^{\gamma} + C_2^{\gamma} \
\Big) \,M_1^{Z^{\cal A}_{v}} \Big)  
,\\\widetilde F_{LT}^{ 2 1 \,+ }( 2 2 \,+ ) &=& 
  -{\frac{4\,\pi}{\sqrt{3}}} \,
   \Big( \Big({\sqrt{2}}\,C_0^{Z^{\cal A}_{v}}- 
      C_2^{Z^{\cal A}_{v}}\Big)\,M_1^{\gamma} + 
     \Big( {\sqrt{2}}\,C_0^{\gamma} - 
        C_2^{\gamma} \Big) \,
      M_1^{Z^{\cal A}_{v}} \Big)  
, \\
\rule{0pt}{30pt}
\widetilde F_{LT}^{\,\prime\, 0 0 \,+ }( 1 1 \,+ ) &=& 
  \frac{2\,{\sqrt{2}} }{3}\,\pi \,\Big( \Big(4\,
        C_0^{Z^{\cal A}_{v}} + {\sqrt{2}}\,C_2^{Z^{\cal A}_{v}}\Big)\,
        M_1^{\gamma} + \Big( 4\,C_0^{\gamma} + 
          {\sqrt{2}}\,C_2^{\gamma} \Big) \, M_1^{Z^{\cal A}_{v}} \Big) 
,\\\widetilde F_{LT}^{\,\prime\, 0 0 \,+ }( 2 1 \,+ ) &=& 
  4\,\pi \,C_2^\gamma\,E_1^{Z^{\cal A}_a} 
,\\\widetilde F_{LT}^{\,\prime\, 1 0 \,+ }( 1 1 \,+ ) &=& 
  -\frac{4\,\pi}{{\sqrt{6}}} \,\Big( {\sqrt{2}}\,C_0^\gamma - 
         C_2^\gamma \Big) \,E_1^{Z^{\cal A}_a} 
,\\\widetilde F_{LT}^{\,\prime\, 1 0 \,+ }( 2 1 \,+ ) &=& 
  2\,{\sqrt{\frac{2}{3}}}\,\pi \,
   \Big( \Big({\sqrt{2}}\,C_0^{Z^{\cal A}_{v}}- C_2^{Z^{\cal A}_{v}}\Big)\,
      M_1^{\gamma} + \Big( {\sqrt{2}}\,C_0^{\gamma} - C_2^{\gamma} \Big) \,
      M_1^{Z^{\cal A}_{v}} \Big)  
,\\\widetilde F_{LT}^{\,\prime\, 1 1 \,+ }( 2 0 \,+ ) &=& 
  -\frac{2\,{\sqrt{2}}}{3}\,\pi \,\Big( \Big({\sqrt{2}}\,
        C_0^{Z^{\cal A}_{v}} + 5\,C_2^{Z^{\cal A}_{v}}\Big)\,M_1^{\gamma} + 
       \Big( {\sqrt{2}}\,C_0^{\gamma} + 5\,C_2^{\gamma} \Big) \,
        M_1^{Z^{\cal A}_{v}} \Big)  
,\\\widetilde F_{LT}^{\,\prime\, 1 1 \,+ }( 2 2 \,+ ) &=& 
  {\frac{4\,\pi}{\sqrt{3}}} \,
   \Big( \Big({\sqrt{2}}\,C_0^{Z^{\cal A}_{v}}- 
      C_2^{Z^{\cal A}_{v}}\Big)\,M_1^{\gamma} + 
     \Big( {\sqrt{2}}\,C_0^{\gamma} - C_2^{\gamma} \Big) \,
      M_1^{Z^{\cal A}_{v}} \Big)  
,\\\widetilde F_{LT}^{\,\prime\, 2 0 \,+ }( 2 1 \,+ ) &=& 
  -2\,\sqrt{2}\,\pi\,\Big( {\sqrt{2}}\,C_0^\gamma + 
         C_2^\gamma \Big) \,E_1^{Z^{\cal A}_a}  
,\\\widetilde F_{LT}^{\,\prime\, 2 1 \,+ }( 2 2 \,+ ) &=& 
  -\frac{4\,\pi}{{\sqrt{3}}}  \,\Big( {\sqrt{2}}\,C_0^\gamma  - 
       C_2^\gamma \Big) \,E_1^{Z^{\cal A}_a}
, \\
\rule{0pt}{30pt}
\widetilde F_{TT}^{ 0 0 \,+ }( 2 2 \,+ ) &=& 
  \frac{4\,\pi}{{\sqrt{3}}} \,M_1^{\gamma}\,
     M_1^{Z^{\cal A}_{v}} 
,\\\widetilde F_{TT}^{ 1 1 \,+ }( 1 1 \,+ ) &=& 
  4\,\pi \,M_1^{\gamma}\,M_1^{Z^{\cal A}_{v}} 
,\\\widetilde F_{TT}^{ 2 0 \,+ }( 2 2 \,+ ) &=& 
  -4\,{\sqrt{\frac{2}{3}}}\,\pi \,M_1^{\gamma}\,
   M_1^{Z^{\cal A}_{v}} 
,\\\widetilde F_{TT}^{ 2 1 \,+ }( 2 1 \,+ ) &=& 
  4\,\pi \,M_1^{\gamma}\,M_1^{Z^{\cal A}_{v}} .
\eeqa

\renewcommand{\theequation}{D\arabic{equation}}
\setcounter{equation}{0}

\section{Listing of various asymmetries}

Here we list all nonvanishing asymmetries 
$A_{d/ed}^{IM{\rm sig}_M}(I'M'{\rm sig}_{M'})$ for ${\rm sig}_M=+$, 
$I'\ge I$, and $M'\ge M$. Those for $I'< I$, and $M'< M$ as well as the ones 
for ${\rm sig}_M=-$ can be obtained from the listed ones using the symmetry 
relations in (\ref{intchangef}), (\ref{asymsiga}) and (\ref{asymsigb}). Note 
that ${\rm sig}_{M'}$ is fixed uniquely with the choice of ${\rm sig}_{M}$.

(A) Asymmetries for $P$- and $T$-conserved contributions:

(i) Scalar asymmetries:
\beqa
S_0\,A_{d}^{ 0 0 \,+ }( 0 0 \,+ ) &=& 
  \frac{4\,\pi}{3} \,\Big( {(C_0^\gamma})^2 + 
        {(C_2^\gamma})^2 \Big) \,v_L\, + 
   \frac{4\,\pi}{3} \,{(M_1^\gamma})^2\,v_T\, 
,\\ S_0\,A_{d}^{ 0 0 \,+ }( 2 0 \,+ ) &=& 
  -\frac{2\,\pi}{3} \,\Big( 4\,C_0^\gamma + 
        {\sqrt{2}}\,{C_2^\gamma} \Big)\,C_2^\gamma \,v_L\, - 
   \frac{{\sqrt{2}}\,\pi}{3} \,{(M_1^\gamma})^2\,v_T\, 
,\\ S_0\,A_{d}^{ 0 0 \,+ }( 2 1 \,+ ) &=& 
  4\,\pi \,C_2^\gamma\,M_1^\gamma\,v_{LT}\, 
,\\ S_0\,A_{d}^{ 0 0 \,+ }( 2 2 \,+ ) &=& 
  \frac{2\,\pi}{{\sqrt{3}}} \,{(M_1^\gamma})^2\,v_{TT}\,, \\
\rule{0pt}{30pt}
S_0\,A_{ed}^{ 0 0 \,+ }( 1 0 \,+ ) &=& 
  {\sqrt{\frac{2}{3}}}\,\pi \,{(M_1^\gamma})^2\,
   v_T^{\,\prime}\, 
,\\ S_0\,A_{ed}^{ 0 0 \,+ }( 1 1 \,+ ) &=& 
  \frac{2\,{\sqrt{2}}\,\pi}{3} \,\Big( 4\,C_0^\gamma+ 
       {\sqrt{2}}\,C_2^\gamma \Big)\,M_1^\gamma \,
     v_{LT}^{\,\prime}\,. 
\eeqa

(ii) Vector asymmetries:
\beqa
S_0\,A_{d}^{ 1 0 \,+ }( 1 0 \,+ ) &=& 
  \frac{2\,\pi}{3} \,\Big( \sqrt{2}\,{C_0^\gamma} - 
       {C_2^\gamma} \Big)^2 \,v_L\, 
,\\ S_0\,A_{d}^{ 1 0 \,+ }( 1 1 \,+ ) &=& 
  -2\,\sqrt{\frac{2}{3}}\,\pi \,\Big( {\sqrt{2}}\,C_0^\gamma- 
         C_2^\gamma\ \Big) M_1^\gamma\,v_{LT}\, 
,\\ S_0\,A_{d}^{ 1 1 \,+ }( 1 1 \,+ ) &=& 
  \frac{4\,\pi}{3} \,\Big( 2\,{(C_0^\gamma})^2 + 
        {\sqrt{2}}\,C_0^\gamma\,C_2^\gamma - 
        2\,{(C_2^\gamma})^2 \Big) \,v_L\, +
   2\,\pi \,{(M_1^\gamma})^2\,(v_T + v_{TT})\,,\\
\rule{0pt}{30pt}
S_0\,A_{ed}^{ 1 0 \,+ }( 2 0 \,+ ) &=& 
  \frac{2\,\pi}
   {{\sqrt{3}}}  \,{(M_1^\gamma})^2\,v_T^{\,\prime}\,
,\\ S_0\,A_{ed}^{ 1 0 \,+ }( 2 1 \,+ ) &=& 
  2\,\sqrt{\frac{2}{3}}\,\pi\,\Big( {\sqrt{2}}\,C_0^\gamma
        - C_2^\gamma \Big)\,M_1^\gamma
 \,v_{LT}^{\,\prime}\, 
,\\ S_0\,A_{ed}^{ 1 1 \,+ }( 2 1 \,+ ) &=& 
  2\,\pi \,{(M_1^\gamma})^2\,v_T^{\,\prime}\, 
,\\ S_0\,A_{ed}^{ 1 1 \,+ }( 2 2 \,+ ) &=& 
  \frac{4\,\pi}{{\sqrt{3}}} \,\Big( \sqrt{2}\,C_0^\gamma- 
         C_2^\gamma\Big)\,M_1^\gamma  \,
       v_{LT}^{\,\prime}\,.  
\eeqa

(iii) Tensor asymmetries:
\beqa
S_0\,A_{d}^{ 2 0 \,+ }( 2 0 \,+ ) &=& 
  \frac{2\,\pi}{3} \,\Big( (\sqrt{2}\,C_0^\gamma + 
        C_2^\gamma)^2 + 
        2\,(C_2^\gamma)^2 \Big) \,v_L\, - 
   \frac{4\,\pi }{3}\,{(M_1^\gamma})^2\,v_T\, 
,\\ S_0\,A_{d}^{ 2 0 \,+ }( 2 1 \,+ ) &=& 
  -2\,\sqrt{2}\,\pi\,\Big( {\sqrt{2}}\,C_0^\gamma + 
            C_2^\gamma \Big) \,M_1^\gamma \,
       v_{LT}\, 
,\\ S_0\,A_{d}^{ 2 0 \,+ }( 2 2 \,+ ) &=& 
  -2\,{\sqrt{\frac{2}{3}}}\,\pi \,{(M_1^\gamma})^2\,v_{TT}\, 
,\\ S_0\,A_{d}^{ 2 1 \,+ }( 2 1 \,+ ) &=& 
  \frac{4\,\pi}{3} \,\Big( 2\,{(C_0^\gamma})^2 + 
        {\sqrt{2}}\,C_0^\gamma\,C_2^\gamma - 
        2\,{(C_2^\gamma})^2 \Big) \,v_L\, -
   2\,\pi \,{(M_1^\gamma})^2\,(v_T - v_{TT})\, 
,\\ S_0\,A_{d}^{ 2 1 \,+ }( 2 2 \,+ ) &=& 
  -\frac{4\,\pi}{{\sqrt{3}}} \,\Big( {\sqrt{2}}\,C_0^\gamma - 
       C_2^\gamma\Big) \,M_1^\gamma \,
     v_{LT}\, 
,\\ S_0\,A_{d}^{ 2 2 \,+ }( 2 2 \,+ ) &=& 
  \frac{4\,\pi }{3}\,\Big( \sqrt{2}\,{C_0^\gamma} - 
       {C_2^\gamma} \Big)^2 \,v_L\,. 
\eeqa

(B) Asymmetries for $P$-violating contributions:

(i) Scalar asymmetries:
\beqa
S_0\,A_{d}^{ 0 0 \,+ }( 0 0 \,+ ) &=& 
  \frac{8\,\pi}{3} \,E_1^{Z^{\cal A}_{a}}\,
     M_1^\gamma\,v_T^{\,\prime}\, 
,\\ S_0\,A_{d}^{ 0 0 \,+ }( 1 0 \,+ ) &=& 
  2\,{\sqrt{\frac{2}{3}}}\,\pi \,
   \Big(E_1^\gamma+E_1^{Z^{\cal V}_{a}}\Big)\,
   M_1^\gamma\,v_T+ 
   2\,{\sqrt{\frac{2}{3}}}\,\pi \,M_1^\gamma\,
    M_1^{Z^{\cal A}_{v}}\,v_T^{\,\prime}\, 
,\\ S_0\,A_{d}^{ 0 0 \,+ }( 1 1 \,+ ) &=& 
  \frac{4\,\pi}{3} \,\Big(E_1^\gamma+E_1^{Z^{\cal 
V}_{a}}\Big)\,\Big( 2\,{\sqrt{2}}\,C_0^\gamma + 
       C_2^\gamma \Big) \,v_{LT} \nonumber\\&&+
   \frac{2\,{\sqrt{2}}}{3}\,\pi \,\Big( \Big(4\,
         C_0^{Z^{\cal A}_{v}} + {\sqrt{2}}\,C_2^{Z^{\cal A}_{v}}\Big)\,
         M_1^\gamma + \Big(4\,C_0^\gamma + {\sqrt{2}}\,C_2^\gamma\Big)\,
         M_1^{Z^{\cal A}_{v}} \Big) \,v_{LT}^{\,\prime} \, 
,\\ S_0\,A_{d}^{ 0 0 \,+ }( 2 0 \,+ ) &=& 
  -\frac{2\,{\sqrt{2}}}{3}\,\pi \,E_1^{Z^{\cal A}_{a}}\,
     M_1^\gamma\,v_T^{\,\prime}\, 
,\\ S_0\,A_{d}^{ 0 0 \,+ }( 2 1 \,+ ) &=& 
  4\,\pi \,C_2^\gamma\,E_1^{Z^{\cal A}_{a}}\,
   v_{LT}^{\,\prime}\,, \\
\rule{0pt}{30pt}
S_0\,A_{ed}^{ 0 0 \,+ }( 0 0 \,+ ) &=& 
  \frac{8\,\pi}{3}\,\Big( C_0^\gamma\, C_0^{Z^{\cal A}_{v}} + 
  C_2^\gamma\,C_2^{Z^{\cal A}_{v}} \Big) \,v_{L} +
  \frac{8\,\pi}{3}\,M_1^\gamma\, M_1^{Z^{\cal A}_{v}}\,v_T 
  +\frac{8\,\pi}{3}\,\Big(E_1^\gamma+E_1^{Z^{\cal V}_{a}}\Big)\,
  M_1^\gamma\,v_T^{\,\prime}\, 
,\\ S_0\,A_{ed}^{ 0 0 \,+ }( 1 0 \,+ ) &=& 
  2\,{\sqrt{\frac{2}{3}}}\,\pi \,E_1^{Z^{\cal A}_{a}}\,
   M_1^\gamma\,v_T\, 
,\\ S_0\,A_{ed}^{ 0 0 \,+ }( 1 1 \,+ ) &=& 
  \frac{4\,\pi}{3} \,\Big( 2\,{\sqrt{2}}\,C_0^\gamma + 
       C_2^\gamma \Big) \,
     E_1^{Z^{\cal A}_{a}}\,v_{LT}\, 
,\\ S_0\,A_{ed}^{ 0 0 \,+ }( 2 0 \,+ ) &=& 
  -\frac{4\,\pi}{3} \,\Big( 2\,C_0^{Z^{\cal A}_{v}}\,
  C_2^\gamma + 2\,C_0^\gamma\,C_2^{Z^{\cal A}_{v}} + 
  {\sqrt{2}}\,C_2^\gamma\, C_2^{Z^{\cal A}_{v}} \Big) \,v_{L}\nonumber\\&&
  -\frac{2\,{\sqrt{2}}}{3} \,\pi\,M_1^\gamma\,
      M_1^{Z^{\cal A}_{v}}\,v_T
  -\frac{2\,{\sqrt{2}}}{3} \,\pi\,\Big(E_1^\gamma+E_1^
{Z^{\cal V}_{a}}\Big)\,M_1^\gamma\,
     v_T^{\,\prime}\, 
,\\ S_0\,A_{ed}^{ 0 0 \,+ }( 2 1 \,+ ) &=& 
  4\,\pi \,\Big( C_2^{Z^{\cal A}_{v}}\,
       M_1^\gamma + C_2^\gamma\,
       M_1^{Z^{\cal A}_{v}} \Big) \,v_{LT}+
  4\,\pi \,\Big(E_1^\gamma+E_1^{Z^{\cal 
V}_{a}}\Big)\,C_2^\gamma\,v_{LT}^{\,\prime}
,\\ S_0\,A_{ed}^{ 0 0 \,+ }( 2 2 \,+ ) &=& 
  \frac{4\,\pi }{{\sqrt{3}}}\,M_1^\gamma\,
     M_1^{Z^{\cal A}_{v}}\,v_{TT}\,. 
\eeqa

(ii) Vector asymmetries:
\beqa
S_0\,A_{d}^{ 1 0 \,+ }( 1 1 \,+ ) &=& 
  - 2\,{\sqrt{\frac{2}{3}}}\,\pi \,
     \Big( {\sqrt{2}}\,C_0^\gamma - C_2^\gamma \Big) \,
     E_1^{Z^{\cal A}_{a}}\,v_{LT}^{\,\prime}\, 
,\\ S_0\,A_{d}^{ 1 0 \,+ }( 2 0 \,+ ) &=& 
  \frac{4\,\pi}{{\sqrt{3}}}\,\Big(E_1^\gamma+E_1^{Z^{\cal V}_{a}}\Big)
  \,M_1^\gamma\,v_T 
  + \frac{4\,\pi}{{\sqrt{3}}}\,M_1^{Z^{\cal A}_{v}}\,M_1^\gamma
  \,v_T^{\,\prime}\,
,\\ S_0\,A_{d}^{ 1 0 \,+ }( 2 1 \,+ ) &=& 
  2\,{\sqrt{\frac{2}{3}}}\,\pi \,\Big(E_1^\gamma+E_1^{Z^{\cal V}_{a}}\Big)\,
   \Big( {\sqrt{2}}\,C_0^\gamma - C_2^\gamma \Big) \, v_{LT}\nonumber\\&&
+ 2\,{\sqrt{\frac{2}{3}}}\,\pi \,\Big( \Big({\sqrt{2}}\,C_0^{Z^{\cal A}_{v}} - 
   C_2^{Z^{\cal A}_{v}}\Big)\, M_1^\gamma + \Big({\sqrt{2}}\,C_0^\gamma- 
      C_2^\gamma\Big)\,M_1^{Z^{\cal A}_{v}} \Big) \,v_{LT}^{\,\prime}\, 
,\\ S_0\,A_{d}^{ 1 1 \,+ }( 1 1 \,+ ) &=& 
  4\,\pi \,E_1^{Z^{\cal A}_{a}}\,M_1^\gamma\,
   v_T^{\,\prime}\, 
,\\ S_0\,A_{d}^{ 1 1 \,+ }( 2 1 \,+ ) &=& 
  4\,\pi \,\Big(E_1^\gamma+E_1^{Z^{\cal 
V}_{a}}\Big)\,M_1^\gamma\,v_T+ 
   4\,\pi \,M_1^\gamma\,M_1^{Z^{\cal A}_{v}}\,
    v_T^{\,\prime}\, 
,\\ S_0\,A_{d}^{ 1 1 \,+ }( 2 2 \,+ ) &=& 
  \frac{4\,\pi}{\sqrt{3}} \,
     \Big(E_1^\gamma+E_1^{Z^{\cal V}_{a}}\Big)\,
     \Big( {\sqrt{2}}\,C_0^\gamma - C_2^\gamma \
\Big) \,v_{LT}\nonumber\\&&+\frac{4\,\pi}{\sqrt{3}}\,
    \Big(\Big({\sqrt{2}}\,C_0^{Z^{\cal A}_{v}}- 
       C_2^{Z^{\cal A}_{v}}\Big)\,M_1^\gamma  +
      \Big({\sqrt{2}}\,C_0^\gamma- 
       C_2^\gamma\Big)\,M_1^{Z^{\cal A}_{v}} \Big) \, v_{LT}^{\,\prime}\,, \\
\rule{0pt}{30pt}
S_0\,A_{ed}^{ 1 0 \,+ }( 1 0 \,+ ) &=& 
  \frac{4\,\pi}{3} \,\Big( {\sqrt{2}}\,C_0^\gamma-C_2^\gamma\Big)
\Big({\sqrt{2}}\, C_0^{Z^{\cal A}_{v}}  -   C_2^{Z^{\cal A}_{v}}\Big) 
  \,v_{L}\, 
,\\ S_0\,A_{ed}^{ 1 0 \,+ }( 1 1 \,+ ) &=& 
  -2\,{\sqrt{\frac{2}{3}}}\,\pi \,
    \Big( \Big({\sqrt{2}}\,C_0^{Z^{\cal A}_{v}}- C_2^{Z^{\cal A}_{v}}\Big)\,
       M_1^\gamma + \Big({\sqrt{2}}\,C_0^\gamma- 
      C_2^\gamma\Big)\,M_1^{Z^{\cal A}_{v}} \
\Big) \,v_{LT}\,  \nonumber\\&&- 2\,{\sqrt{\frac{2}{3}}}\,\pi \,
     \Big(E_1^\gamma+E_1^{Z^{\cal V}_{a}}\Big)\,
     \Big( {\sqrt{2}}\,C_0^\gamma - C_2^\gamma \Big) \,
     v_{LT}^{\,\prime}\,
,\\ S_0\,A_{ed}^{ 1 0 \,+ }( 2 0 \,+ ) &=& 
  \frac{4\,\pi}{{\sqrt{3}}} \,E_1^{Z^{\cal A}_{a}}\,
     M_1^\gamma\,v_T\, 
,\\ S_0\,A_{ed}^{ 1 0 \,+ }( 2 1 \,+ ) &=& 
  2\,{\sqrt{\frac{2}{3}}}\,\pi \,
   \Big( {\sqrt{2}}\,C_0^\gamma - C_2^\gamma \Big) \,
   E_1^{Z^{\cal A}_{a}}\,v_{LT}\, 
,\\ S_0\,A_{ed}^{ 1 1 \,+ }( 1 1 \,+ ) &=& 
  \frac{4\,\pi}{3}\,\Big( C_0^\gamma\Big(4\, C_0^{Z^{\cal A}_{v}} 
+ {\sqrt{2}}\, C_2^{Z^{\cal A}_{v}}\Big)+
  \,C_2^\gamma\, {\Big(\sqrt{2}}\,C_0^{Z^{\cal A}_{v}}  
  - 4\,C_2^{Z^{\cal A}_{v}}\Big)  \Big) \,v_{L}\nonumber\\&&
  +4\,\pi \,\Big(E_1^\gamma+E_1^{Z^{\cal V}_{a}}\Big)\,M_1^\gamma\,
  v_T^{\,\prime}+ 4\,\pi \,M_1^\gamma\,M_1^{Z^{\cal A}_{v}}\,(v_T+v_{TT})\, 
,\\ S_0\,A_{ed}^{ 1 1 \,+ }( 2 1 \,+ ) &=& 
  4\,\pi \,E_1^{Z^{\cal A}_{a}}\,M_1^\gamma\,
   v_T\,
,\\ S_0\,A_{ed}^{ 1 1 \,+ }( 2 2 \,+ ) &=& 
  \frac{4\,\pi}{\sqrt{{3}}} \,
   \Big( {\sqrt{2}}\,C_0^\gamma - C_2^\gamma \
\Big) \,E_1^{Z^{\cal A}_{a}}\,v_{LT}\,. 
\eeqa

(iii) Tensor asymmetries:
\beqa
S_0\,A_{d}^{ 2 0 \,+ }( 2 0 \,+ ) &=& 
  -\frac{8\,\pi}{3}  \,E_1^{Z^{\cal A}_{a}}\,
     M_1^\gamma\,v_T^{\,\prime}\,
,\\ S_0\,A_{d}^{ 2 0 \,+ }( 2 1 \,+ ) &=& 
  -2\,\pi \,\Big( 2\,C_0^\gamma + 
       {\sqrt{2}}\,C_2^\gamma \Big) \,
     E_1^{Z^{\cal A}_{a}}\,v_{LT}^{\,\prime}\, \
,\\ S_0\,A_{d}^{ 2 1 \,+ }( 2 1 \,+ ) &=& 
  -4\,\pi \,E_1^{Z^{\cal A}_{a}}\,M_1^\gamma\,
   v_T^{\,\prime}\, 
,\\ S_0\,A_{d}^{ 2 1 \,+ }( 2 2 \,+ ) &=& 
  \frac{4\,\pi}{{\sqrt{3}}}  \,\Big( -{\sqrt{2}}\,C_0^\gamma   + 
       C_2^\gamma \Big) \,
     E_1^{Z^{\cal A}_{a}}\,v_{LT}^{\,\prime}\,,\\
\rule{0pt}{30pt}
S_0\,A_{ed}^{ 2 0 \,+ }( 2 0 \,+ ) &=& 
  \frac{4\,\pi}{3} \,\Big(C_0^\gamma\,\Big(2\,C_0^{Z^{\cal A}_{v}} 
  + {\sqrt{2}}\,C_2^{Z^{\cal A}_{v}}\Big)+ 
  C_2^\gamma\,\Big({\sqrt{2}}\,C_0^{Z^{\cal A}_{v}} +  3\,C_2^{Z^{\cal A}_{v}} 
  \Big)\Big) \,v_{L}
  - \frac{8\,\pi}{3} \,M_1^\gamma\, M_1^{Z^{\cal A}_{v}}\,v_T\nonumber\\&&
  -\frac{8\,\pi}{3} \,\Big(E_1^\gamma+E_1^{Z^{\cal V}_{a}}\Big)
\,M_1^\gamma\,v_T^{\,\prime}\, 
,\\ S_0\,A_{ed}^{ 2 0 \,+ }( 2 1 \,+ ) &=& 
  -2\,{\sqrt{2}}\,\pi \,\Big( \Big( {\sqrt{2}}\,
          C_0^{Z^{\cal A}_{v}} + 
         C_2^{Z^{\cal A}_{v}} \Big) \,
       M_1^\gamma + \Big({\sqrt{2}}\,C_0^\gamma + 
      C_2^\gamma\Big)\,M_1^{Z^{\cal A}_{v}}\Big) \,v_{LT}\nonumber\\&&  
   -2\,\pi \,\Big(E_1^\gamma+E_1^{Z^{\cal V}_{a}}\Big)\,\Big( 2\,C_0^\gamma + 
       {\sqrt{2}}\,C_2^\gamma \Big) \,
     v_{LT}^{\,\prime}\,  
,\\ S_0\,A_{ed}^{ 2 0 \,+ }( 2 2 \,+ ) &=& 
  -4\,{\sqrt{\frac{2}{3}}}\,\pi \,M_1^\gamma\,
   M_1^{Z^{\cal A}_{v}}\,v_{TT}\, 
,\\ S_0\,A_{ed}^{ 2 1 \,+ }( 2 1 \,+ ) &=& 
  \frac{4\,\pi }{3}\,\Big( C_0^\gamma\Big(4\, C_0^{Z^{\cal A}_{v}} 
+ {\sqrt{2}}\, C_2^{Z^{\cal A}_{v}}\Big)+
  \,C_2^\gamma\, {\Big(\sqrt{2}}\,C_0^{Z^{\cal A}_{v}}  
  - 4\,C_2^{Z^{\cal A}_{v}}\Big)  \Big)\,v_{L}\nonumber\\&&
  -4\,\pi \,\Big(E_1^\gamma+E_1^{Z^{\cal V}_{a}}\Big)\,M_1^\gamma\,
  v_T^{\,\prime}- 4\,\pi \,M_1^\gamma\,M_1^{Z^{\cal A}_{v}}\,(v_T-v_{TT})\, 
,\\ S_0\,A_{ed}^{ 2 1 \,+ }( 2 2 \,+ ) &=& 
  -\frac{4\,\pi }{{\sqrt{3}}}\,
    \Big( \Big(\sqrt{2}\,C_0^{Z^{\cal A}_{v}}- 
       C_2^{Z^{\cal A}_{v}}\Big)\,M_1^\gamma + 
      \Big(\sqrt{2}\,C_0^\gamma - C_2^\gamma\Big)\,
       M_1^{Z^{\cal A}_{v}} \Big) \,v_{LT}\, \nonumber\\&& 
  -\frac{4\,\pi }{{\sqrt{3}}} \,\Big(E_1^\gamma+E_1^{Z^{\cal 
V}_{a}}\Big)\,\Big( {\sqrt{2}}\,C_0^\gamma  - 
       C_2^\gamma \Big) \,v_{LT}^{\,\prime}\,
,\\ S_0\,A_{ed}^{ 2 2 \,+ }( 2 2 \,+ ) &=& 
  \frac{8\,\pi}{3} \,\Big( {\sqrt{2}}\,C_0^\gamma- C_2^\gamma\Big)
  \Big({\sqrt{2}}\,C_0^{Z^{\cal A}_{v}} 
  -  C_2^{Z^{\cal A}_{v}}\Big)\,v_{L}\,.
\eeqa

(C) Asymmetries for $T$-violating contributions:

(i) Scalar asymmetries:
\beqa
S_0\,A_{d}^{ 0 0 \,+ }( 1 1 \,- ) &=& 
  4\,\pi \,C_2^\gamma\,E_2^\gamma\,v_{LT}\,, \\
\rule{0pt}{30pt}
S_0\,A_{ed}^{ 0 0 \,+ }( 2 1 \,- ) &=& 
  \frac{4\,\pi}{3} \,\Big( 2\,{\sqrt{2}}\,C_0^\gamma + 
       C_2^\gamma \Big) \,E_2^\gamma\,
     v_{LT}^{\,\prime}\,. 
\eeqa

(ii) Vector asymmetries:

\beqa
S_0\,A_{d}^{ 1 0 \,+ }( 2 1 \,- ) &=& 
  - 2\,{\sqrt{\frac{2}{3}}}\,\pi \,
     \Big( {\sqrt{2}}\,C_0^\gamma - C_2^\gamma \Big) \,
     E_2^\gamma\,v_{LT}\,  
,\\ S_0\,A_{d}^{ 1 1 \,+ }( 2 1 \,- ) &=& 
  4\,\pi \,E_2^\gamma\,M_1^\gamma\,(v_T +v_{TT})\, 
,\\ S_0\,A_{d}^{ 1 1 \,+ }( 2 2 \,- ) &=& 
  \frac{4\,\pi }{{\sqrt{3}}}\,
     \Big( {\sqrt{2}}\,C_0^\gamma - C_2^\gamma \
\Big) \,E_2^\gamma\,v_{LT}\,, \\
\rule{0pt}{30pt}
S_0\,A_{ed}^{ 1 0 \,+ }( 1 1 \,- ) &=& 
  2\,{\sqrt{\frac{2}{3}}}\,\pi \,\Big( {\sqrt{2}}\,C_0^\gamma - 
     C_2^\gamma \Big) \,E_2^\gamma\,
   v_{LT}^{\,\prime}\, 
,\\ S_0\,A_{ed}^{ 1 1 \,+ }( 1 1 \,- ) &=& 
  4\,\pi \,E_2^\gamma\,M_1^\gamma\,
   v_T^{\,\prime}\,. 
\eeqa

(iii) Tensor asymmetries:
\beqa
S_0\,A_{ed}^{ 2 0 \,+ }( 2 1 \,- ) &=& 
  -\frac{ 2\,\pi  }{3}\,\Big( 2\,C_0^\gamma + 
         5\,{\sqrt{2}}\,C_2^\gamma \Big) \,E_2^\gamma\,
       v_{LT}^{\,\prime}\,  
,\\ S_0\,A_{ed}^{ 2 1 \,+ }( 2 1 \,- ) &=& 
  -4\,\pi \,E_2^\gamma\,M_1^\gamma\,
   v_T^{\,\prime}\, 
,\\ S_0\,A_{ed}^{ 2 1 \,+ }( 2 2 \,- ) &=& 
  -\frac{4\,\pi}{{\sqrt{3}}} \,\Big( {\sqrt{2}}\,C_0^\gamma - 
       C_2^\gamma \Big) \,E_2^\gamma\,
     v_{LT}^{\,\prime}\,. 
\eeqa

\renewcommand{\theequation}{E\arabic{equation}}
\setcounter{equation}{0}

\section{Listing of nonvanishing asymmetries in terms of invariant 
form factors as in Appendix D}\label{appE}

(A) Asymmetries for $P$- and $T$-conserved contributions:

(i) Scalar asymmetries:
\beqa
 S_0\,A_{d}^{ 0 0 \,+ }( 0 0 \,+ ) &=& 
  {G_{C}}^2 + \frac{8}{9}\,{\eta }^2\,{G_{Q}}^2 + 
   \frac{2}{3} \,\eta \,
      \Big( 1 + 2\,( 1 + \eta  ) \,
         {\tan^2 \frac{\theta }{2}} \Big)\,{G_{M}}^2 
,\\ S_0\,A_{d}^{ 0 0 \,+ }( 2 0 \,+ ) &=& 
  -\frac{\eta}{3\,{\sqrt{2}}}\,
      \Big( 8(G_{C} + \frac{\eta}{3} \,{G_{Q}})\,\,G_{Q}  + 
   ( 1 + 2\,( 1 + \eta ) \,{\tan^2 \frac{\theta }{2}})
   \,{G_{M}}^2 \Big)\,
,\\ S_0\,A_{d}^{ 0 0 \,+ }( 2 1 \,+ ) &=& 
  \frac{4}{{\sqrt{3}}} \,\sec \frac{\theta }{2}\,{\eta }\,
     {\sqrt{\eta \,\Big( 1 + \eta\,{\sin^2 \frac{\theta }{2}} 
      \Big) }}\,G_{M}\,G_{Q}
,\\ S_0\,A_{d}^{ 0 0 \,+ }( 2 2 \,+ ) &=& 
  -\frac{\eta }{{\sqrt{3}}}\,{G_{M}}^2 \,,\\
\rule{0pt}{30pt}
 S_0\,A_{ed}^{ 0 0 \,+ }( 1 0 \,+ ) &=& 
  {\sqrt{\frac{2}{3}}}\,\sec \frac{\theta }{2}\,
   \tan \frac{\theta }{2} \,{\eta }\,
   {\sqrt{( 1 + \eta ) \,
       \Big( 1 + \eta\,{\sin^2 \frac{\theta }{2}}  \Big) }}\,{G_{M}}^2
,\\ S_0\,A_{ed}^{ 0 0 \,+ }( 1 1 \,+ ) &=& 
  \frac{4}{{\sqrt{3}}} \,
     \tan \frac{\theta }{2}\,{\sqrt{\eta \,( 1 + \eta ) }}\,
     \Big( G_{C} + \frac{\eta}{3} \,G_{Q} \Big)\,G_{M}\,. 
\eeqa

(ii) Vector asymmetries:
\beqa
 S_0\,A_{d}^{ 1 0 \,+ }( 1 0 \,+ ) &=& 
  \Big(G_{C} - \frac{2}{3}\,\eta \,G_{Q} \Big)^2 
,\\ S_0\,A_{d}^{ 1 0 \,+ }( 1 1 \,+ ) &=& 
  -{\sqrt{2}}\sec \frac{\theta }{2}\,
     {\sqrt{{\eta }\,
         \Big( 1 + \eta\,{\sin^2 \frac{\theta }{2}}  \Big) }}\,
     \Big( {G_{C}} - \frac{2}{3}\,\eta \,G_{Q} \Big)\,G_{M}  
,\\ S_0\,A_{d}^{ 1 1 \,+ }( 1 1 \,+ ) &=& 
   2\,\Big( (G_{C} +\frac{\eta}{3} \,G_{Q})^2-\eta^2\, G_{Q}^2\Big)  + 
   2\,\eta \,( 1 + \eta ) \,{\tan^2 \frac{\theta }{2}}\,{G_{M}}^2 \,
,\\
\rule{0pt}{30pt}
S_0\,A_{ed}^{ 1 0 \,+ }( 2 0 \,+ ) &=& 
  \frac{2}{{\sqrt{3}}}\,\sec \frac{\theta }{2}\,\tan \frac{\theta }{2}
  \,{\eta }\, {\sqrt{( 1 + \eta ) \,
         \Big( 1 + \eta\,{\sin^2 \frac{\theta }{2}}  \Big) }}\,
     {G_{M}}^2 
,\\ S_0\,A_{ed}^{ 1 0 \,+ }( 2 1 \,+ ) &=& 
  \sqrt{2} \,
     \tan \frac{\theta }{2}\,{\sqrt{\eta \,( 1 + \eta ) }}\,
     \Big( G_{C} - \frac{2}{3}\,\eta \,G_{Q} \Big)\,G_{M} 
,\\ S_0\,A_{ed}^{ 1 1 \,+ }( 2 1 \,+ ) &=& 
  2\,\sec \frac{\theta }{2}\,
   \tan \frac{\theta }{2} \,{\eta }\,
   {\sqrt{( 1 + \eta ) \,
       \Big( 1 + \eta\,{\sin^2 \frac{\theta }{2}}  \Big) }}\,{G_{M}}^2
,\\ S_0\,A_{ed}^{ 1 1 \,+ }( 2 2 \,+ ) &=& 
     2\,\tan \frac{\theta }{2}\, 
  {\sqrt{\eta \,( 1 + \eta ) }}\,
     \Big( G_{C} - \frac{2}{3}\,\eta \,G_{Q} \Big)\,G_{M}\,. 
\eeqa

(iii) Tensor asymmetries:
\beqa
 S_0\,A_{d}^{ 2 0 \,+ }( 2 0 \,+ ) &=& 
  \frac{1}{3}\,\Big( (G_{C} + 2\,\eta \,G_{Q})^2 + 2\,G_{C}^2 \Big) - 
   \frac{2 }{3}\,\eta \,
      \Big( 1 + 2\,( 1 + \eta ) \,
         {\tan^2 \frac{\theta }{2}} \Big)\,{G_{M}}^2 
,\\ S_0\,A_{d}^{ 2 0 \,+ }( 2 1 \,+ ) &=& 
  -{{\sqrt{6}}} \, \sec \frac{\theta }{2}\,
   {\sqrt{{\eta }\,\Big( 1 + \eta\,{\sin^2 \frac{\theta }{2}}  \Big) }}\,
       \Big( G_{C} + \frac{2}{3}\,\eta \,G_{Q} \Big)\,G_{M}  
,\\ S_0\,A_{d}^{ 2 0 \,+ }( 2 2 \,+ ) &=& 
  {\sqrt{\frac{2}{3}}}\,\eta \,{G_{M}}^2 
,\\ S_0\,A_{d}^{ 2 1 \,+ }( 2 1 \,+ ) &=& 
   2\,\Big( (G_{C} +\frac{\eta}{3} \,G_{Q})^2-\eta^2\, G_{Q}^2\Big) - 
   2\,\eta \,\Big( 1 + 
      ( 1 + \eta ) \,{\tan^2 \frac{\theta }{2}} \Big)\,{G_{M}}^2  
,\\ S_0\,A_{d}^{ 2 1 \,+ }( 2 2 \,+ ) &=& 
  -2\,\sec \frac{\theta }{2}\,
     {\sqrt{{\eta }\,
         \Big( 1 + \eta\,{\sin^2 \frac{\theta }{2}}  \Big) }}\,
     \Big( {G_{C}} - \frac{2}{3}\,\eta \,G_{Q} \Big)\,G_{M}
,\\ S_0\,A_{d}^{ 2 2 \,+ }( 2 2 \,+ ) &=& 
  2\,{\Big( G_{C} - \frac{2}{3}\,\eta \,G_{Q} \Big) }^2\,.
\eeqa

(B) Asymmetries for $P$-violating contributions:

(i) Scalar asymmetries:
\beqa
 S_0\,A_{d}^{ 0 0 \,+ }( 0 0 \,+ ) &=& 
  \frac{8}{3}\,\sec \frac{\theta }{2}\,
     \tan \frac{\theta }{2} \,{\eta }\,
     {\sqrt{( 1 + \eta ) \,
         \Big( 1 + \eta\,{\sin^2 \frac{\theta }{2}}  \Big) }}\,
     G_{E1}^{Z^{\cal A}_{a}}\,G_{M}
,\\ S_0\,A_{d}^{ 0 0 \,+ }( 1 0 \,+ ) &=& 
  {\sqrt{\frac{2}{3}}} \,\eta \,
   \Big( 1 + 2\,( 1 + \eta ) \,{\tan^2 \frac{\theta }{2}} \Big) 
   \,\Big(G_{E1}^\gamma+G_{E1}^{Z^{\cal V}_{a}}\Big)\,G_{M} \nonumber\\&&+ 
   2\,{\sqrt{\frac{2}{3}}}\,\sec\frac{\theta }{2}\,
   \tan\frac{\theta }{2}\,\eta\,
   \sqrt{(1+\eta)(1+\eta\,\sin^2\frac{\theta }{2})}\,
   G_{M}\,G_{M}^{Z^{\cal A}_{v}}\, 
,\\ S_0\,A_{d}^{ 0 0 \,+ }( 1 1 \,+ ) &=& 
  \frac{4}{{\sqrt{3}}}\,\sec \frac{\theta }{2}\,
     {\sqrt{\eta \,\Big( 1 + \eta\,{\sin^2 \frac{\theta }{2}}  \Big) }}\,
     {\Big(G_{E1}^\gamma+G_{E1}^{Z^{\cal V}_{a}}\Big)}\,
     \Big( G_{C} + \frac{\eta}{3} \,G_{Q} \Big) \nonumber\\&&+ 
   \frac{4}{{\sqrt{3}}}\, \tan\frac{\theta }{2}\,\sqrt{\eta\,(1+\eta)}\, 
    \Big( \Big(G_{C}^{Z^{\cal A}_{v}} + 
        \frac{\eta}{3} \,G_{Q}^{Z^{\cal A}_{v}}\Big)\,G_{M} + 
    \Big(G_{C} + \frac{\eta}{3}\,G_{Q}\Big)\,G_{M}^{Z^{\cal A}_{v}} \Big) \,
,\\ S_0\,A_{d}^{ 0 0 \,+ }( 2 0 \,+ ) &=& 
  -\frac{2\,{\sqrt{2}}}{3}\,\sec \frac{\theta }{2}\,
     \tan \frac{\theta }{2} \,{\eta }\,
     {\sqrt{( 1 + \eta ) \,
         \Big( 1 + \eta\,{\sin^2 \frac{\theta }{2}}  \Big) }}\,
     G_{E1}^{Z^{\cal A}_{a}}\,G_{M}\,
,\\ S_0\,A_{d}^{ 0 0 \,+ }( 2 1 \,+ ) &=& 
  \frac{4}{{\sqrt{3}}}\,\eta \,{\sqrt{\eta \,( 1 + \eta ) }}\,
     \tan \frac{\theta }{2}\, 
     G_{E1}^{Z^{\cal A}_{a}}\,G_{Q} \,,\\
\rule{0pt}{30pt}
 S_0\,A_{ed}^{ 0 0 \,+ }( 0 0 \,+ ) &=& 
  2\,G_{C}\,G_{C}^{Z^{\cal A}_{v}} + 
   \frac{16}{9}\,{\eta }^2\,G_{Q}^{Z^{\cal A}_{v}}\,G_{Q} + 
   \frac{4}{3}\,\eta\,\Big(1+2\,(1+\eta)\,\tan^2\frac{\theta }{2}\Big)
   \,G_{M}\,G_{M}^{Z^{\cal A}_{v}}\,
 \nonumber\\&&+\frac{8}{3}\,\sec \frac{\theta }{2}\,
     \tan \frac{\theta }{2} \,
     \eta \,{\sqrt{( 1 + \eta ) \,
     \Big( 1 + \eta\,{\sin^2 \frac{\theta }{2}}  \Big) }}\,
     {\Big(G_{E1}^\gamma+G_{E1}^{Z^{\cal V}_{a}}\Big)}\,
G_{M}
,\\ S_0\,A_{ed}^{ 0 0 \,+ }( 1 0 \,+ ) &=& 
  {\sqrt{\frac{2}{3}}}\,\eta \,\Big( 1 + 2\,( 1 + \eta ) \,
      {\tan^2 \frac{\theta }{2}} \Big)\,G_{E1}^{Z^{\cal A}_{a}}\,
   G_{M}  
,\\S_0\,A_{ed}^{ 0 0 \,+ }( 1 1 \,+ ) &=& 
  \frac{4}{{\sqrt{3}}} \,G_{E1}^{Z^{\cal A}_{a}}\,
     \Big( G_{C} + \frac{\eta}{3} \,G_{Q} \Big) \,\sec\frac{\theta}{2}\,
 \sqrt{\eta\,(1+\eta\,\sin^2\frac{\theta}{2})}\,
,\\ S_0\,A_{ed}^{ 0 0 \,+ }( 2 0 \,+ ) &=& 
   - \frac{4\,{\sqrt{2}}}{3}\, 
      \Big( G_{C}\,G_{Q}^{Z^{\cal A}_{v}} + 
        G_{C}^{Z^{\cal A}_{v}}\,G_{Q} + 
     \frac{2}{3}\,\eta \,G_{Q}^{Z^{\cal A}_{v}}\,G_{Q} \Big) - 
   \frac{{\sqrt{2}}}{3}\,
      \eta\,\Big(1+2\,(1+\eta)\,\tan^2\frac{\theta }{2}\Big)\,
  G_{M}\,G_{M}^{Z^{\cal A}_{v}}\,
  \nonumber\\&&-\frac{2\,{\sqrt{2}}}{3}\,\sec \frac{\theta }{2}\,
     \tan \frac{\theta }{2} \,\eta \,{\sqrt{( 1 + \eta ) \,
         \Big( 1 + \eta\,{\sin^2 \frac{\theta }{2}}  \Big) }}\,
     {\Big(G_{E1}^\gamma+G_{E1}^{Z^{\cal V}_{a}}\Big)}\,G_{M}
,\\ S_0\,A_{ed}^{ 0 0 \,+ }( 2 1 \,+ ) &=& 
  \frac{4}{{\sqrt{3}}}\,\eta \,\Big( G_{Q}^{Z^{\cal A}_{v}}\,
         G_{M} + G_{M}^{Z^{\cal A}_{v}}\,G_{Q} \
\Big) \,\sec\frac{\theta}{2}\,\sqrt{\eta\,(1+\eta\,\sin^2\frac{\theta}{2})}\,
 \nonumber\\&& 
  +\frac{4}{{\sqrt{3}}}\,\eta \,
     \tan \frac{\theta }{2} \,{\sqrt{\eta \,( 1 + \eta ) }}
  \,\Big(G_{E1}^\gamma+G_{E1}^{Z^{\cal V}_{a}}\Big)\,G_{Q}\,
,\\ S_0\,A_{ed}^{ 0 0 \,+ }( 2 2 \,+ ) &=& 
  -\frac{2}{{\sqrt{3}}}\,\eta\,G_{M}\,G_{M}^{Z^{\cal A}_{v}}\,.
\eeqa

(ii) Vector asymmetries:
\beqa
S_0\,A_{d}^{ 1 0 \,+ }( 1 1 \,+ ) &=& 
  -\sqrt{2}\,\tan \frac{\theta }{2}\,
   {\sqrt{\eta \,( 1 + \eta ) }}\,
       G_{E1}^{Z^{\cal A}_{a}}\,
       \Big( G_{C} - \frac{2}{3}\,\eta \,G_{Q} \Big) 
,\\ S_0\,A_{d}^{ 1 0 \,+ }( 2 0 \,+ ) &=& 
  \frac{2}{{\sqrt{3}}} \,\eta \,
  \Big( 1 + 2\,( 1 + \eta ) \,{\tan^2 \frac{\theta }{2}} \Big)\,
  \Big(G_{E1}^\gamma+G_{E1}^{Z^{\cal V}_{a}}\Big)\,G_{M} \nonumber\\&&+ 
   \frac{4}{{\sqrt{3}}}\,
      \sec\frac{\theta}{2}\,\tan\frac{\theta}{2}\,\eta\,
\sqrt{(1+\eta)(1+\eta\,\sin^2\frac{\theta}{2})}
\,G_{M}\,G_{M}^{Z^{\cal A}_{v}}\, 
,\\ S_0\,A_{d}^{ 1 0 \,+ }( 2 1 \,+ ) &=& 
  \sqrt{2}\,\sec \frac{\theta }{2}\,{\sqrt{\eta \,
         \Big( 1 + \eta\,{\sin^2 \frac{\theta }{2}}  \Big) }}\,
     {\Big(G_{E1}^\gamma+G_{E1}^{Z^{\cal V}_{a}}\Big)}\,
     \Big( G_{C} - \frac{2}{3}\,\eta \,G_{Q} \Big) \nonumber\\&&+ 
   \sqrt{2}\,\tan\frac{\theta }{2}\,\sqrt{\eta\,(1+\eta)}\,  
   \Big( \Big(G_{C}^{Z^{\cal A}_{v}} - 
        \frac{2}{3}\,\eta \,G_{Q}^{Z^{\cal A}_{v}}\Big)\,G_{M} + 
        \Big(G_{C} - 
        \frac{2}{3}\,\eta \,G_{Q}\Big)\,G_{M}^{Z^{\cal A}_{v}} \Big) \,
,\\ S_0\,A_{d}^{ 1 1 \,+ }( 1 1 \,+ ) &=& 
  4\,\sec \frac{\theta }{2}\,{\eta }\,
   \tan \frac{\theta }{2} \,
   {\sqrt{( 1 + \eta ) \,
       \Big( 1 + \eta\,{\sin^2 \frac{\theta }{2}}  \Big) }}\,
   G_{E1}^{Z^{\cal A}_{a}}\,G_{M}
,\\ S_0\,A_{d}^{ 1 1 \,+ }( 2 1 \,+ ) &=& 
  2\,\eta \,
  \Big( 1 +  2\,( 1 + \eta )\,{\tan^2 \frac{\theta }{2}} \Big)
  \,\Big(G_{E1}^\gamma+G_{E1}^{Z^{\cal V}_{a}}\Big)\,G_{M} \nonumber\\&&+ 
   4\,\sec\frac{\theta }{2}\,
\tan\frac{\theta }{2}\,\eta\,
\sqrt{(1+\eta)(1+\eta\,\sin^2\frac{\theta }{2})}\,
G_{M}\,G_{M}^{Z^{\cal A}_{v}}\,  
,\\ S_0\,A_{d}^{ 1 1 \,+ }( 2 2 \,+ ) &=& 
  2\,\sec \frac{\theta }{2}\,{\sqrt{{\eta }\,
       \Big( 1 + \eta\,{\sin^2 \frac{\theta }{2}}  \Big) }}\,
   {\Big(G_{E1}^\gamma+G_{E1}^{Z^{\cal V}_{a}}\Big)}\,\Big( G_{C} - 
   \frac{2 }{3}\,\eta\,\,G_{Q} \Big)
\nonumber\\&&+ 
  2\,\tan\frac{\theta }{2}\,\sqrt{\eta\,(1+\eta)}\,
   \Big( \Big(G_{C}^{Z^{\cal A}_{v}} - 
      \frac{2}{3}\,\eta \,G_{Q}^{Z^{\cal A}_{v}}\Big)\,G_{M} + 
   \Big(G_{C} - \frac{2}{3}\,\eta \,G_{Q}\Big)\,G_{M}^{Z^{\cal A}_{v}}\Big) 
\,,\\
\rule{0pt}{30pt}
S_0\,A_{ed}^{ 1 0 \,+ }( 1 0 \,+ ) &=& 
   2\,\Big( G_{C}^{Z^{\cal A}_{v}} - 
        \frac{2}{3}\,\eta \,G_{Q}^{Z^{\cal A}_{v}} \Big) \,
      \Big( G_{C} - \frac{2}{3}\,\eta \,G_{Q} \Big) \, 
,\\S_0\,A_{ed}^{ 1 0 \,+ }( 1 1 \,+ ) &=& 
  -{\sqrt{2}}\,\,\sec\frac{\theta}{2}\,
\sqrt{\eta\,(1+\eta\,\sin^2\frac{\theta}{2})}
\Big( \Big(G_{C}^{Z^{\cal A}_{v}} - 
        \frac{2}{3}\,\eta \,G_{Q}^{Z^{\cal A}_{v}}\Big)\,G_{M} + 
        \Big(G_{C} - 
        \frac{2}{3}\,\eta \,G_{Q}\Big)\,G_{M}^{Z^{\cal A}_{v}} \Big) \, 
  \nonumber\\&&
  -\sqrt{2}\,\tan \frac{\theta }{2}\,
   {\sqrt{\eta \,( 1 + \eta ) }}\,
    \Big(G_{E1}^\gamma+G_{E1}^{Z^{\cal V}_{a}}\Big)\,
       \Big( G_{C} - \frac{2}{3}\,\eta \,G_{Q} \Big) 
,\\ S_0\,A_{ed}^{ 1 0 \,+ }( 2 0 \,+ ) &=& 
  \frac{2}{{\sqrt{3}}}\,\eta \,\Big( 1 + 2\,( 1 + \eta ) \,
  {\tan^2 \frac{\theta }{2}} \Big)\,G_{E1}^{Z^{\cal A}_{a}}\,G_{M}  
,\\S_0\,A_{ed}^{ 1 0 \,+ }( 2 1 \,+ ) &=& 
  {\sqrt{2}} \,\sec\frac{\theta}{2}\,
  \sqrt{\eta\,(1+\eta\,\sin^2\frac{\theta}{2})}\,G_{E1}^{Z^{\cal A}_{a}}\,
     \Big( G_{C} - \frac{2}{3}\,\eta \,G_{Q} \Big) \,
,\\ S_0\,A_{ed}^{ 1 1 \,+ }( 1 1 \,+ ) &=& 
   4\,\Big(  G_{C}^{Z^{\cal A}_{v}}\,
         \Big( G_{C} + \frac{\eta}{3} \,G_{Q} \Big) +
  \frac{\eta}{3} \,G_{Q}^{Z^{\cal A}_{v}}\,
         \Big( G_{C} - \frac{8}{3}\,\eta \,G_{Q} \Big)  
 \Big) \, + 
   4\,\eta\,(1+\eta)\,\tan^2\frac{\theta }{2}\,
   G_{M}\,G_{M}^{Z^{\cal A}_{v}}\,\nonumber\\&& 
  +4\,\sec \frac{\theta }{2}\,
   \tan \frac{\theta }{2} \,\eta \,{\sqrt{( 1 + \eta ) \,
   \Big( 1 + \eta\,{\sin^2 \frac{\theta }{2}}  \Big) }}\,
   {\Big(G_{E1}^\gamma+G_{E1}^{Z^{\cal V}_{a}}\Big)}\,G_{M}\, 
,\\ S_0\,A_{ed}^{ 1 1 \,+ }( 2 1 \,+ ) &=& 
  2\,\eta \,\Big( 1 + 2\,( 1 + \eta ) \,{\tan^2 \frac{\theta }{2}} \Big) 
  \,G_{E1}^{Z^{\cal A}_{a}}\,G_{M}\, 
,\\S_0\,A_{ed}^{ 1 1 \,+ }( 2 2 \,+ ) &=& 
  2\,\sec\frac{\theta}{2}\,\sqrt{\eta\,(1+\eta\,\sin^2\frac{\theta}{2})}
  \Big( G_{C} - 
     \frac{2}{3}\,\eta \,G_{Q}\Big)\,G_{E1}^{Z^{\cal A}_{a}}  \,.
\eeqa

(iii) Tensor asymmetries:
\beqa
S_0\,A_{d}^{ 2 0 \,+ }( 2 0 \,+ ) &=& 
  -\frac{8}{3}\,\sec \frac{\theta }{2}\,
     \tan \frac{\theta }{2}\,{\eta } \,
     {\sqrt{( 1 + \eta ) \,
         \Big( 1 + \eta\,{\sin^2 \frac{\theta }{2}}  \Big) }}\,
     G_{E1}^{Z^{\cal A}_{a}}\,G_{M}
,\\ S_0\,A_{d}^{ 2 0 \,+ }( 2 1 \,+ ) &=& 
- {{\sqrt{6}}}\,\tan \frac{\theta }{2}
    \,{\sqrt{\eta \,( 1 + \eta ) }}\,
       G_{E1}^{Z^{\cal A}_{a}}\,
       \Big( G_{C} + \frac{2}{3}\,\eta \,G_{Q} \Big)\, 
,\\ S_0\,A_{d}^{ 2 1 \,+ }( 2 1 \,+ ) &=& 
  -4\,\sec \frac{\theta }{2}\,
   \tan \frac{\theta }{2} \,{\eta }\,
   {\sqrt{( 1 + \eta ) \,
       \Big( 1 + \eta\,{\sin^2 \frac{\theta }{2}}  \Big) }}\,
   G_{E1}^{Z^{\cal A}_{a}}\,G_{M}
,\\ S_0\,A_{d}^{ 2 1 \,+ }( 2 2 \,+ ) &=& 
  -2\,\tan \frac{\theta }{2}\,{\sqrt{\eta \,( 1 + \eta ) }}\,
   \Big(  G_{C} - \frac{2}{3} \,\eta\,G_{Q} \Big)\,
   G_{E1}^{Z^{\cal A}_{a}}\,,\\
\rule{0pt}{30pt}
S_0\,A_{ed}^{ 2 0 \,+ }( 2 0 \,+ ) &=& 
  2\,\Big( G_{C}^{Z^{\cal A}_{v}}\,
         \Big( G_{C} + \frac{2}{3}\,\eta \,G_{Q} \Big) 
   + \frac{2}{3}\,\eta \,G_{Q}^{Z^{\cal A}_{v}}\,
         \Big( G_{C} + 2\,\eta \,G_{Q} \Big) \Big)  \nonumber\\&&- 
   \frac{4}{3}\,\eta\,\Big(1+2\,(1+\eta)\,\tan^2\frac{\theta}{2}\Big)\,
   G_{M}\,G_{M}^{Z^{\cal A}_{v}}\,
       \nonumber\\&&- \frac{8}{3}\,
      \sec\frac{\theta}{2}\,\tan\frac{\theta}{2}\,\eta\,
   \sqrt{(1+\eta)(1+\eta\,\sin^2\frac{\theta}{2})}\,
   \Big(G_{E1}^\gamma+G_{E1}^{Z^{\cal V}_{a}}\Big)\,G_{M}\, 
,\\ S_0\,A_{ed}^{ 2 0 \,+ }( 2 1 \,+ ) &=& 
  - {\sqrt{6}}
  \,\sec\frac{\theta}{2}\,\sqrt{\eta\,(1+\eta\,\sin^2\frac{\theta}{2})}\,
      \Big( \Big(G_{C}^{Z^{\cal A}_{v}} + 
        \frac{2}{3}\,\eta \,G_{Q}^{Z^{\cal A}_{v}}\Big)\,G_{M} + 
        \Big(G_{C}\,G_{M}^{Z^{\cal A}_{v}} + 
        \frac{2}{3}\,\eta \,G_{Q}\Big)\,G_{M}^{Z^{\cal A}_{v}} \Big) \, 
 \nonumber\\&&  -{{\sqrt{6}}}\,\tan \frac{\theta }{2}
   \,{\sqrt{\eta \,( 1 + \eta ) }}
   \,\Big(G_{E1}^\gamma+G_{E1}^{Z^{\cal V}_{a}}\Big)\,
       \Big( G_{C} + \frac{2}{3}\,\eta \,G_{Q} \Big) 
,\\ S_0\,A_{ed}^{ 2 0 \,+ }( 2 2 \,+ ) &=& 
   2\,{\sqrt{\frac{2}{3}}}\,\eta\,G_{M}\,
   G_{M}^{Z^{\cal A}_{v}}\, 
,\\ S_0\,A_{ed}^{ 2 1 \,+ }( 2 1 \,+ ) &=& 
   4\,\Big(         G_{C}^{Z^{\cal A}_{v}}\,
         \Big( G_{C} + \frac{\eta}{3} \,G_{Q} \Big)
   +\frac{\eta}{3} \,G_{Q}^{Z^{\cal A}_{v}}\,
         \Big( G_{C} - \frac{8}{3}\,\eta \,G_{Q} \Big)\Big) \, \nonumber\\&&- 
   4\,\eta\,\Big(1+(1+\eta)\,\tan^2\frac{\theta }{2}\Big)\,
  G_{M}\,G_{M}^{Z^{\cal A}_{v}}
   +2\,{\sqrt{\frac{2}{3}}}\,G_{M}\,
   G_{M}^{Z^{\cal A}_{v}}\,\eta\,\nonumber\\&& 
  -4\,\sec \frac{\theta }{2}\,
   \tan \frac{\theta }{2} \,\eta \,{\sqrt{( 1 + \eta ) \,
       \Big( 1 + \eta\,{\sin^2 \frac{\theta }{2}}  \Big) }}\,
   {\Big(G_{E1}^\gamma+G_{E1}^{Z^{\cal V}_{a}}\Big)}\,G_{M}\, 
,\\ S_0\,A_{ed}^{ 2 1 \,+ }( 2 2 \,+ ) &=& 
  -2\,\,\sec\frac{\theta}{2}\,\sqrt{\eta\,(1+\eta\,\sin^2\frac{\theta}{2})}
  \Big( \Big(G_{C}^{Z^{\cal A}_{v}} 
   - \frac{2}{3}\,\eta \,G_{Q}^{Z^{\cal A}_{v}}\Big)\,
         G_{M} + \Big(G_{C}\,G_{M}^{Z^{\cal A}_{v}} - 
        \frac{2}{3}\,\eta \,G_{Q}\Big)\,G_{M}^{Z^{\cal A}_{v}} \
\Big) \, \nonumber\\&&
  -2\,\tan \frac{\theta }{2}\,{\sqrt{\eta \,( 1 + \eta ) }}\,
   \Big(G_{E1}^\gamma+G_{E1}^{Z^{\cal V}_{a}}\Big)\,\Big(  G_{C}   - 
     \frac{2}{3} \,\eta\,G_{Q} \Big) \, 
,\\ S_0\,A_{ed}^{ 2 2 \,+ }( 2 2 \,+ ) &=& 
   4\,\Big( G_{C}^{Z^{\cal A}_{v}} - 
        \frac{2}{3}\,\eta \,G_{Q}^{Z^{\cal A}_{v}} \Big) \,
      \Big( G_{C} - \frac{2}{3}\,\eta \,G_{Q} \Big) \,. 
\eeqa

(C) Asymmetries for $T$-violating contributions:

(i) Scalar asymmetries:
\beqa
 S_0\,A_{d}^{ 0 0 \,+ }( 1 1 \,- ) &=& 
  \frac{4}{{\sqrt{3}}}\,
     \sec \frac{\theta }{2}\,{\eta }\,
     {\sqrt{\eta \,\Big( 1 + \eta\,{\sin^2 \frac{\theta }{2}} \Big) }}
    \,G_{E2}^\gamma\,G_{Q}\,,\\
\rule{0pt}{30pt}
 S_0\,A_{ed}^{ 0 0 \,+ }( 2 1 \,- ) &=& 
  \frac{4}{{\sqrt{3}}}\,
     \tan \frac{\theta }{2} \,{\sqrt{\eta \,( 1 + \eta ) }}\,
     G_{E2}^\gamma\,\Big( G_{C} + \frac{1}{3}\,\eta \,G_{Q} \Big)\,. 
\eeqa

(ii) Vector asymmetries:
\beqa
 S_0\,A_{d}^{ 1 0 \,+ }( 2 1 \,- ) &=& 
  -\sqrt{2}\,\sec \frac{\theta }{2}\,
       {\sqrt{{\eta }\,
           \Big( 1 + \eta\,{\sin^2 \frac{\theta }{2}}  \Big) }}\,
       {G_{E2}^\gamma}\,
       \Big( G_{C} - \frac{2}{3}\,\eta \,G_{Q} \Big) 
,\\ S_0\,A_{d}^{ 1 1 \,+ }( 2 1 \,- ) &=& 
  4\eta \,( 1 + \eta ) \,{\tan^2 \frac{\theta }{2}}\,
    G_{E2}^\gamma\,G_{M}  
,\\ S_0\,A_{d}^{ 1 1 \,+ }( 2 2 \,- ) &=& 
  2\,\sec \frac{\theta }{2}\,{\sqrt{{\eta }\,
       \Big( 1 + \eta\,{\sin^2 \frac{\theta }{2}}  \Big) }}\,
   G_{E2}^\gamma\, \Big( {G_{C}} - \frac{2}{3}\,\eta \,G_{Q} \Big)\,,\\
\rule{0pt}{30pt}
 S_0\,A_{ed}^{ 1 0 \,+ }( 1 1 \,- ) &=& 
  \sqrt{2}\,\tan \frac{\theta }{2}
  \,{\sqrt{\eta \,( 1 + \eta ) }}\,
  G_{E2}^\gamma\,\Big( G_{C} - \frac{2}{3}\,\eta \,G_{Q} \Big) 
,\\ S_0\,A_{ed}^{ 1 1 \,+ }( 1 1 \,- ) &=& 
  4\,\sec \frac{\theta }{2}\,\tan \frac{\theta }{2}\,{\eta }\,
   {\sqrt{( 1 + \eta ) \,
       \Big( 1 + \eta\,{\sin^2 \frac{\theta }{2}}  \Big) }}\,
   G_{E2}^\gamma\,G_{M} \,.
\eeqa

(iii) Tensor asymmetries:
\beqa
 S_0\,A_{ed}^{ 2 0 \,+ }( 2 1 \,- ) &=& 
  -\sqrt{\frac{2}{{3}}}\,
    \tan \frac{\theta }{2}\, {\sqrt{\eta \,( 1 + \eta ) }}\,
    G_{E2}^\gamma\,\Big( G_{C} +\frac{10}{3}\,\eta \,G_{Q} \Big) 
,\\ S_0\,A_{ed}^{ 2 1 \,+ }( 2 1 \,- ) &=& 
  -4\,\sec \frac{\theta }{2}\,\tan \frac{\theta }{2}\,{\eta }\,
   {\sqrt{( 1 + \eta ) \,
       \Big( 1 + \eta\,{\sin^2 \frac{\theta }{2}}  \Big) }}\,
   G_{E2}^\gamma\,G_{M} 
,\\ S_0\,A_{ed}^{ 2 1 \,+ }( 2 2 \,- ) &=& 
  - 2\,\tan \frac{\theta }{2} \,{\sqrt{\eta \,( 1 + \eta ) }}\,
   \Big( G_{C}- \frac{2\,\eta}{3} \,G_{Q} \Big)\,G_{E2}^\gamma \,.
\eeqa

\end{appendix}

\end{document}